\documentclass[]{aa}     
\usepackage{graphicx}
\usepackage{times}
\usepackage{natbib}
\bibpunct{(}{)}{;}{a}{}{,} 
\def\cm2{\,{\rm cm^{-2}}}

\def\13co{\,{\rm ^{13}CO}}
\def\h2{\,{\rm H$_{2}$}}
\def\hi{H\,{\sc i~}}

\def\apj{{\rm ApJ} }

\def\apjs{{\rm ApJS} }
\def\apjl{{\rm ApJL} }
\def\araa{{\rm ARAA} }

\begin{document}
 
\title{Diagnostics of irradiated gas in galaxy nuclei}
 
   \subtitle{I: A Far-ultraviolet and X-ray dominated region code}
 
\author{Rowin Meijerink
          \inst{1}
        and Marco Spaans
          \inst{2}
           }
 
\authorrunning{Rowin Meijerink \& Marco Spaans}

   \offprints{Rowin Meijerink}
 
  \institute{Sterrewacht Leiden, P.O. Box 9513, 2300 RA Leiden,
             The Netherlands
  \and       Kapteyn Astronomical Institute, P.O. Box 800, 9700 AV Groningen,  
             The Netherlands}

\date{Received ????; accepted ????}
 
\abstract{We present a far-ultraviolet (PDR) and an X-ray dominated
region (XDR) code. We include and discuss thermal and chemical
processes that pertain to irradiated gas. An elaborate chemical
network is used and a careful treatment of PAHs and H$_2$ formation,
destruction and excitation is included. For both codes we calculate
four depth-dependent models for different densities and radiation
fields, relevant to conditions in starburst galaxies and active
galactic nuclei. A detailed comparison between PDR and XDR physics is
made for total gas column densities between $\sim 10^{20}$ and $\sim
10^{25}$ cm$^{-2}$. We show cumulative line intensities for a number
of fine-structure lines (e.g., [CII], [OI], [CI], [SiII], [FeII]), as
well as cumulative column densities and column density ratios for a
number of species (e.g., CO/H$_2$, CO/C, HCO$^+$/HCN, HNC/HCN). The
comparison between the results for the PDRs and XDRs shows that column
density ratios are almost constant up to $N_{\rm H}=10^{22}\ {\rm
cm^{-2}}$ for XDRs, unlike those in PDRs. For example, CO/C in PDRs
changes over four orders of magnitude from the edge to $N_{\rm
H}=10^{22}\ {\rm cm^{-2}}$. The CO/C and CO/H$_2$ ratios are lower in
XDRs at low column densities and rise at $N_{\rm H}>10^{23}\ {\rm
cm^{-2}}$. At most column densities $N_{\rm H} > 10^{21.5}\ {\rm
cm^{-2}}$, HNC/HCN ratios are lower in XDRs too, but they show a more
moderate increase at higher $N_{\rm H}$.}

\maketitle
 
\section{Introduction}

Gas clouds in the inner kpc of many galaxies are exposed to intense
radiation, which can originate from an active galactic nucleus (AGN),
starburst regions or both. O and B stars dominate the radiation from
starbursts, which is mostly in the far-ultraviolet ($6.0 < \rm E <
13.6$~eV), turning cloud {\it surfaces} into Photon Dominated Regions
(PDRs, Tielens \& Hollenbach \citeyear{Tielens1985}).  Hard X-rays
($\rm E > 1$~keV) from black hole environments (AGN) penetrate deep
into cloud {\it volumes} creating X-ray Dominated Regions (XDRs,
Maloney et al. \citeyear{Maloney1996}).  For each X-ray energy there
is a characteristic depth where photon absorption occurs. So for
different spectral shapes, one has different thermal and chemical
structures through the cloud. Although one source can dominate over
the other energetically, (e.g., an AGN in NGC~1068 or a starburst in
NGC~253), the very different physics (surface vs. volume) require that
both should be considered simultaneously in every galaxy.

In PDRs and XDRs, the chemical structure and thermal balance are
completely determined by the radiation field. Therefore, PDRs and XDRs
are direct manifestations of the energy balance of interstellar gas
and their study allows one to determine how the ISM survives the
presence of stars and AGN (Tielens \& Hollenbach
\citeyear{Tielens1985}; Boland \& de Jong \citeyear{Boland1982}; van
Dishoeck \& Black \citeyear{vanDishoeck1988}; Le Bourlot et
al. \citeyear{leBourlot1993}; Wolfire et al. \citeyear{Wolfire1993};
Spaans et al. \citeyear{Spaans1994}; Sternberg \& Dalgarno
\citeyear{Sternberg1995}; Stoerrzer et al.\citeyear{Stoerzer1998};
Spaans \citeyear{Spaans1996}; Bertoldi \& Draine
\citeyear{Bertoldi1996}; Maloney et al. \citeyear{Maloney1996}; Lee et
al. \citeyear{Lee1996}; Kaufman et al. \citeyear{Kaufman1999}; Le
Petit et al. \citeyear{lePetit2002} and references therein).

PDRs and XDRs have become increasingly important as diagnostic tools
of astrophysical environments with the advent of infrared and
(sub-)millimetre telescopes. PDRs emit fine-structure lines of [CI]
609, [CII] 158 and [OI] 63 $\mu$m; rotational lines of CO;
ro-vibrational and pure rotational lines of H$_2$; many H$_2$O lines
as well as many broad mid-IR features associated with Polycylic
Aromatic Hydrocarbons (PAHs). In PDRs, the bulk of H$_2$ is converted
into atomic hydrogen at the edge and CO to neutral carbon into ionised
carbon. XDRs emit brightly in the [OI] 63, [CII] 158, [SiII] 35, and
the [FeII] 1.26, 1.64 $\mu$m lines as well as the 2 $\mu$m
ro-vibrational H$_2$ transitions. The abundance of neutral carbon in
XDRs is elevated compared to that in PDRs and the chemical transitions
from H to H$_2$ and C$^+$ to C to CO are smoother \citep{Maloney1996}.

In this paper, we compare a far-ultraviolet and X-ray dominated region
code. For the PDR and XDR, we discuss the cooling, heating and
chemical processes. Then we show four models with different radiation
fields and densities, for a semi-infinite slab geometry and
irradiation from one side without geometrical dilution. We conclude
with a comparison between the column densities, integrated line fluxes
and abundance ratios. We will apply this tool to the centres of nearby
active galaxies in subsequent papers. We would like to point out that
these codes can be used over a broad range of physical situations and
scales, e.g., young stellar objects, planetary nebulae or gas outflow
in galaxy clusters.

\section{The Photon Dominated Region model}

The global properties of PDRs are determined by a number of physical
processes:
\begin{enumerate}
\item[{\it i}.]{Heating through photo-electric emission by dust
grains and PAHs (c.f. Bakes \& Tielens \citeyear{Bakes1994},
Weingartner \& Draine \citeyear{Weingartner2001}).}
\item[{\it ii}] Heating by FUV pumping of H$_2$, followed by
collisional de-excitation (c.f. Hollenbach \& McKee
\citeyear{Hollenbach1979}).
\item[{\it iii}.]{Heating by cosmic rays (c.f. Field \citeyear{Field1969}).}
\item[{\it iv}.]{Fine-structure line cooling of [CI] 609, [CII] 158,
[OI] 146 and 63 $\mu$m (c.f. Tielens \& Hollenbach
\citeyear{Tielens1985}; Spaans et al. \citeyear{Spaans1994}).}
\item[{\it v.}]{Molecular line cooling by warm molecular gas
containing CO, H$_2$, H$_2$O, OH and CH (c.f. Neufeld et
al. \citeyear{Neufeld1995}; Spaans \& Silk \citeyear{Spaans2000}).}
\item[{\it vi}.]{Ion-molecule reactions driven by the ionisation
degree of $\sim 10^{-4}$ maintained by the ionisation of carbon in the
FUV (c.f. Black \& Dalgarno \citeyear{Black1977}; van Dishoeck \&
Black \citeyear{vanDishoeck1986}).}
\item[{\it vii}.]{The ionisation balance of atomic gas under the
influence of photo-ionisation reactions driven by FUV photons and
counteracting recombination and charge transfer reactions with metals
and particularly PAHs (c.f. Lepp \& Dalgarno \citeyear{Lepp1988};
Bakes \& Tielens \citeyear{Bakes1994}).}
\end{enumerate}

\noindent As one moves into a PDR the extinction along the line of
sight increases and the impinging radiation field is
attenuated. Consequently, there are two zones over which the chemical
composition of the PDR changes in a fundamental way. The first
fundamental change occurs at the very edge of the PDR as atomic
hydrogen is converted into H$_2$ because the Lyman and Werner
electronic bands that lead to dissociation of the H$_2$ molecule in
the FUV become optically thick (so-called self-shielding). Deeper into
the PDR, at about 3 mag of extinction, ionised carbon is quickly
converted into neutral form as the FUV flux decreases due to dust
absorption. C is subsequently transformed into CO, since the FUV field
is reduced by grain opacity, H$_2$ shielding and some CO
self-shielding.

The first few magnitudes of extinction of the PDR are usually referred
to as the radical region since many carbon hydrides and their ions,
e.g., CH, CH$^+$, CN, HCN, HCO$^+$ (and also CO$^+$), reach their peak
abundance there, caused by the presence of both C$^+$ and H$_2$ and
the high ($\sim 10^2 - 10^3$~K) temperatures. Ion-molecule reactions
take place that lead to the formation of a large number of different
molecular species. Many of the atoms and molecules in (the radical
region of) a PDR are collisionally excited at the ambient densities
and temperatures, and emit brightly in the mid-IR, FIR, millimetre and
sub-millimetre.

The global characteristics of any PDR are defined by a few key parameters:
\begin{enumerate}
\item[{\it i}.]{The strength of the impinging radiation field, $G_0$ or
$I_{UV}$, in units of the \citet{Habing1969} or \citet{Draine1978}
radiation field, respectively, determines the total available
radiative flux at the edge of the PDR.}
\item[{\it ii}.]{The temperature and the ambient hydrogen density,
$n_{\rm H} = n({\rm H}) + 2n({\rm H_2})$, sets to a large extent the
pace of the chemical reactions and the excitation rates of the
coolants.}
\item[{\it iii}.]{The metallicity $Z$, in units of the solar value
$Z_\odot$, constrains the total abundances possible for carbon- and
oxygen-bearing species and hence influences the chemical and thermal
structure.}
\item[{\it iv}.]{The spectral shape of the impinging radiation field,
parameterised by the colour temperature $T_{\rm eff}$ for black bodies
or the frequency slope for power laws, fixes the distribution of
photon flux over energy.}
\end{enumerate}


\noindent The details about the thermal and chemical processes we use
in the code are discussed in the appendices. In the rest of the paper
we use $G_0$, the Habing flux, as the normalisation in which we
express the incident FUV radiation field, where $G_0=1$ corresponds to
a flux of $1.6\times 10^{-3}\ {\rm erg\ cm^{-2}\ s^{-1}}$.




\section{PDR test models}\label{testmodels}

In this section, we discuss the results for four PDR models in which we
have varied the radiation field $G_0$ and the density $n_{\rm H}$. The
models are for a semi-infinite slab geometry, but the code also allows
for two-sided slab geometries.  The adopted model parameters are
listed in Table \ref{models}. Models 2 and 4 will be shown in a
paper by R\"ollig et al (in prep.), where they are used to compare 12
different PDR codes that are commonly used. The parameters are listed
in Table \ref{models}. These values are typical for the high density,
strong radiation field conditions we want to investigate in, e.g., a
starburst.

\begin{table}[!h]
\caption[]{Adopted model parameters}
\begin{flushleft}
\begin{tabular}{cccc}
\hline
\hline
\noalign{\smallskip}
Model & $G_0$ & $F_{FUV}$ & $n_{\rm H}$ \\
      &       & [$\rm erg\ cm^{-2} \ s^{-1}$] & [$\rm cm^{-3}$]  \\
\noalign{\smallskip}
\hline
\noalign{\smallskip}
1     & $10^3$ & 1.6 & $10^3$ \\
2     & $10^5$ & 160 & $10^3$ \\
3     & $10^3$ & 1.6 & $10^{5.5}$ \\
4     & $10^5$ & 160 & $10^{5.5}$ \\
\noalign{\smallskip}
\hline
\noalign{\smallskip}
$\delta v_{\rm d}$ (km s$^{-1}$)& $2.7$ & &\\
$\delta_{\rm d}$                & $1.0$ & &\\
\noalign{\smallskip}
\hline
\hline
\end{tabular}
\end{flushleft}
\label{models}
\end{table}

\noindent The fixed gas-phase and total abundances we use are given in
Table \ref{XDRabundances}. The total abundances are the average values
of \citet{Asplund2004} and \citet{Jenkins2004}. To calculate the
gas-phase abundances, we use the depletion factors calculated by
\citet{Jenkins2004}. 

\begin{table}[!h]
\caption[]{Abundances}
\begin{flushleft}
\begin{tabular}{llll}
\hline
\hline
\noalign{\smallskip}
Species & $\mathcal{A}_i({\rm gas})$ & $\delta_i$ & $\mathcal{A}_i({\rm total})$\\
\noalign{\smallskip}
\hline
\noalign{\smallskip}
He$^{a,c}$   & $8.5 \times 10^{-2} $ & 1.0   &  $8.5 \times 10^{-2} $\\
C        & $1.4 \times 10^{-4} $ & 0.6   &  $2.5 \times 10^{-4} $\\
N        & $5.2 \times 10^{-5} $ & 0.7   &  $7.2 \times 10^{-5} $\\
O        & $3.4 \times 10^{-4} $ & 0.7   &  $4.7 \times 10^{-4} $\\
Si       & $1.7 \times 10^{-6} $ & 0.05  &  $3.4 \times 10^{-5} $\\
S        & $6.9 \times 10^{-6} $ & 0.5   &  $1.4 \times 10^{-5} $\\
Cl       & $5.4 \times 10^{-8} $ & 0.2   &  $2.4 \times 10^{-7} $\\
Fe       & $2.0 \times 10^{-7} $ & 0.007 &  $2.8 \times 10^{-5} $\\
\noalign{\smallskip}
\hline
\noalign{\smallskip}
P$^{b,c}$& $3.9 \times 10^{-8} $ & 0.1   &  $2.9 \times 10^{-7} $ \\
Na       & $5.9 \times 10^{-7} $ & 0.4   &  $1.5 \times 10^{-6} $\\
Mg       & $2.7 \times 10^{-7} $ & 0.08  &  $3.4 \times 10^{-6} $\\
\noalign{\smallskip}
\hline
\noalign{\smallskip}
Ne$^c$   &                       &       &  $6.9 \times 10^{-5} $\\
Al       &                       &       &  $2.3 \times 10^{-6} $\\
Ar       &                       &       &  $1.5 \times 10^{-6} $\\
Ca       &                       &       &  $2.0 \times 10^{-6} $\\
Cr       &                       &       &  $4.4 \times 10^{-7} $\\
Ni       &                       &       &  $1.7 \times 10^{-6} $\\
\noalign{\smallskip}
\hline
\hline
\end{tabular}
\end{flushleft}
Notes to Table \ref{XDRabundances}:\\
$^a$: Present in both PDR and XDR chemical network\\
$^b$: Present in PDR chemical network \\
$^c$: Used to calculate $\sigma_{\rm pa}$ for XDR\\
\label{XDRabundances}
\end{table}

\subsection{Heating and cooling}

For both radiation fields and densities, the dominant heating source
to a column density $N_{\rm H} \approx 10^{22}\ {\rm cm^{-2}}$ is
photo-electric emission from grains. In the moderately low density,
low radiation-field Model 1, viscous heating is about equally
important in this range. For Model 2 where the radiation field is
increased, it contributes somewhat more than 10 percent. For the low
radiation-field, high density Model 3, carbon ionisation is the second
important heating source. In Model 4, where the radiation field is
increased compared to Model 3, this is H$_2$ pumping. At high column
densities ($N_{\rm H} > 22.5\ {\rm cm^{-2}}$), [OI] 63 $\mu$m
absorption and gas-grain heating are important. For the low density
PDRs Model 1 and 2, only [OI] 63 $\mu$m dominates. When the density is
increased in Models 3 and 4, gas-grain heating is equally important if
not dominant. Other heating processes contribute less than 10 percent,
but are sometimes important in determining the thermal balance.

In all models [OI] 63 $\mu$m cooling dominates to $N_{\rm H} =
10^{21.5}\ {\rm cm^{-2}}$. In the low density PDRs, [CII] 158 $\mu$m
cooling contributes more than ten percent of the cooling in this
range, where at high densities, gas-grain cooling is the second most
important coolant. In the high density, high radiation-field Model 4,
this contribution can be almost forty percent. Deeper into the cloud,
[CI] 610~$\mu$m and CO line cooling become important. H$_2$ line
cooling can contribute up to 10 percent to the total cooling rate at
some point, but is always a minor coolant.

\subsection{Chemical and thermal structure}

The H $\rightarrow$ H$_2$ and C$^+$ $\rightarrow$ C $\rightarrow$ CO
transitions are quite sharp. Their actual location greatly varies,
since this is strongly dependent on density and radiation
field. Exposed to stronger radiation fields, the transitions occur
deeper into the cloud, since the photo-dissociation rates are
larger. At higher densities, the transitions occur closer to the
surface of the cloud, since the recombination rates scale as $n^2$.
For the same reason, the H$^+$ and O$^+$ 
fractional
abundances are systematically higher in the low density models. SiO
and CS are more abundant and formed closer to the surface in the high
density models, which is also the case for HCO$^+$, HCN, HNC and
C$_2$H. 

The edge temperatures (see Fig. \ref{Lines1}) are affected most by the
strength of the radiation field when the density is largest. At a
density of $n_{\rm H} = 10^{5.5}\ {\rm cm^{-3}}$, the difference is a
factor of thirty for an increase from $G_0=10^3$ to $G_0=10^5$. In the
low density case this is only a factor of two. Because of optical
depth effects, CO cooling is less effective at high column
densities. For this reason, temperatures rise again at $N_{\rm H}
\approx 10^{22}\ {\rm cm^{-2}}$ in the low density models.

\section{The X-ray Dominated Region model}

Unlike PDRs, XDRs are mostly heated by direct photo-ionisation of the
gas, which produces fast electrons that lose energy through collisions
with other electrons, as well as H and H$_2$. These fast electrons
collisionally excite H and H$_2$, which subsequently emit Lyman
$\alpha$ and Lyman-Werner band photons, respectively. These photons in
turn are capable of ionising atoms such as C and Si or ionise and
dissociate molecules such as H$_2$ and CO.

Compared to PDRs, the following processes play a role in XDRs
(c.f. Maloney et al. \citeyear{Maloney1996}), in part because of the
production of UV photons as described above:
\begin{enumerate}
\item[{\it i.}]{Photo-ionisation heating (i.e., Coulomb heating with thermal
electrons) dominates by a large factor over the heating through
photo-electric emission by dust grains and PAHs (c.f. Maloney et
al. \citeyear{Maloney1996}; Bakes \& Tielens \citeyear{Bakes1994}).}
\item[{\it ii.}]{Emission from meta-stable lines of [CI] 9823, 9850 \AA\ and [OI]
6300 \AA; fine-structure line cooling of [CII] 158 and [OI] 63 and 146
$\mu$m as well as Lyman $\alpha$ emission (c.f. Maloney et
al. \citeyear{Maloney1996}; Tielens \& Hollenbach
\citeyear{Tielens1985}; Spaans et al. \citeyear{Spaans1994}).}
\item[{\it iii.}]{Molecular line cooling by warm molecular gas containing CO,
H$_2$, H$_2$O and OH as well as gas-grain cooling where warm gas is
cooled at the surfaces of lower temperature dust grains (c.f. Neufeld
et al. \citeyear{Neufeld1995}; Spaans \& Silk \citeyear{Spaans2000}).}
\item[{\it iv.}]{Ion-molecule reactions driven by the ionisation degree of $\sim
10^{-4}$ maintained by the ionisation of carbon in the FUV (c.f. Black
\& Dalgarno \citeyear{Black1977}; van Dishoeck \& Black
\citeyear{vanDishoeck1986}).}
\item[{\it v.}]{The ionisation balance of atomic gas under the influence of
photo-ionisation reactions driven by X-ray photons and charge
transfer. Recombination of ions on grain surfaces is a major ionic
loss route at electron fractions less than $10^{-3}$ (c.f. Lepp \&
Dalgarno \citeyear{Lepp1988}; Bakes \& Tielens \citeyear{Bakes1994};
Maloney et al. \citeyear{Maloney1996}).}
\end{enumerate}

\noindent The global structure of any XDR is defined by a few key
parameters, the density $n_{\rm H}$ and the energy deposition rate
$H_X$ (see Appendix \ref{energy_dep}) per hydrogen atom. Because the heating in XDRs is driven by
photo-ionisation, the heating efficiency is close to unity as opposed
to that in PDRs where the photo-electric heating efficiency is of the
order of $0.3-1.0\%$ \citep{Maloney1996, Bakes1994}. Unlike PDRs, XDRs
are exposed to X-rays as well as FUV photons.

As one moves into the XDR, X-ray photons are attenuated due to atomic
electronic absorptions. The lowest energy photons are attenuated
strongest, which leads to a dependence of the X-ray heating and
ionisation rates at a given point on the slope of the X-ray
spectrum. We assume, for energies between 0.1 and 10 keV, that the
primary ionisation rate of hydrogen is negligible compared to the
secondary ionisation rate and that Auger electrons contribute an
energy that is equal to the photo-ionisation threshold energy
\citep{Void1991}.

The treatment is described in the appendics and follows, in part, the
unpublished and little known work by \citet{Yan1997}. Also, we extend
the work of \citet{Maloney1996} in terms of depth dependence, H$_2$
excitation and extent of the chemical network.




\section{XDR test models}

In this section, we consider four models with the same energy inputs
and densities as the PDRs in Table \ref{models}. The spectral energy
distribution is of the form $\exp(-E/1\ \rm keV)$. The energy is
emitted between 1 and 10 keV and $F_{FUV}$ should be replaced by $F_X$
in Table \ref{models}. This spectral shape and spectral range can be
changed depending on the application. We take the parameters for the 1
keV electron to determine the electron energy deposition, since these
parameters do not change for higher energies.  When the spectral
energy distribution is shifted towards higher energies, the X-rays
will dominate a larger volume, since the absorption cross sections are
smaller for higher energies. $H_X/n$ is the most important parameter
for the chemical and thermal balance, where $H_X$ is the energy
deposition rate per hydrogen nucleus. The abundances used are given in
Table \ref{XDRabundances}. The elements H, He, C, N, O, Si, S, Cl and
Fe are included in the chemical network. The other elements listed are
only used to calculate the photoelectric absorption cross section,
$\sigma_{\rm pa}$.

\subsection{Heating and cooling}

\begin{figure*}[!ht]
\unitlength1cm
\begin{minipage}[b]{8.8cm}
\resizebox{8.5cm}{!}{\includegraphics*{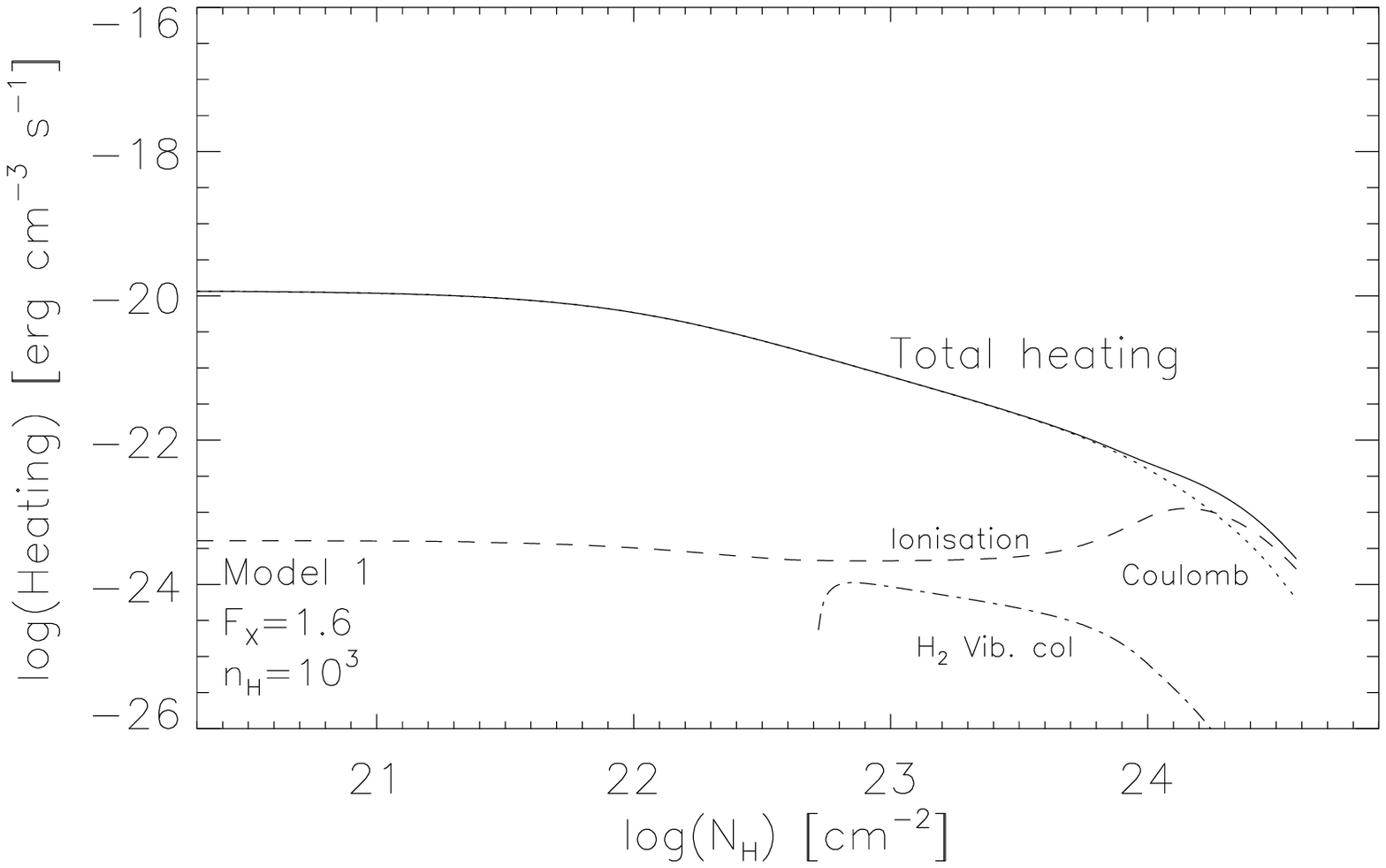}}
\end{minipage}
\hfill
\begin{minipage}[t]{8.8cm}
\resizebox{8.5cm}{!}{\includegraphics*{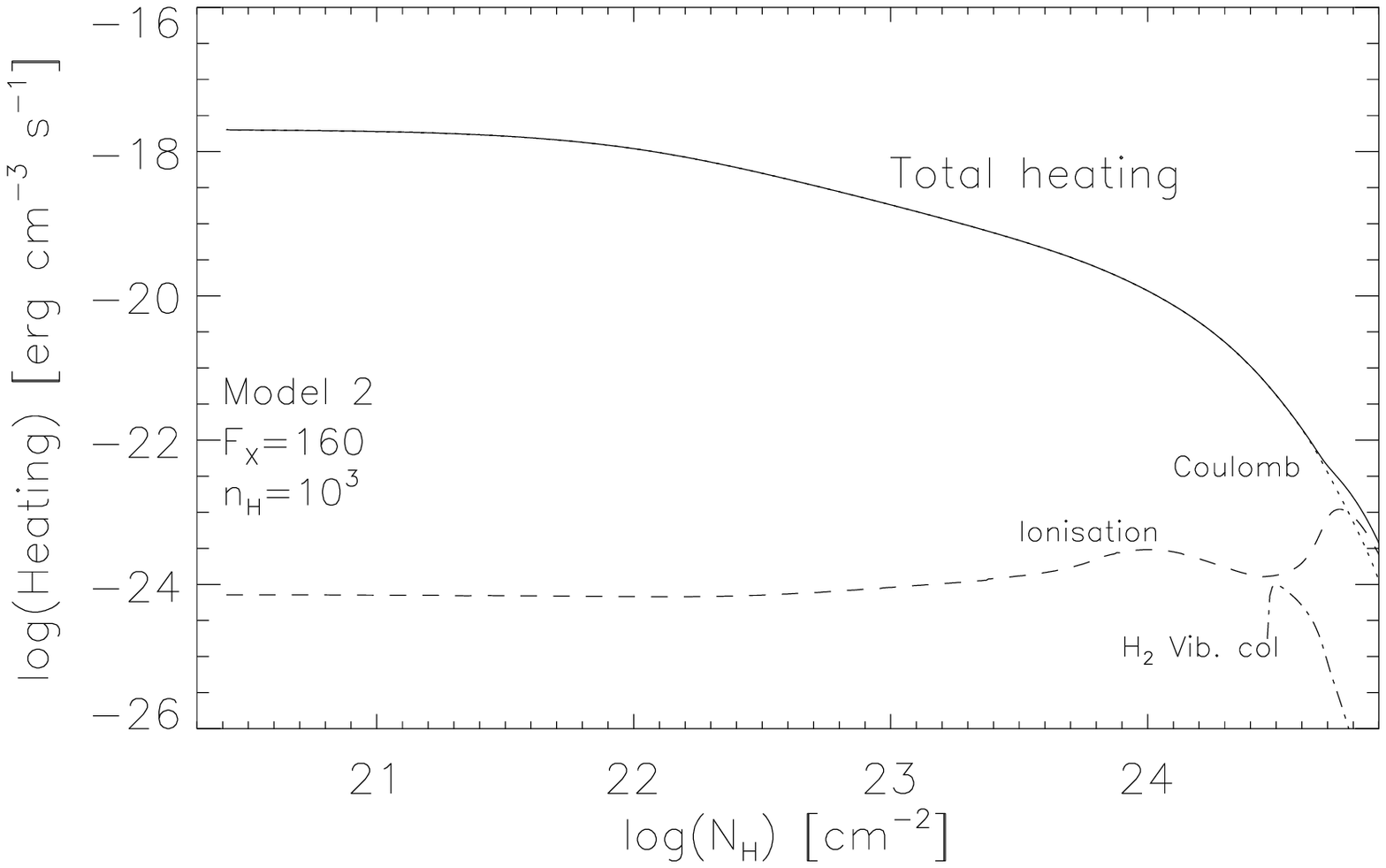}}
\end{minipage}
\hfill
\begin{minipage}[b]{8.8cm}
\resizebox{8.5cm}{!}{\includegraphics*{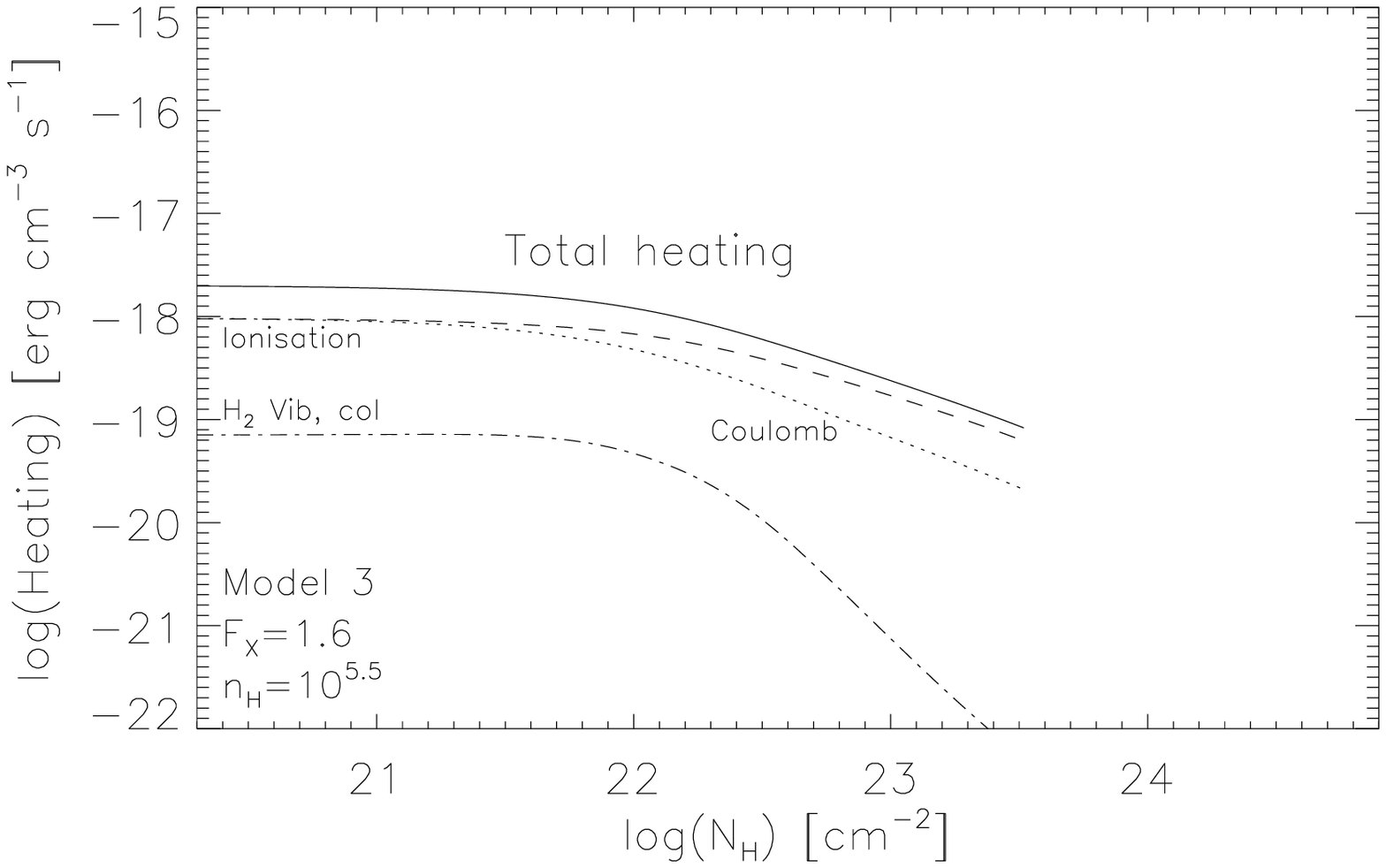}}
\end{minipage}
\hfill
\begin{minipage}[t]{8.8cm}
\resizebox{8.5cm}{!}{\includegraphics*{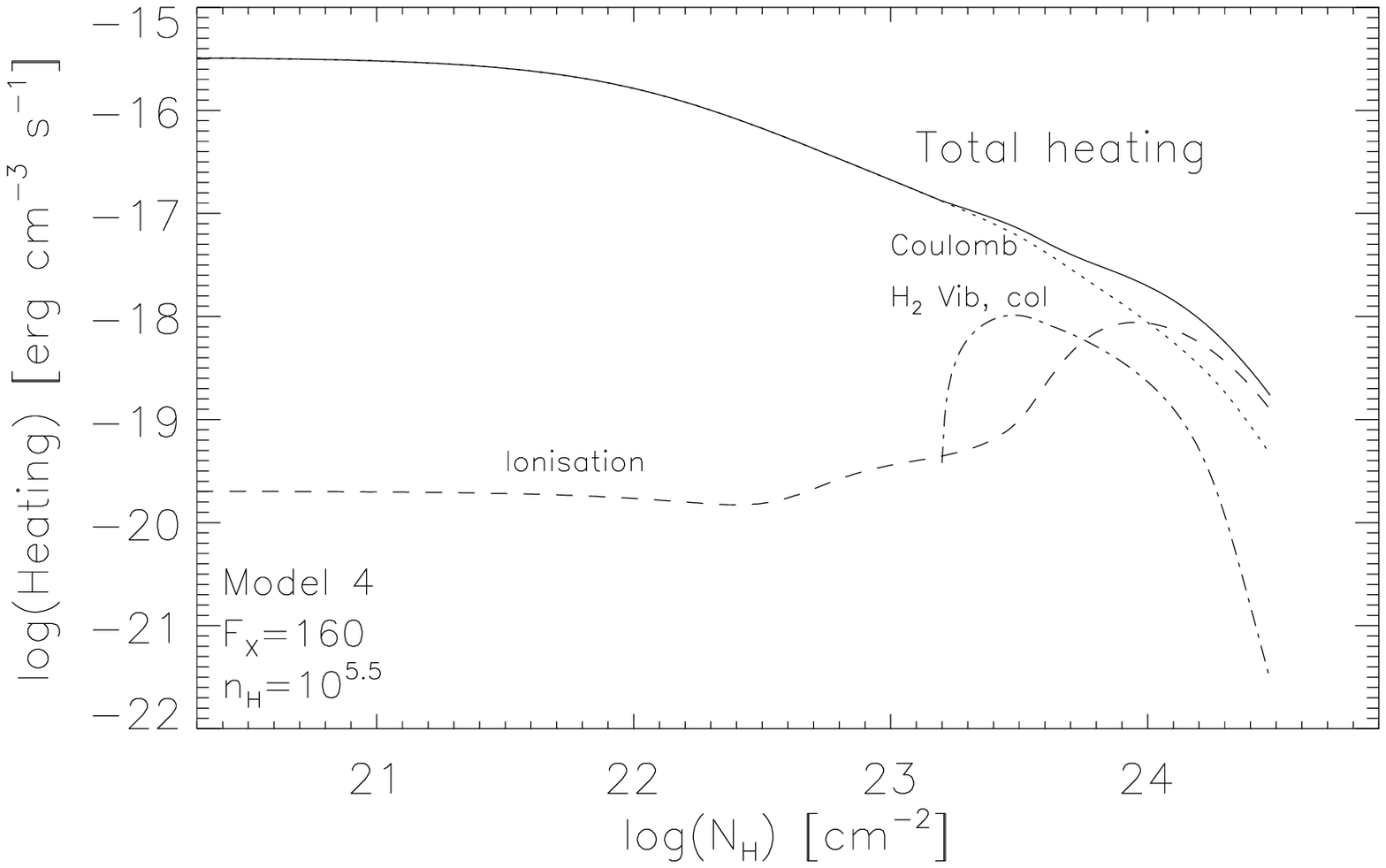}}
\end{minipage}
\caption[] {Important heating processes for Model 1 (top left), 2
(top right), 3 (bottom left) and 4 (bottom right).
}
\label{XDRheating}
\vspace{0.6cm}
\unitlength1cm
\begin{minipage}[b]{8.8cm}
\resizebox{8.5cm}{!}{\includegraphics*{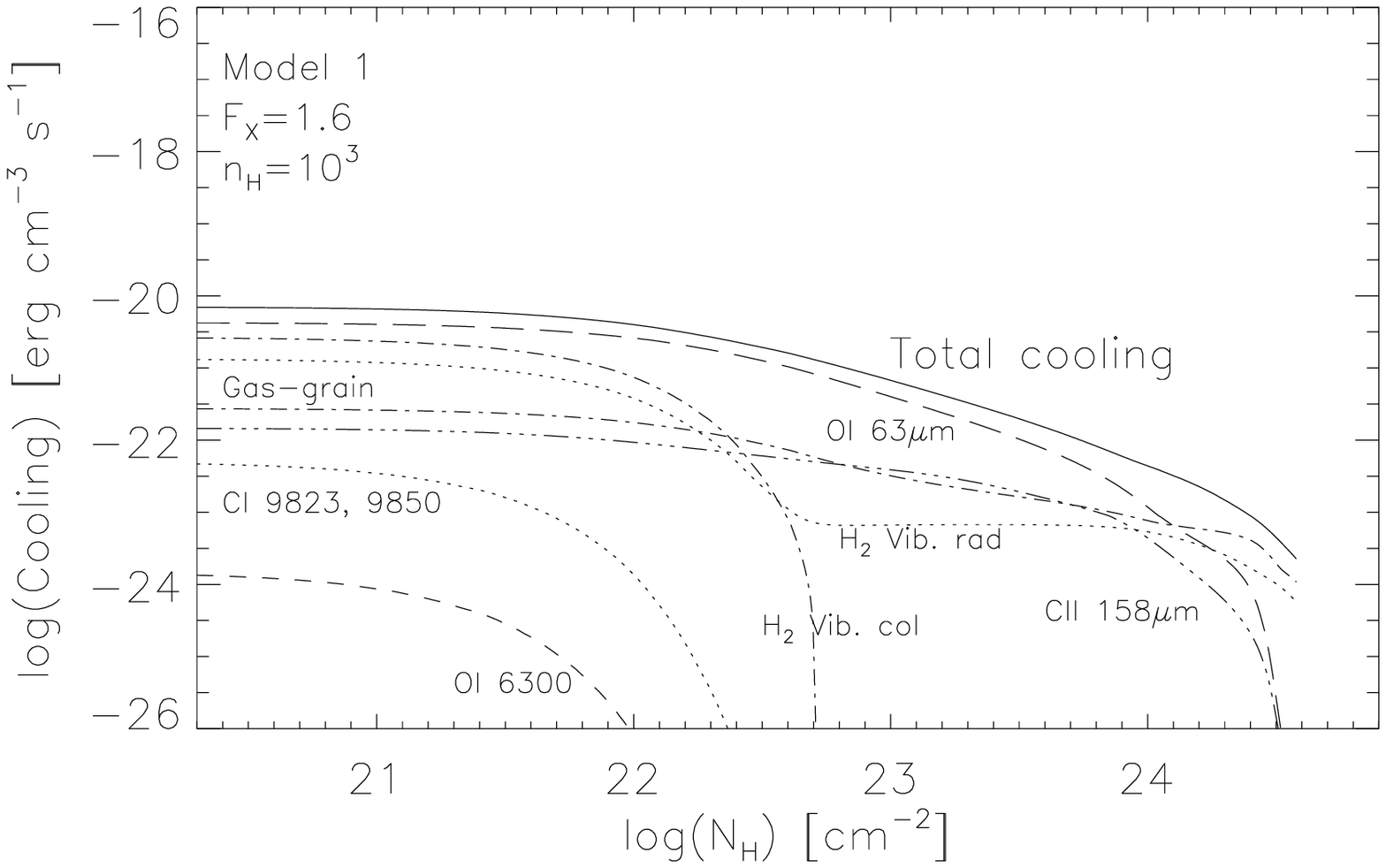}}
\end{minipage}
\hfill
\begin{minipage}[t]{8.8cm}
\resizebox{8.5cm}{!}{\includegraphics*{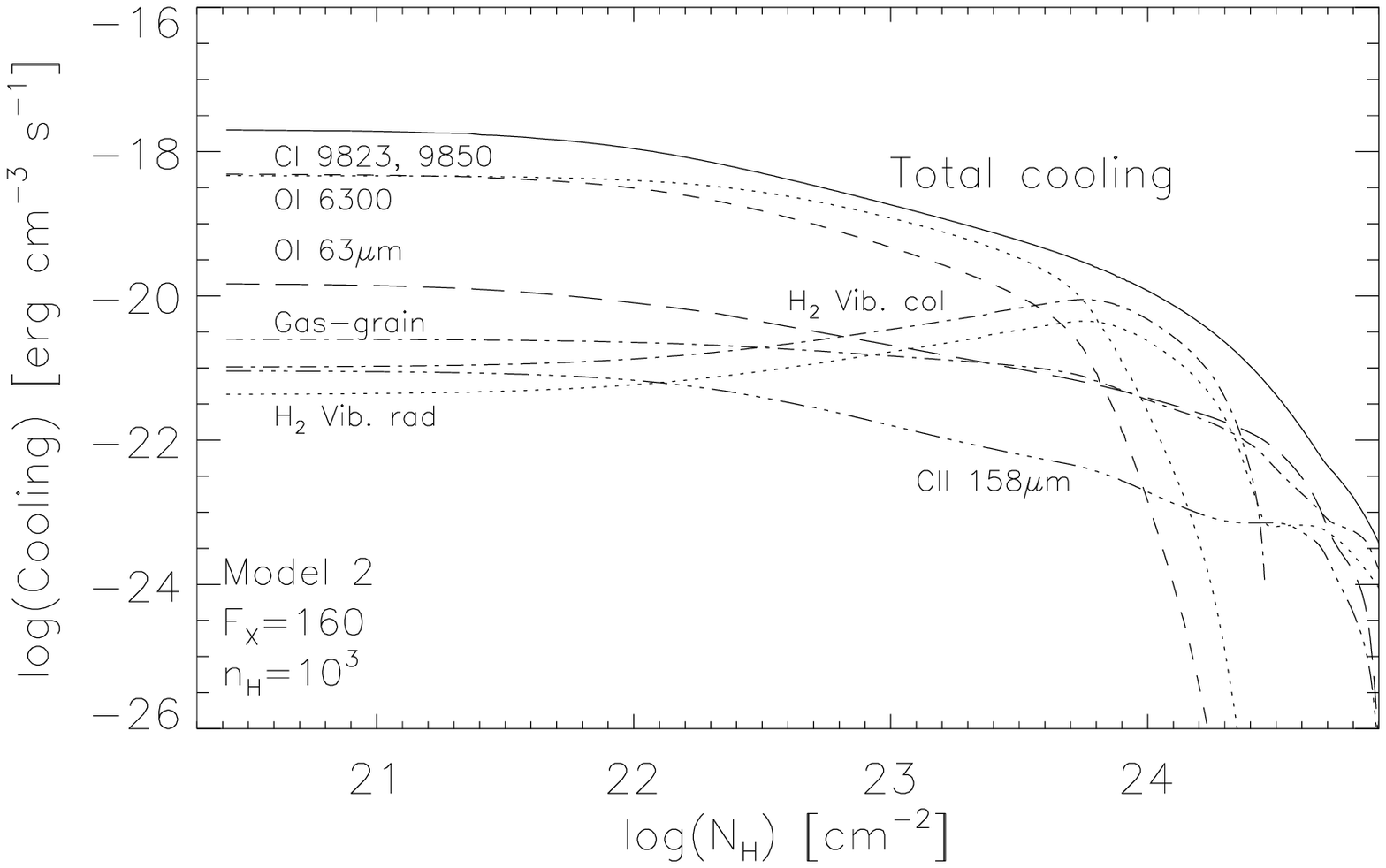}}
\end{minipage}
\hfill
\begin{minipage}[b]{8.8cm}
\resizebox{8.5cm}{!}{\includegraphics*{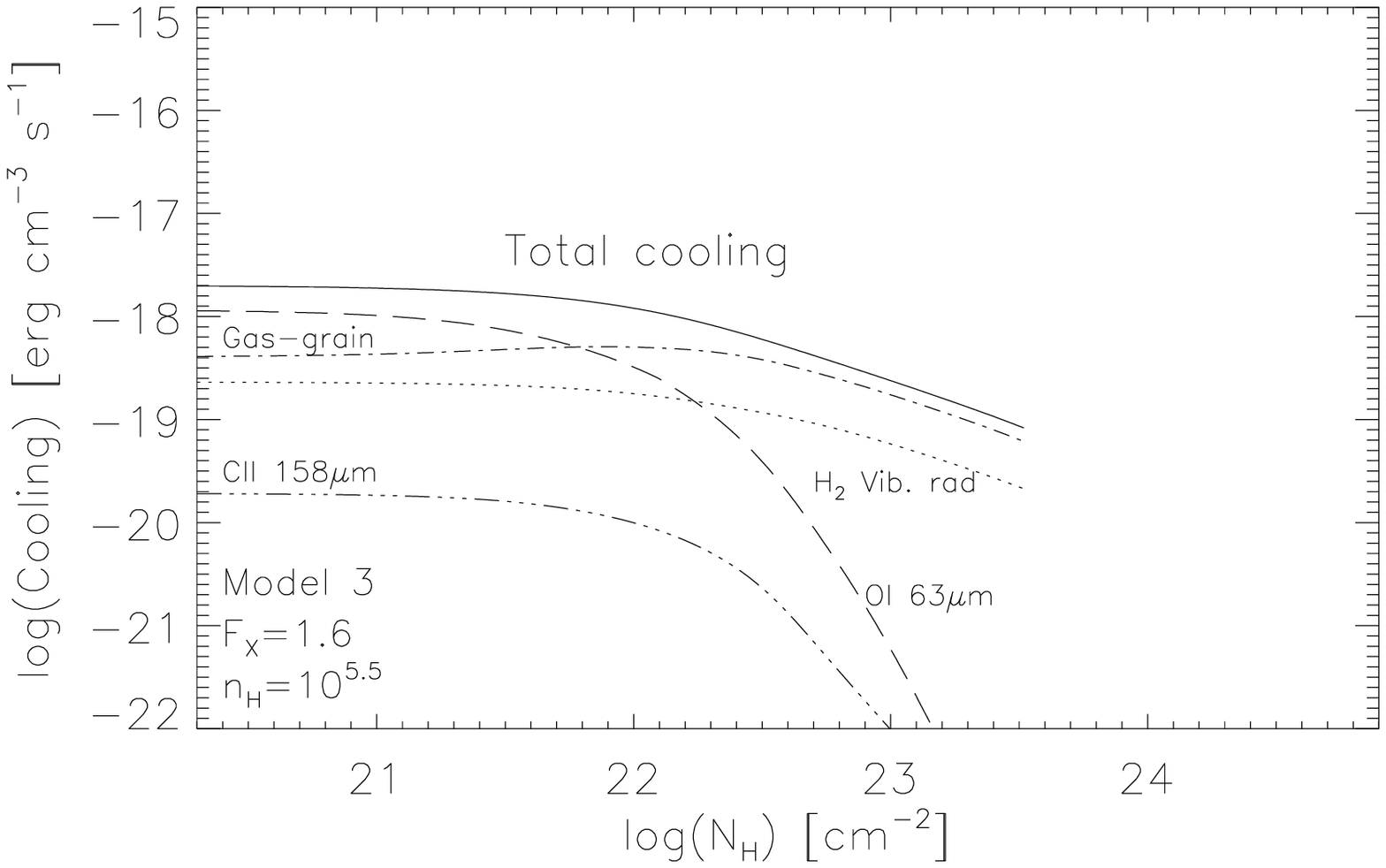}}
\end{minipage}
\hfill
\begin{minipage}[t]{8.8cm}
\resizebox{8.5cm}{!}{\includegraphics*{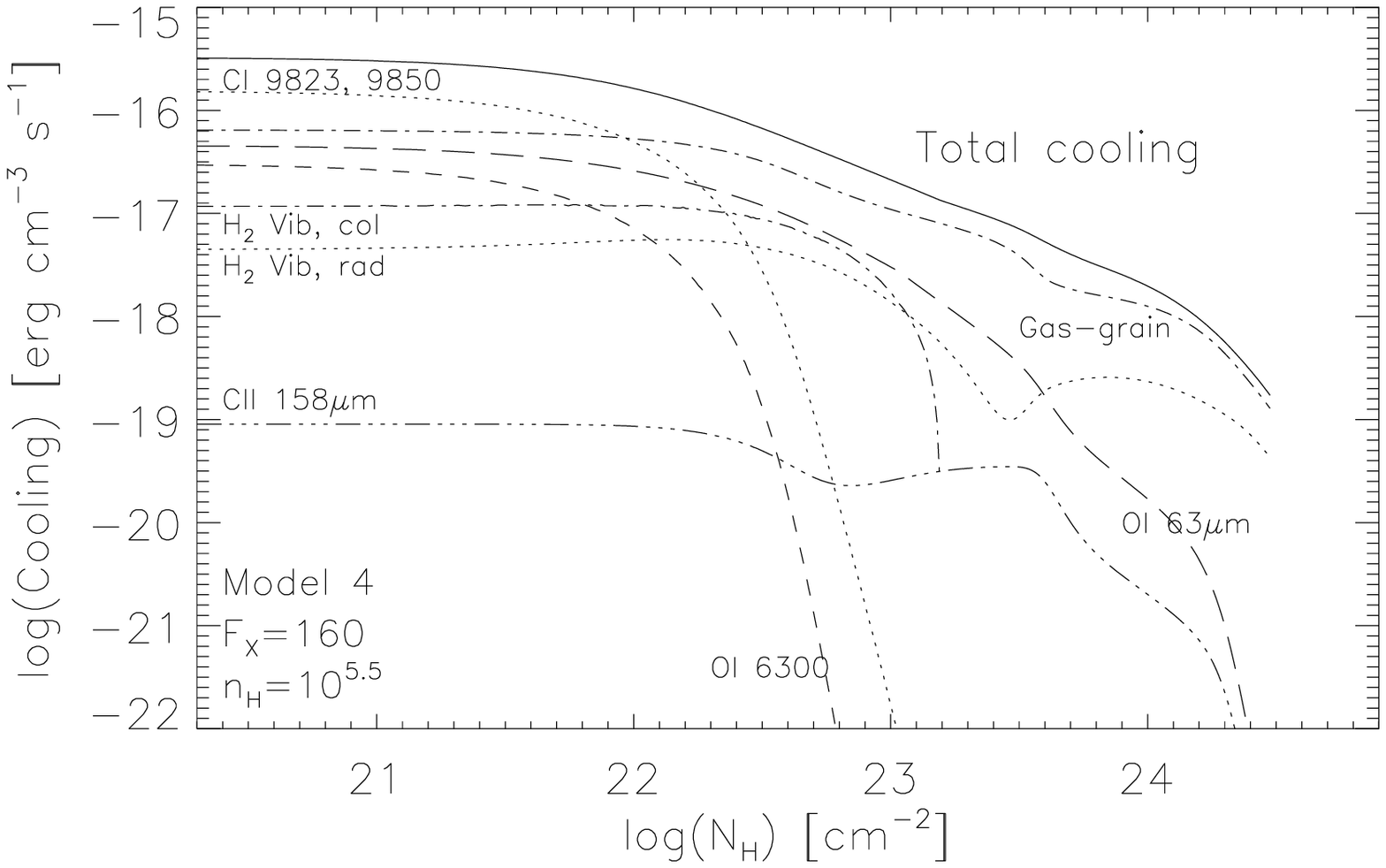}}
\end{minipage}
\caption[] {Important cooling processes for Model 1 (top left), 2
(top right), 3 (bottom left) and 4 (bottom right).
}
\label{XDRcooling}
\end{figure*}

In Fig. \ref{XDRheating}, the different heating sources are shown as a
function of the total hydrogen column density, $N_{\rm H}$. All
heating is done by X-rays, but the way it is transfered to the gas
depends on the ionisation fraction. When the gas is highly ionised,
$x_e \sim 0.1$, most ($\sim 70\%$) of the kinetic energy of the
non-thermal electrons goes into Coulomb heating, which is the case in
Models 1, 2 and 4 where $H_X/n$ is high to $N_{\rm H} > 10^{23}\ {\rm
cm^{-2}}$. For smaller ionisation fractions, $x_{\rm e} \sim 10^{-4}$,
ionisation heating as discussed in Sect. \ref{ion_heat} is important
or even dominant. In Model 3, ionisation heating and Coulomb heating
are equally important at $N_{\rm H} < 10^{21.8}\ {\rm cm^{-2}}$. In
all models ionisation heating dominates especially at high column
densities. When the excitation of H$_2$ is dominated by non-thermal
processes, collisional quenching of H$_2$ can heat the gas. Naively,
one would expect this dominance to occur where most of the X-rays are
absorbed, but for high energy deposition rates $H_X/n$, the
temperature is high and thermal collisions dominate the population of
the vibrational levels. Non-thermal excitation is dominant at low
temperature, i.e., low $H_X/n$.


In Fig. \ref{XDRcooling}, the important cooling processes are shown as
a function of total hydrogen column density, $N_{\rm H}$. At high
temperatures (see Fig. \ref{XDRchem1}), cooling by [CI] 9823, 9850
\AA\ and [OI] 6300 \AA\ metastable lines dominates, as is the case in
the models with high radiation fields, Models 2 and 4. At lower
temperatures, most of the cooling is provided by the fine-structure
line [OI] 63$\mu$m (90$\%$), e.g., at the edge in the low-radiation
field Models 1 and 3. In each model, gas-grain cooling dominates for
low $H_X/n$. In addition, specific cooling processes can be important
in special cases. H$_2$ vibrational cooling dominates at large depths
in Model 2, but in Models 1, 3 and 4 it contributes no more than
10$\%$. H$_2$ vibrational cooling is split into a radiative and a
collisional part. When the excitation of H$_2$ is dominated by
non-thermal electrons, the gas is heated by collisional de-excitation
of H$_2$.

\subsection{Thermal and chemical structure}

\begin{figure*}[!ht]
\unitlength1cm
\begin{minipage}[b]{8.8cm}
\resizebox{9.1cm}{!}{\includegraphics*{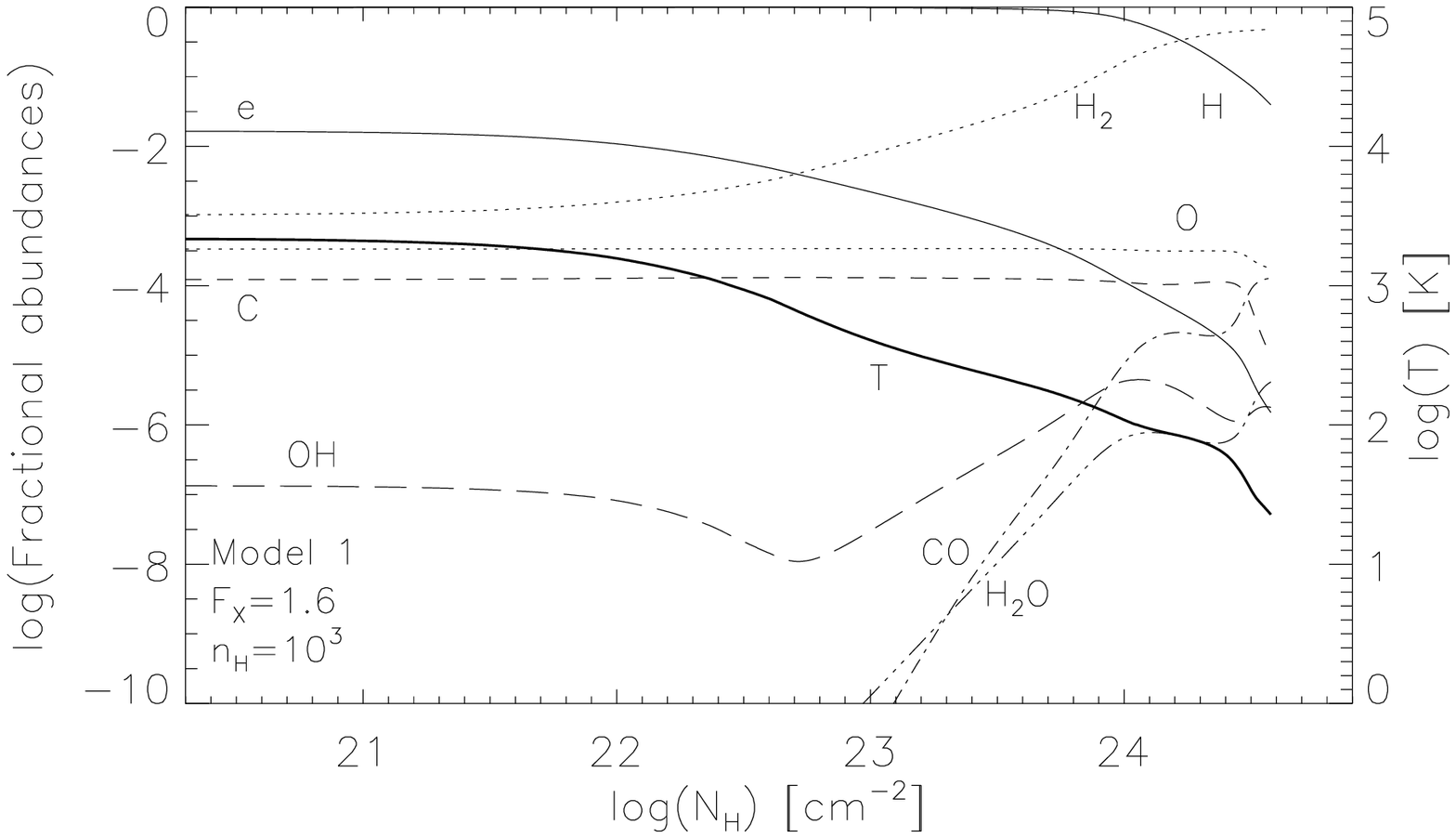}}
\end{minipage}
\hfill
\begin{minipage}[t]{8.8cm}
\resizebox{9.1cm}{!}{\includegraphics*{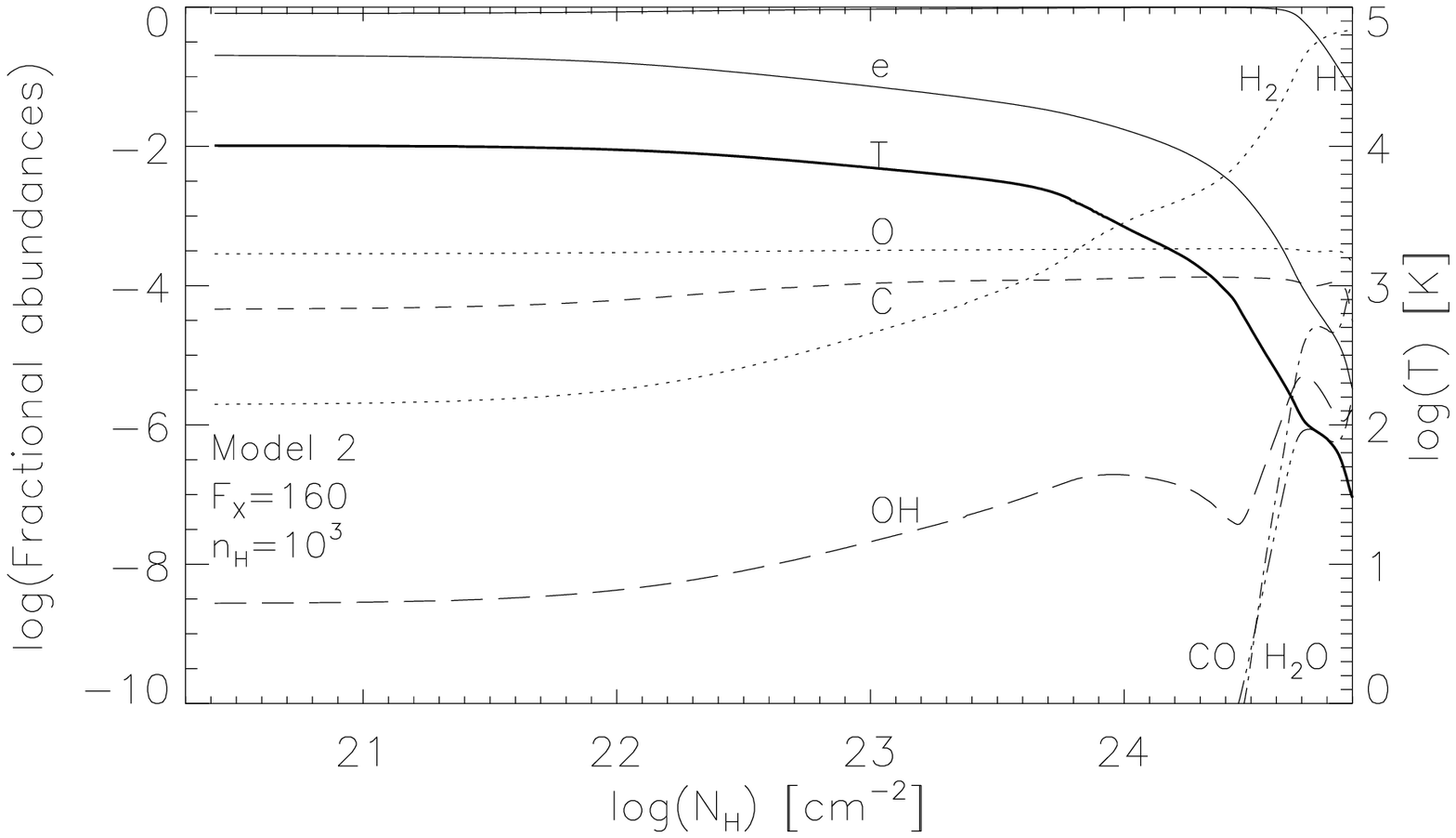}}
\end{minipage}
\hfill
\begin{minipage}[b]{8.8cm}
\resizebox{9.1cm}{!}{\includegraphics*{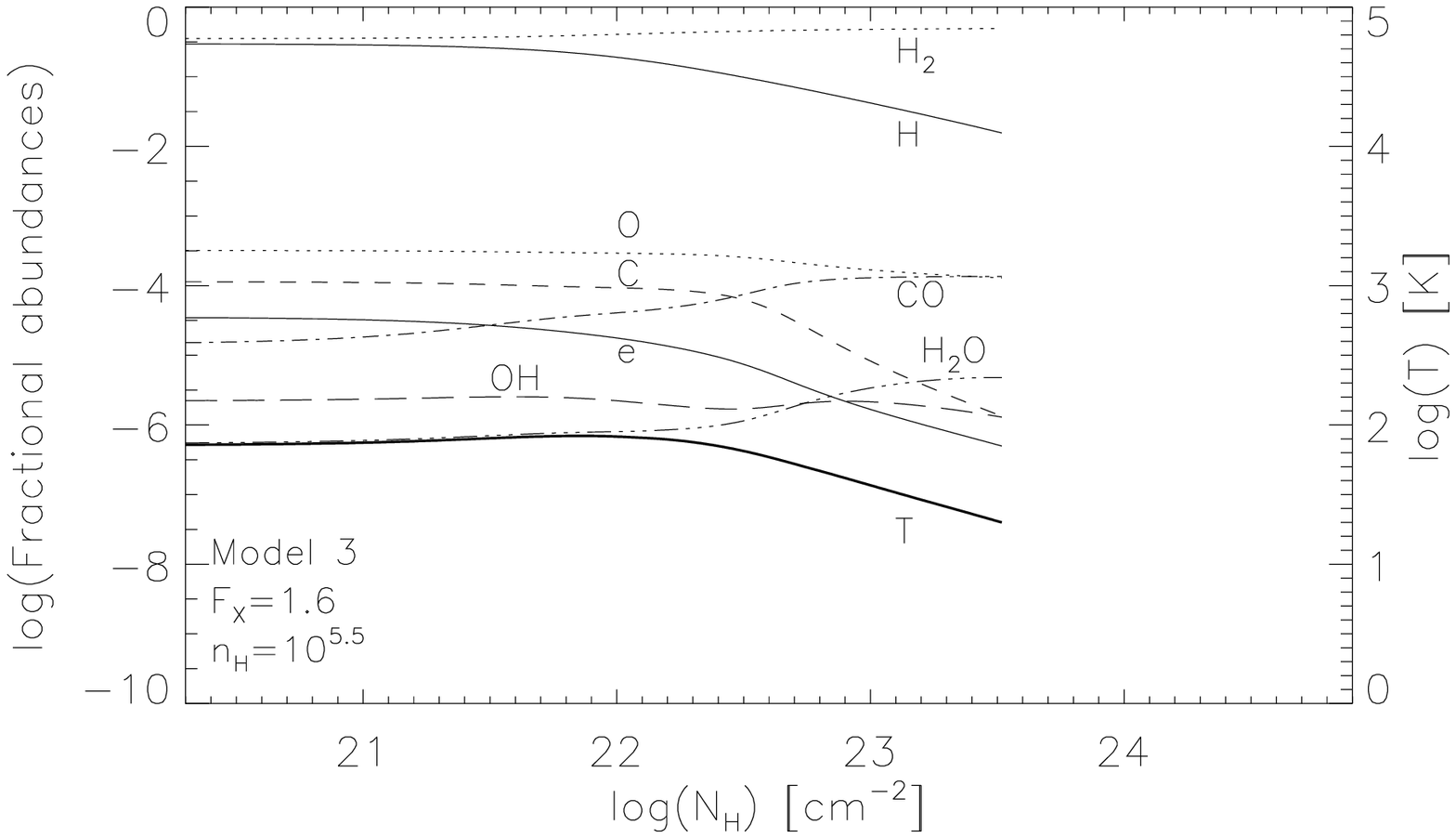}}
\end{minipage}
\hfill
\begin{minipage}[t]{8.8cm}
\resizebox{9.1cm}{!}{\includegraphics*{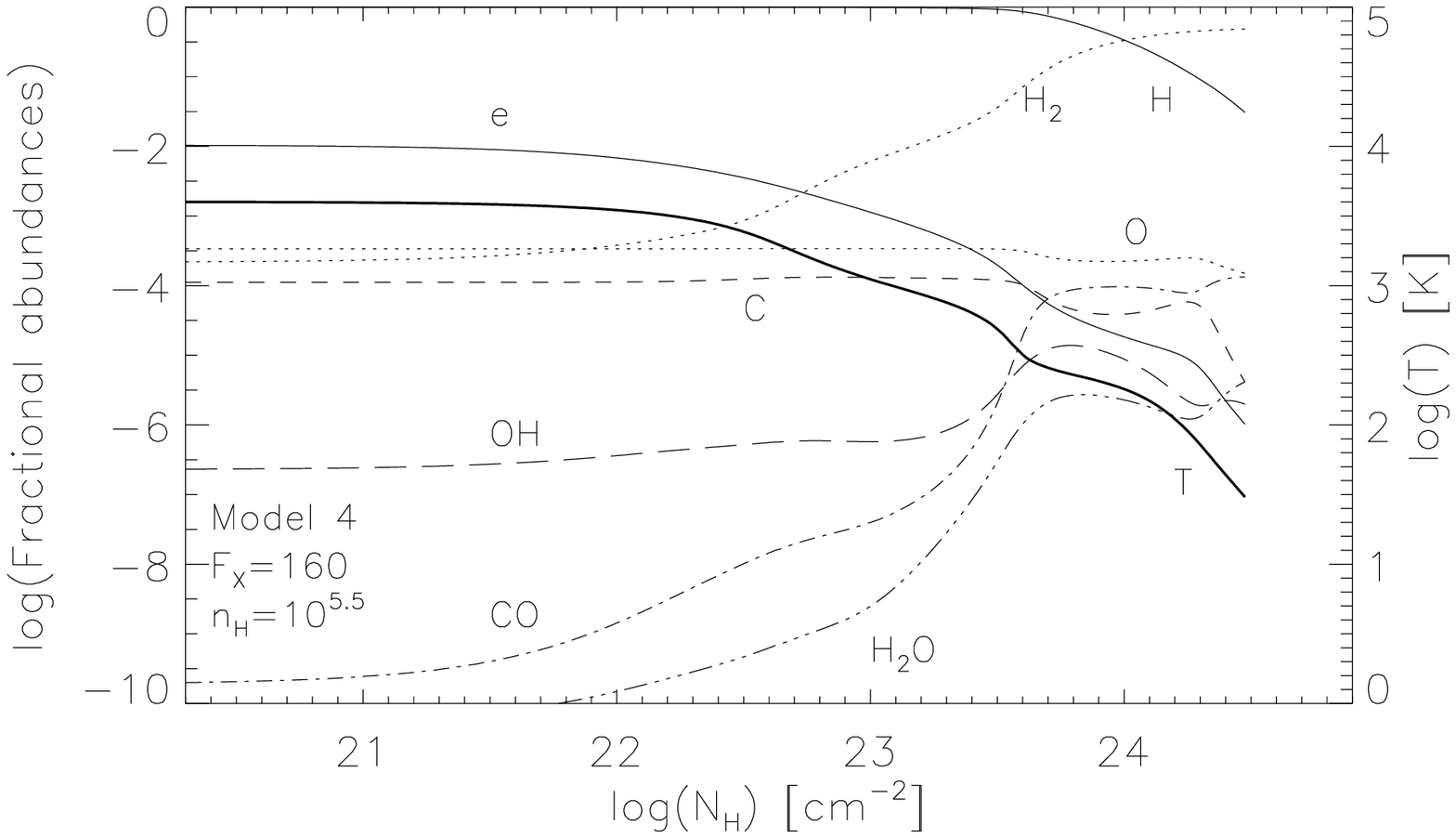}}
\end{minipage}
\caption[] {Fractional abundances and temperature for Model 1 (top
left), 2 (top right), 3 (bottom left) and 4 (bottom right).}
\label{XDRchem1}
\vspace{0.6cm}
\unitlength1cm
\begin{minipage}[b]{8.8cm}
\resizebox{8.5cm}{!}{\includegraphics*{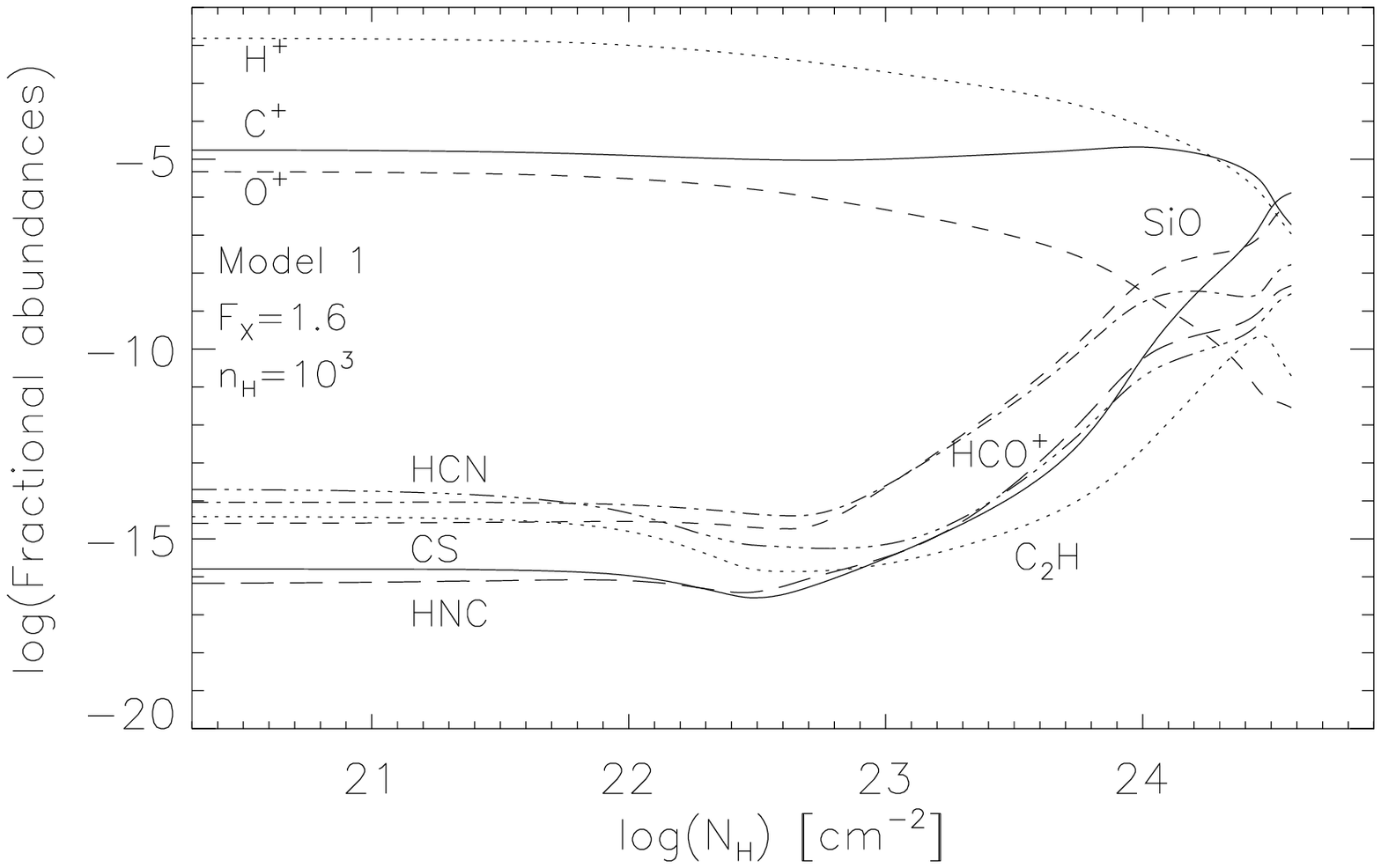}}
\end{minipage}
\hfill
\begin{minipage}[t]{8.8cm}
\resizebox{8.5cm}{!}{\includegraphics*{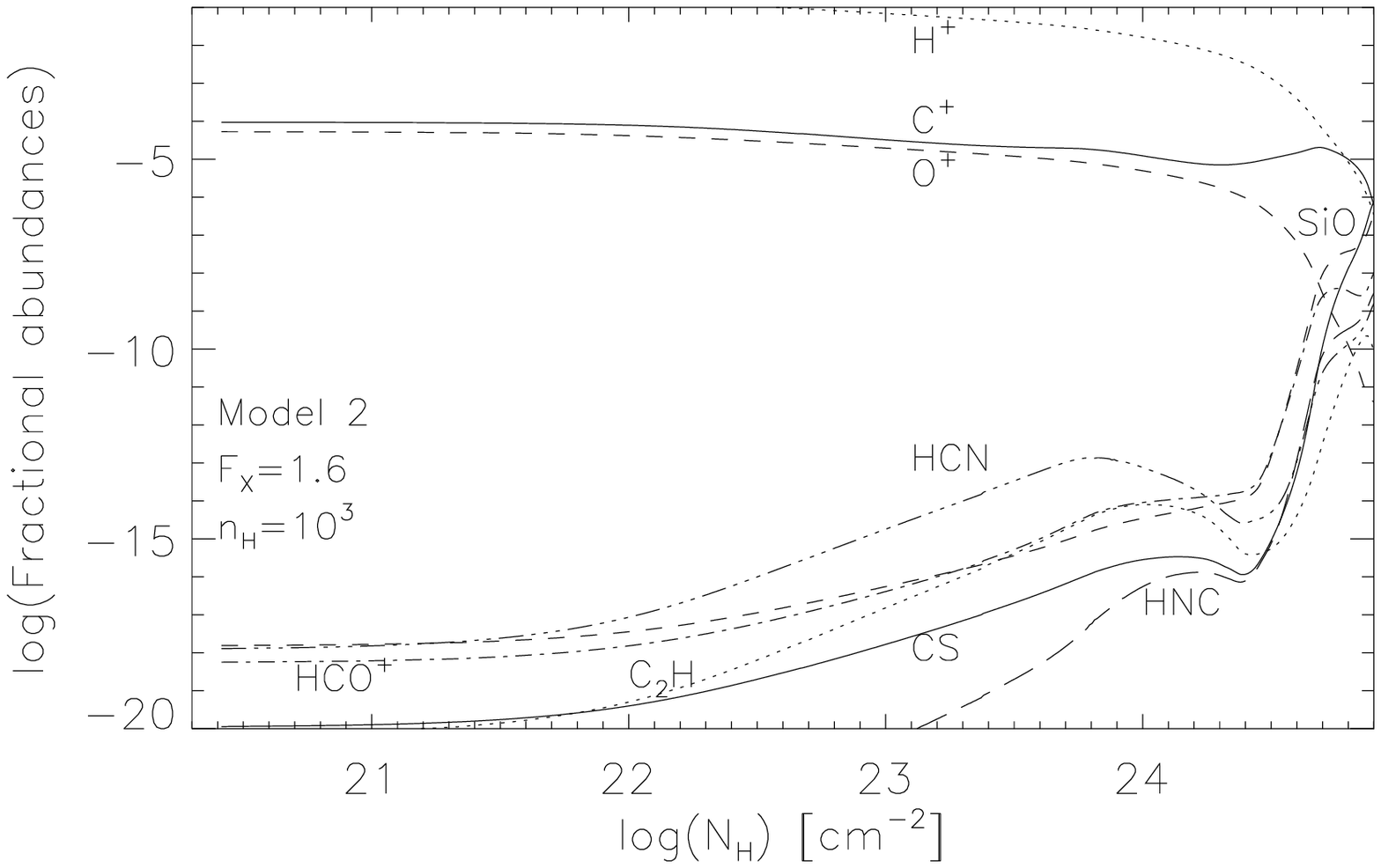}}
\end{minipage}
\hfill
\begin{minipage}[b]{8.8cm}
\resizebox{8.5cm}{!}{\includegraphics*{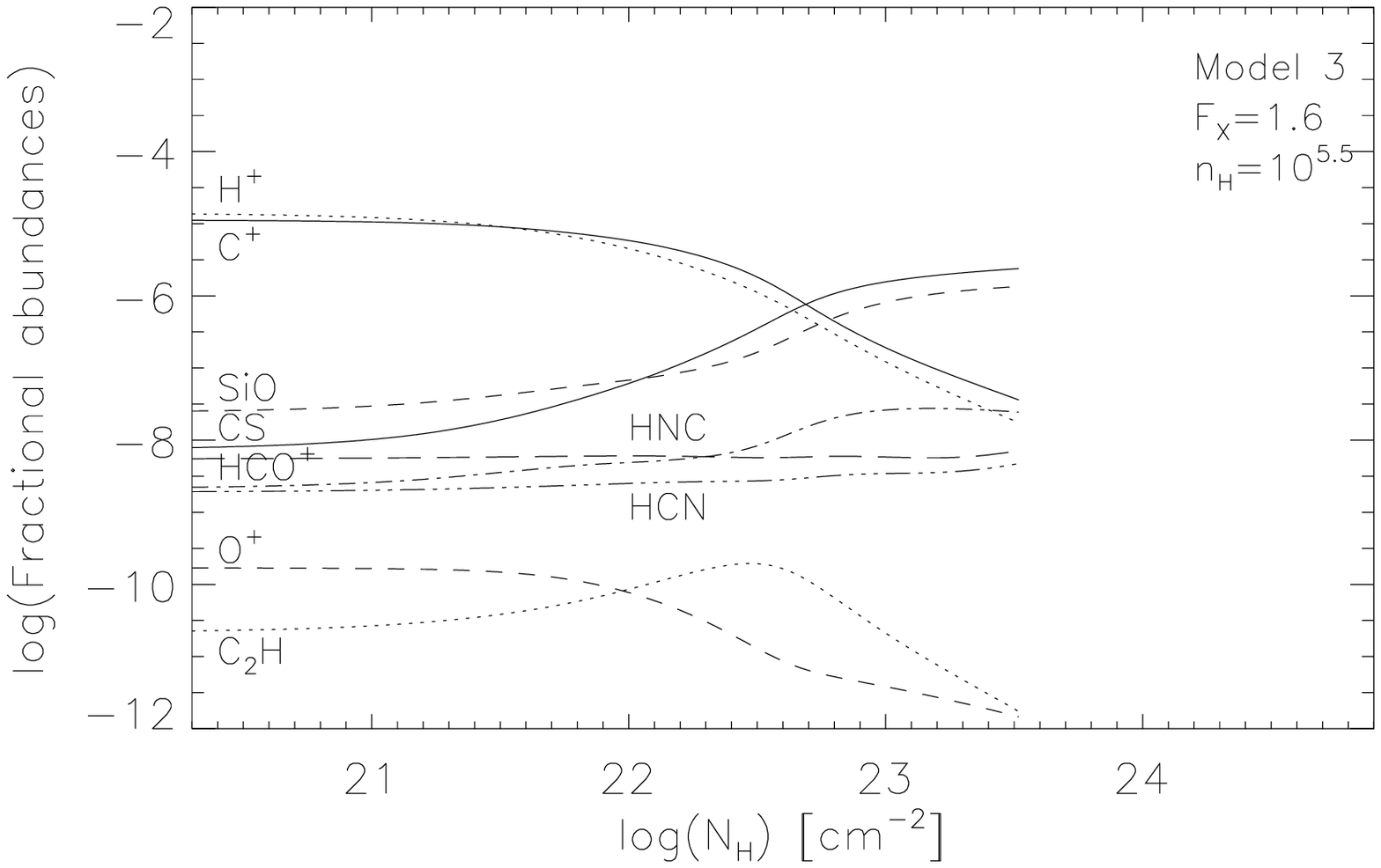}}
\end{minipage}
\hfill
\begin{minipage}[t]{8.8cm}
\resizebox{8.5cm}{!}{\includegraphics*{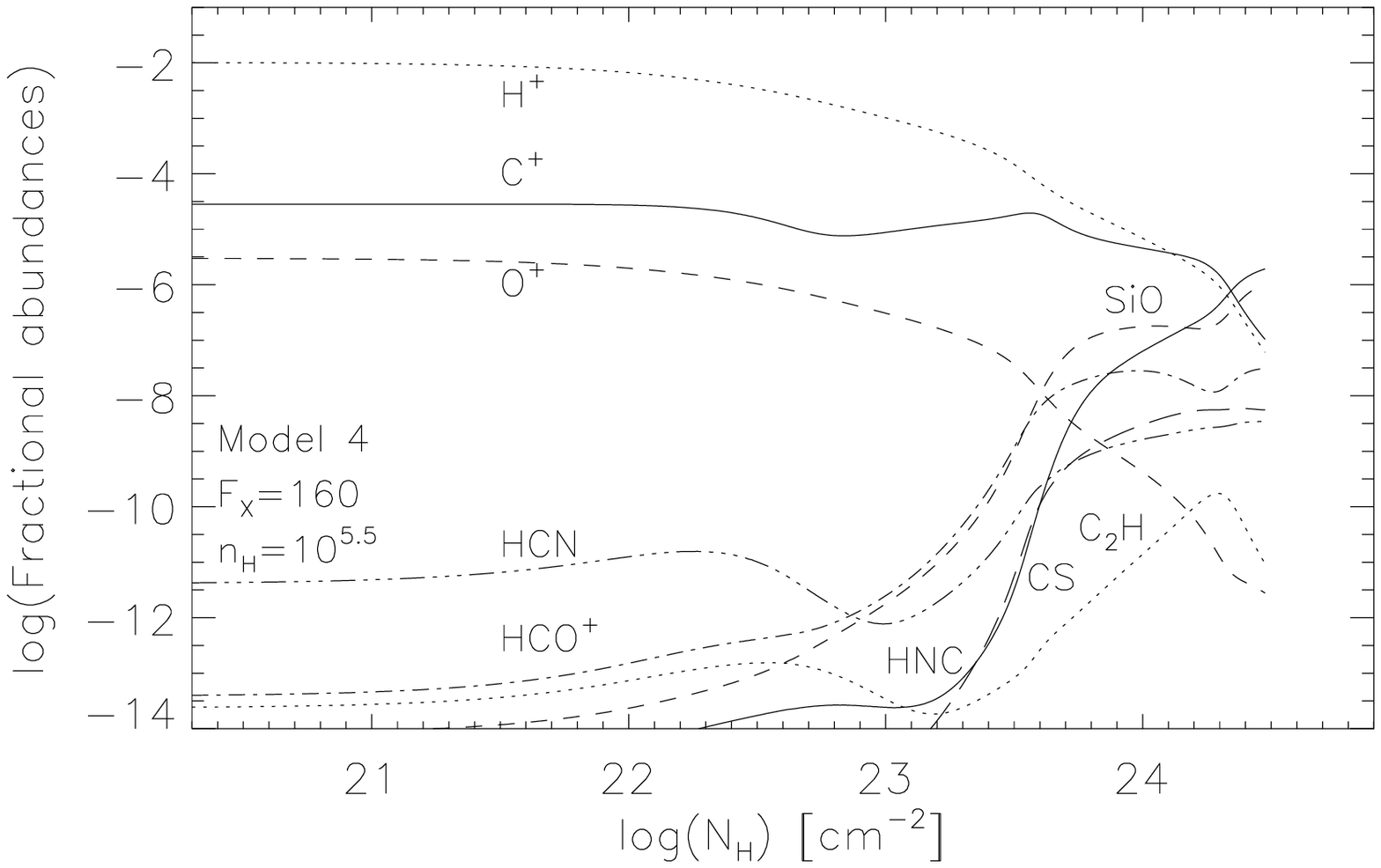}}
\end{minipage}
\caption[] {Fractional abundances for model 1 (top left), 2 (top
right), 3 (bottom left) and 4 (bottom right).}
\label{XDRchem2}
\end{figure*}

In Fig. \ref{XDRchem1}, we show the temperature as a function of total
hydrogen column density, $N_{\rm H}$. Variations in radiation field
strength most strongly affect the high-density models. The temperature
at the edge differs a factor of 30 in the high-density case. Since
X-rays penetrate much deeper into a cloud than FUV photons, high
temperatures are maintained to much greater depths into the
clouds. $H_X/n$ is very important in determining the thermal
balance. When $H_X/n$ is larger, this results in a higher
temperature. Therefore, Model 2 has the largest temperature throughout
the cloud. Density turns out to be important as well. Note that models
1 and 4 have similar incident $H_X/n$ and therefore have about the
same temperature throughout the cloud.


In Figs. \ref{XDRchem1} and \ref{XDRchem2}, we show the fractional
abundances of selected species. $H_X/n$ is not only important in the
thermal balance, but also in the chemistry. Therefore Models 1 and 4
with about the same incident $H_X/n$, show similar abundances. The
most striking difference with the PDR models is that there is no
longer a well-defined transition layer C$^+$ $\rightarrow$ C
$\rightarrow$ CO present. On the contrary, both C and C$^+$ are
present throughout most of the cloud having fractional abundances of
$\sim 10^{-5}-10^{-4}$. Only at very low $H_X/n$, which results in a
low temperature, there is a partial transition to CO. The transition
from atomic to molecular hydrogen is much more gradual than in the PDR
models. A considerable amount of OH is present in all models at all
column densities. The temperature determined by $H_X/n$ is
important. In Model 3, OH has the largest abundance ($> 10^{-6}$) at
all column densities. In other models such large fractions are seen
only at very high depths into the cloud. The formation of CO and
H$_2$O is most efficient at high densities and low $H_X/n$. Therefore,
these species have large abundances throughout the high-density,
low-radiation field Model 3. In Model 4, where the radiation field is
somewhat higher, CO and H$_2$O reach large abundances only at high
$N_{\rm H}$. At low densities, they are only formed at large depths
into the cloud (Model 1 and 2). Secondary ionisations are most
important for the production of H$^+$. Recombination is slower at
lower densities. Therefore, the H$^+$ fractional abundance is highest
in Model 4. HCN, HCO$^+$, HNC, C$_2$H, CS and SiO have much larger
abundances at high temperatures than in the PDR models.


\section{Conclusion}

We conclude this paper by a direct comparison between the PDR and XDR
models. To emphasise that XDRs penetrate much deeper into cloud
volumes than PDRs, we use the same scale for all models. Then, it is
also possible to distinguish between gradients in abundance,
cumulative intensity, column density and column density ratios. XDR
Model 3 is only plotted to $N_{\rm H}\approx 10^{23.5}\ {\rm
cm^{-2}}$, since $H_X/n$ becomes too small and no reliable results are
obtained at higher column densities.

\begin{figure*}[!ht]
\unitlength1cm
\begin{minipage}[b]{5cm}
\resizebox{5.05cm}{!}{\includegraphics*{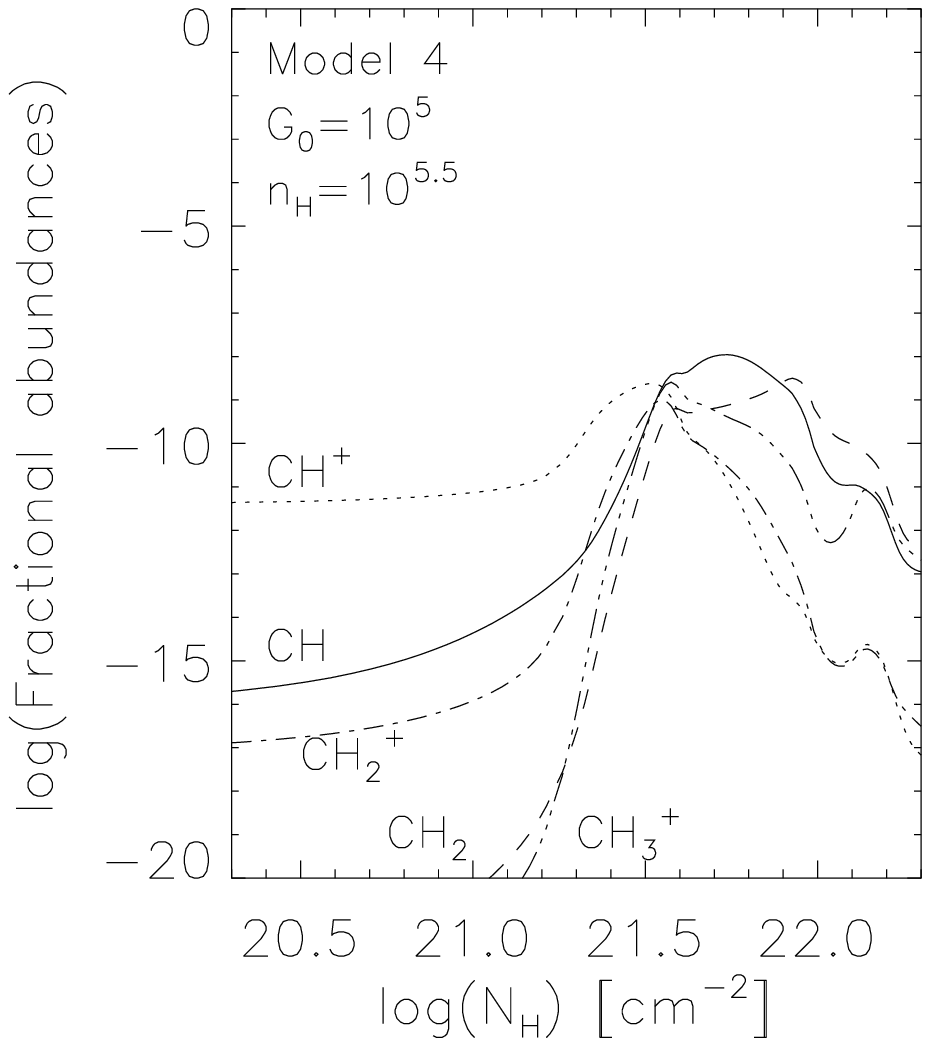}}
\end{minipage}
\hfill
\begin{minipage}[t]{11cm}
\resizebox{9.05cm}{!}{\includegraphics*{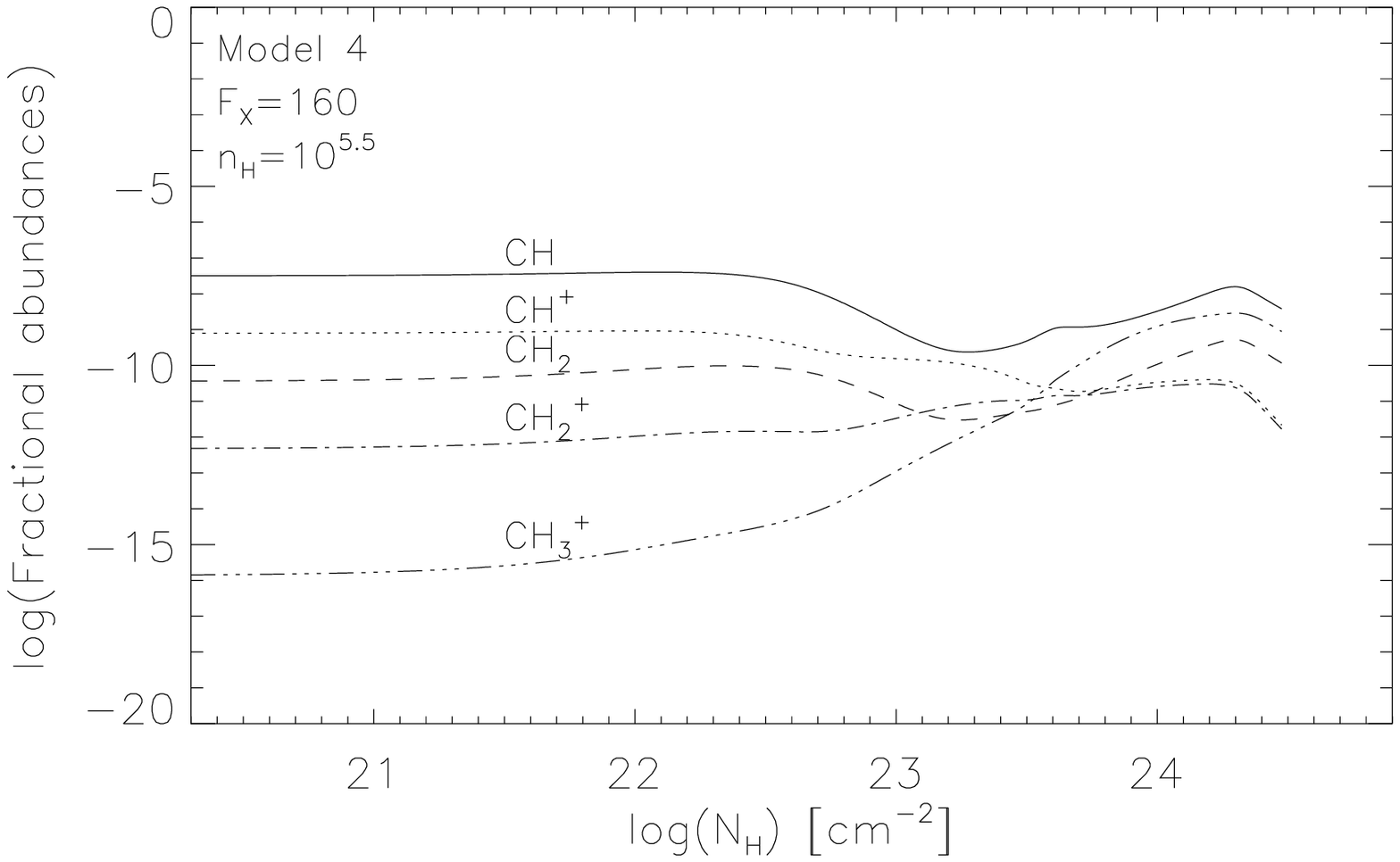}}
\end{minipage}
\hfill
\begin{minipage}[b]{5cm}
\resizebox{5.05cm}{!}{\includegraphics*{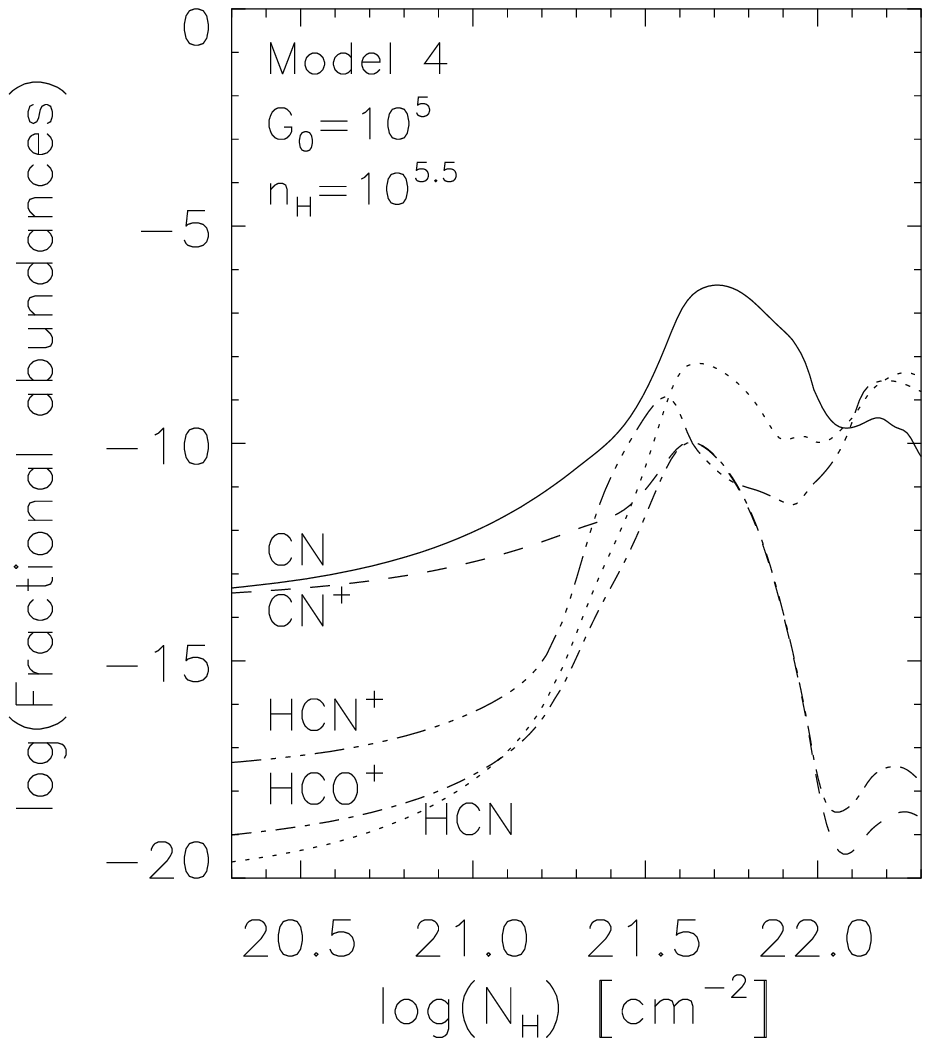}}
\end{minipage}
\hfill
\begin{minipage}[t]{11cm}
\resizebox{9.05cm}{!}{\includegraphics*{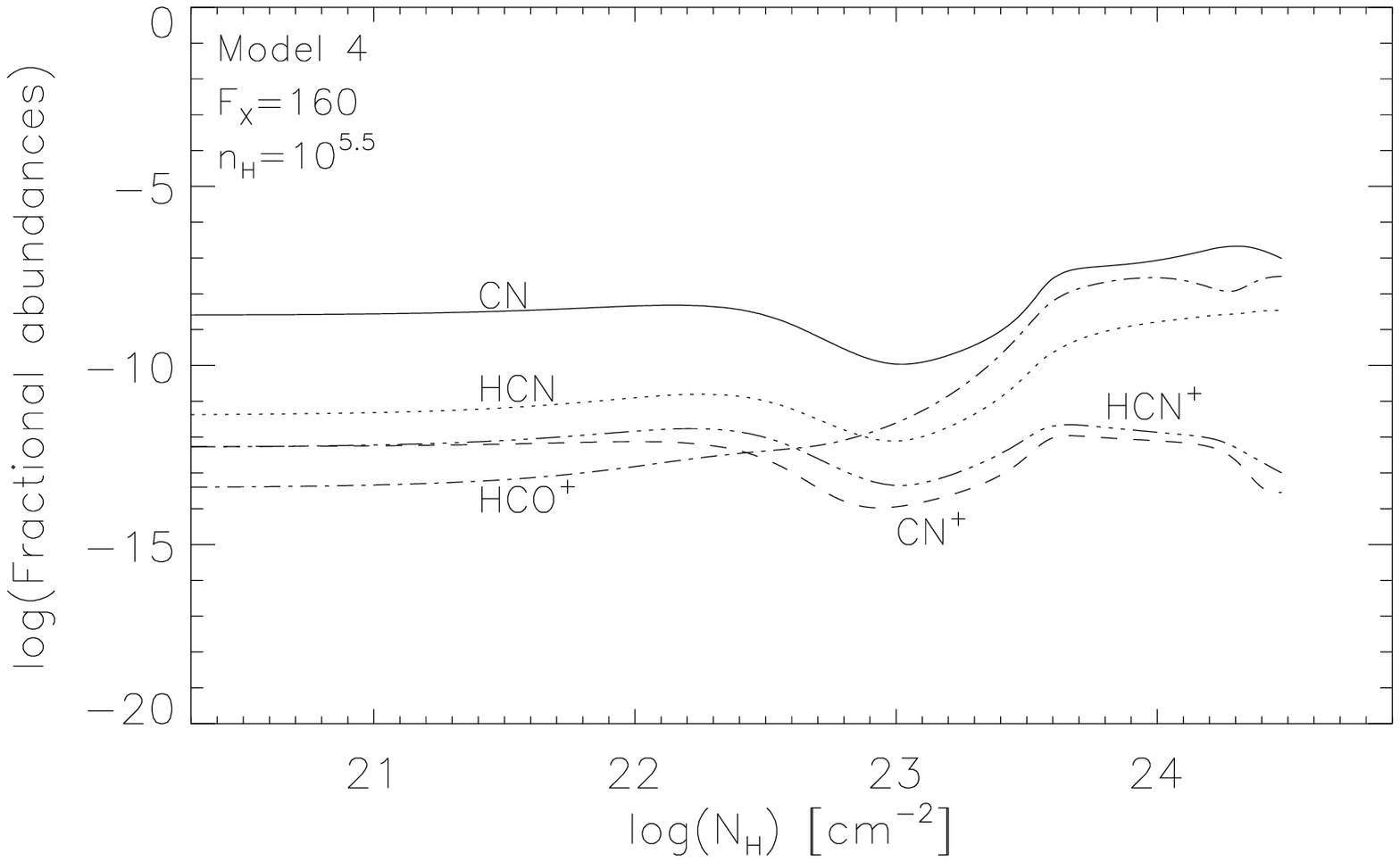}}
\end{minipage}
\hfill
\begin{minipage}[b]{5cm}
\resizebox{5.05cm}{!}{\includegraphics*{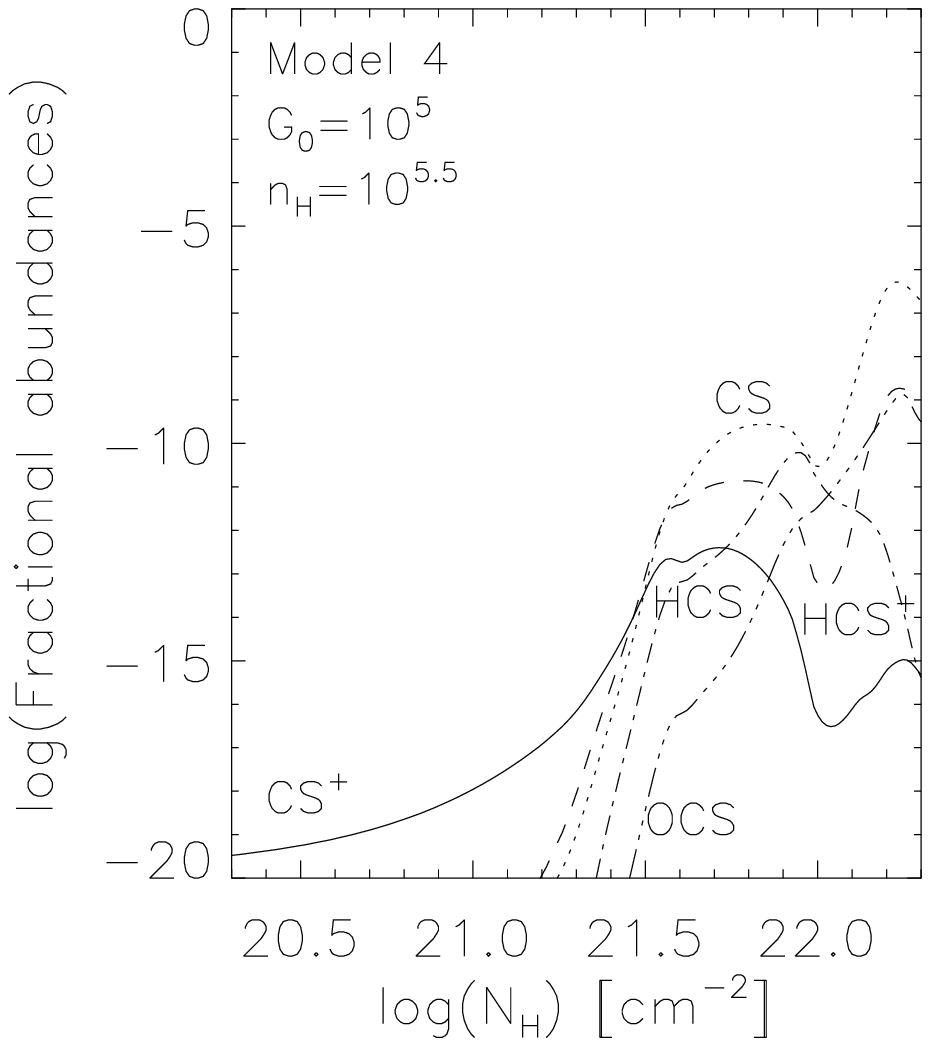}}
\end{minipage}
\hfill
\begin{minipage}[t]{11cm}
\resizebox{9.05cm}{!}{\includegraphics*{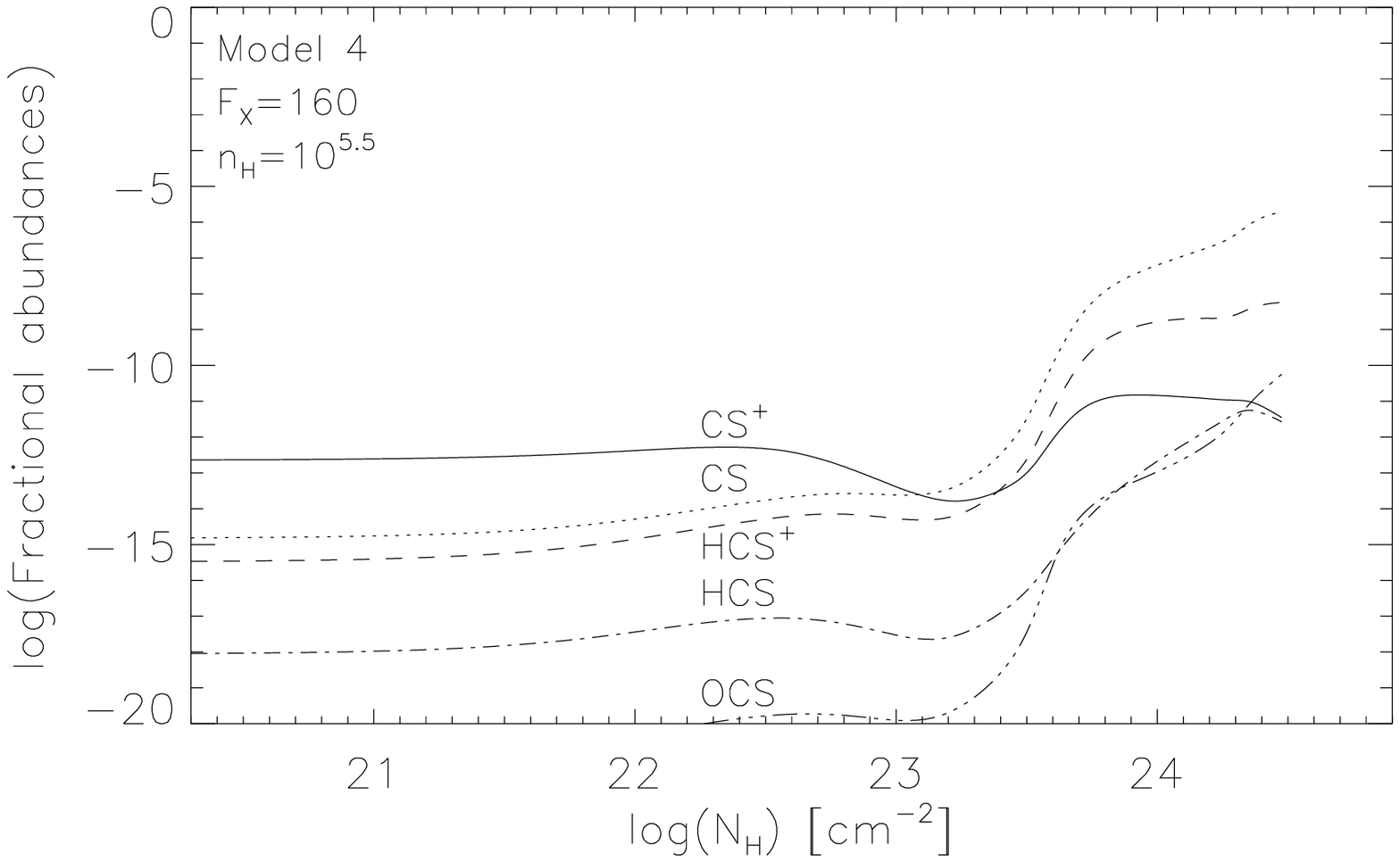}}
\end{minipage}
\hfill
\begin{minipage}[b]{5cm}
\resizebox{5.05cm}{!}{\includegraphics*{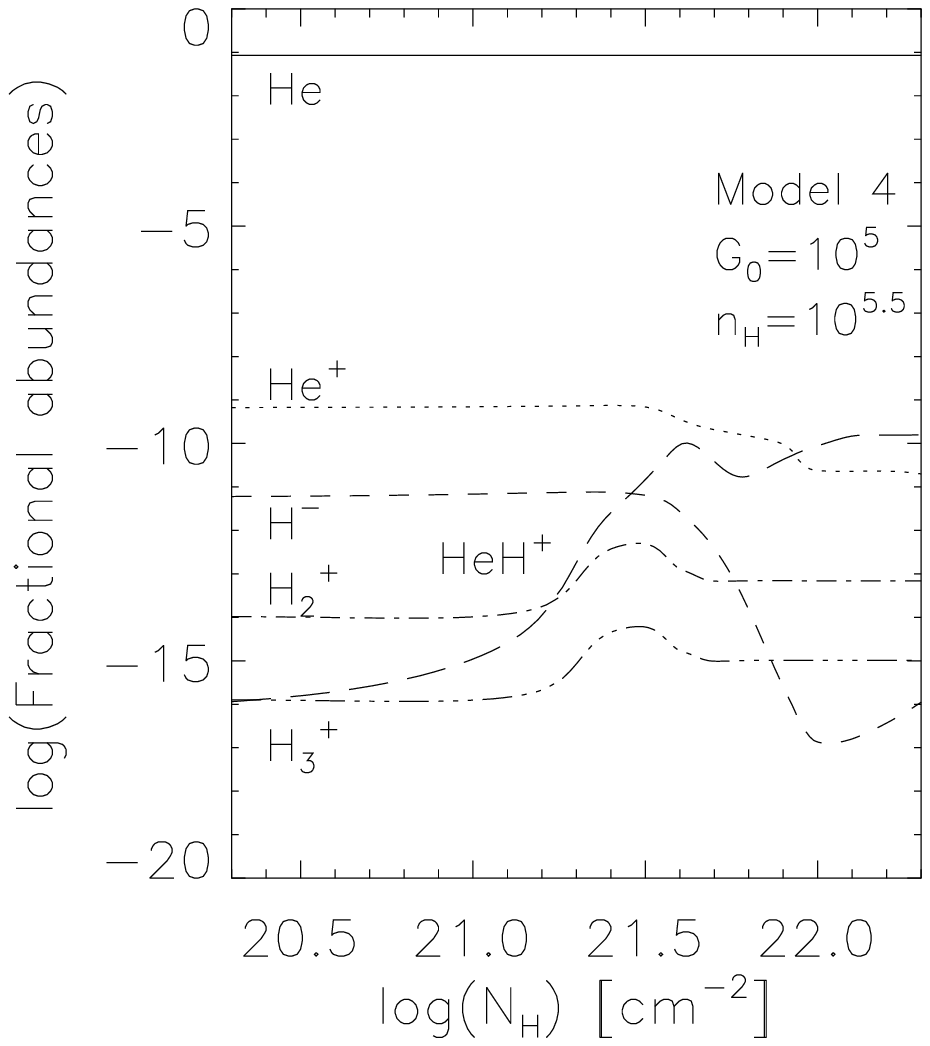}}
\end{minipage}
\hfill
\begin{minipage}[t]{11cm}
\resizebox{9.05cm}{!}{\includegraphics*{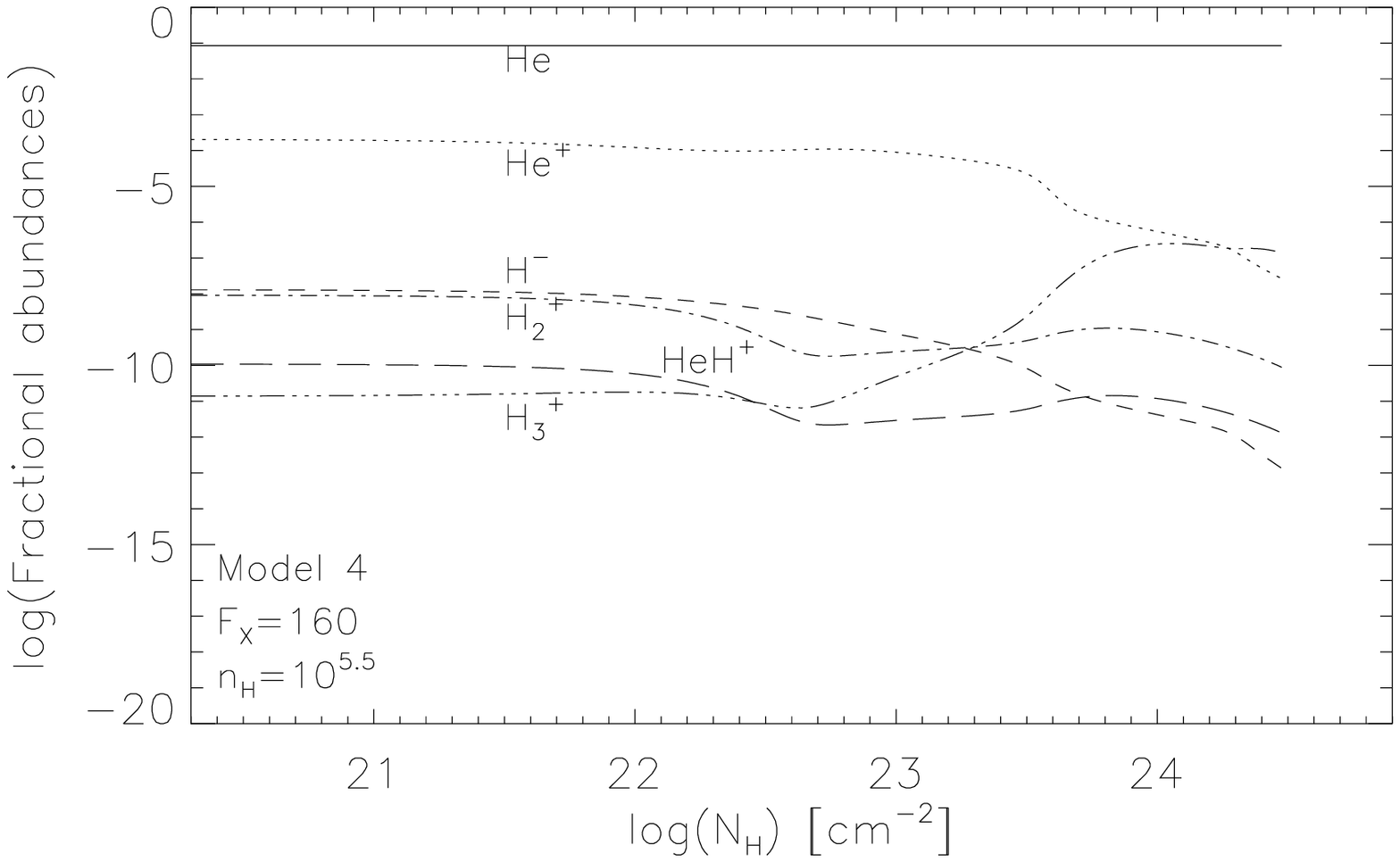}}
\end{minipage}
\caption[] {Comparison between the fractional abundances in the PDR
(left) and XDR (right) for Model 4.
}
\label{Selection}
\end{figure*}

In Fig. \ref{Selection}, we show for Model 4 the abundances of
selected species. At the edge, both neutral and ionised species are
more abundant in the XDR models, and the relative abundances also
differ with respect to one another. In the XDR for example, the
neutral species CH and CH$_2$ are more abundant than CH$^+$ and
CH$_2^+$, respectively. In the PDR, this is the other way around. CN
and CN$^+$ are almost equally abundant at the edge in the PDR, while
CN exceeds CN$^+$ by three orders of magnitude in the XDR. Although
the amounts of CS$^+$ and HCS$^+$ are larger than those for CS and
HCS, respectively, at the edge of the cloud in the XDR, the abundance
difference is less than in the PDR. The abundance of He$^+$ is five
orders of magnitude larger in the XDR, due to secondary
ionisations. H$^-$ is enhanced by three orders of magnitude, due to
the higher ionisation degree. It is also easily seen that in PDRs the
fractional abundances vary over many orders of magnitude, while the
abundances in XDR Model 4 stay almost constant to a column density of
$N_{\rm H}\approx 10^{22}\ {\rm cm^{-2}}$, where the transition from H
to H$_2$ starts.

\begin{figure*}[!ht]
\unitlength1cm
\begin{minipage}[b]{5cm}
\resizebox{5.7cm}{!}{\includegraphics*{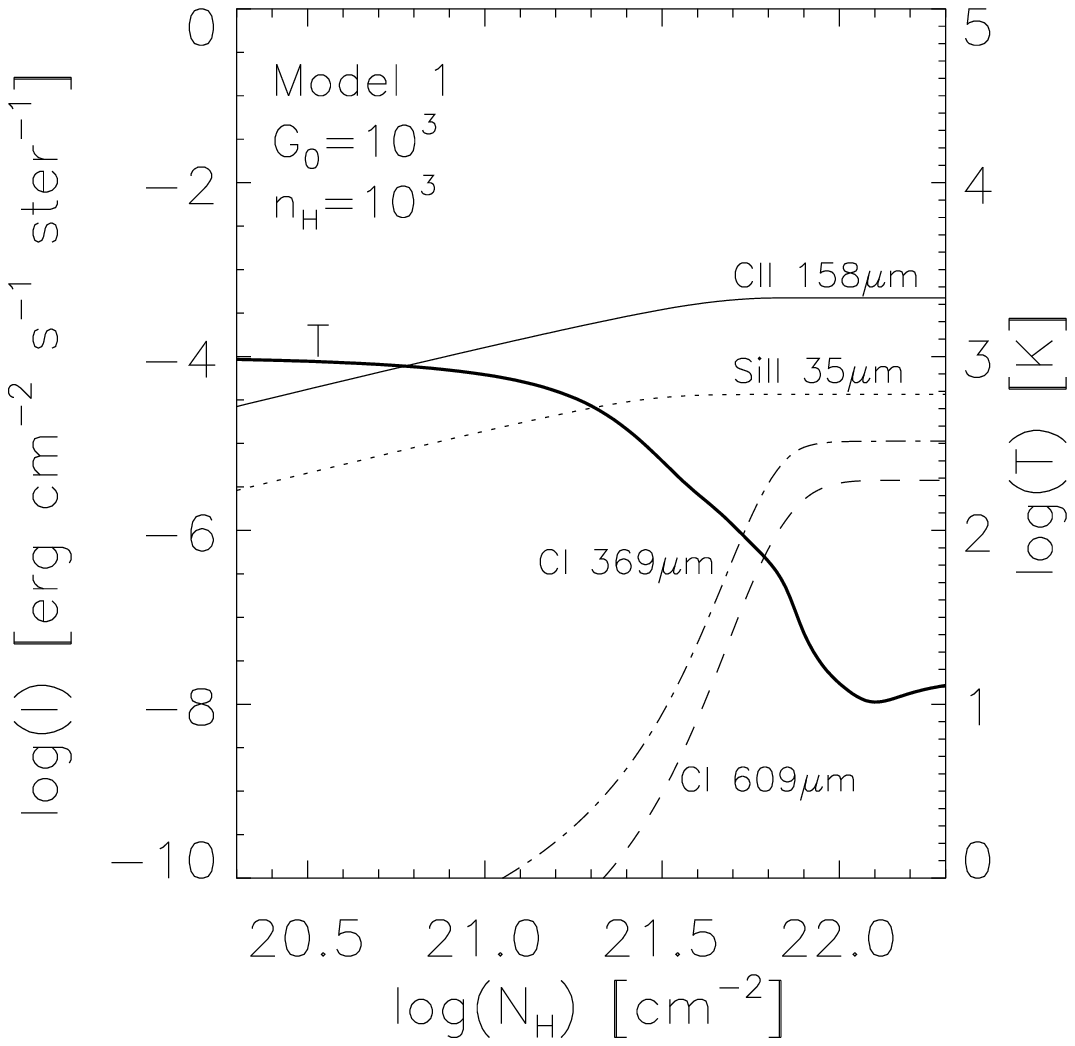}}
\end{minipage}
\hfill
\begin{minipage}[t]{11cm}
\resizebox{9.4cm}{!}{\includegraphics*{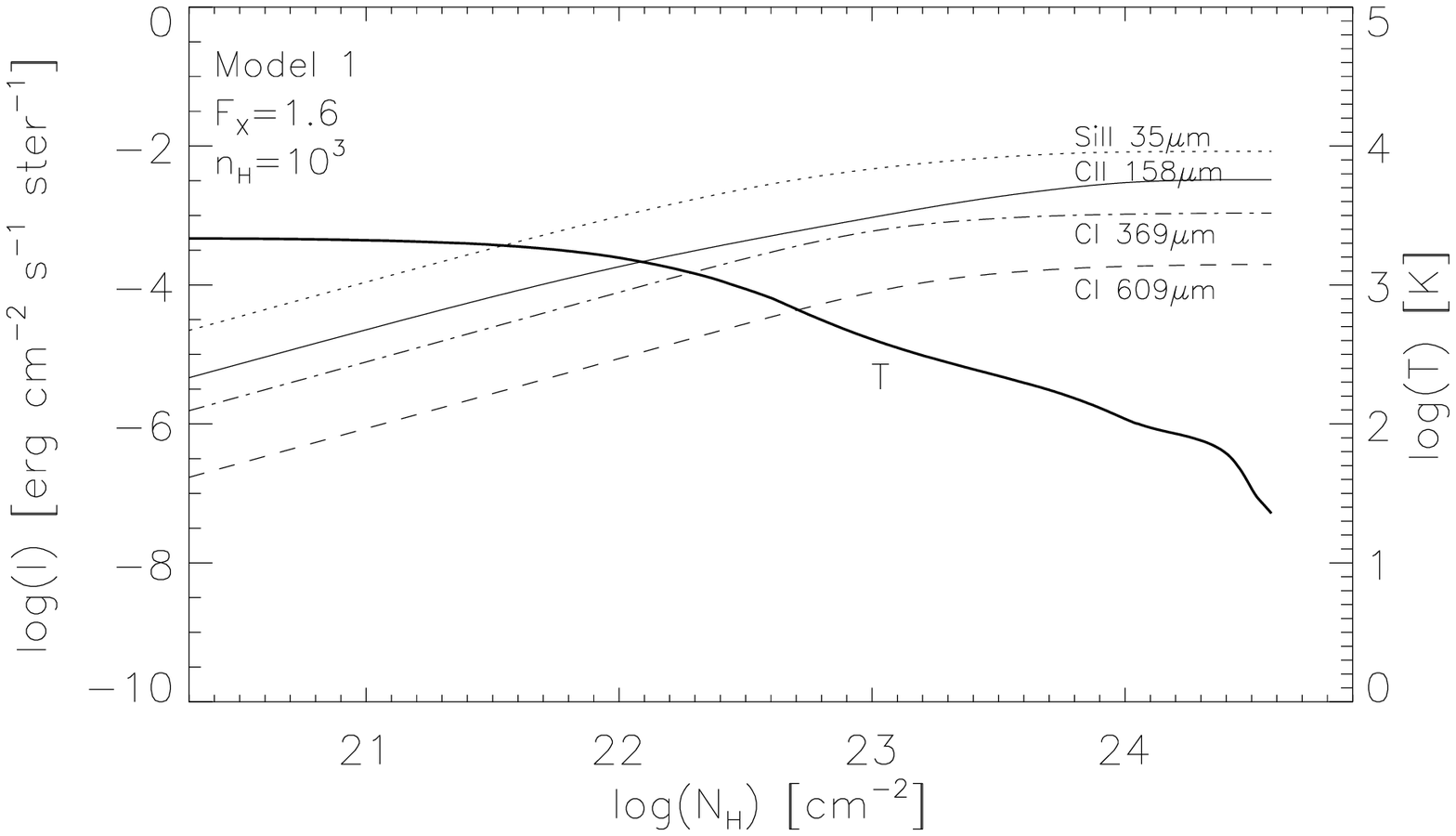}}
\end{minipage}
\hfill
\begin{minipage}[b]{5cm}
\resizebox{5.7cm}{!}{\includegraphics*{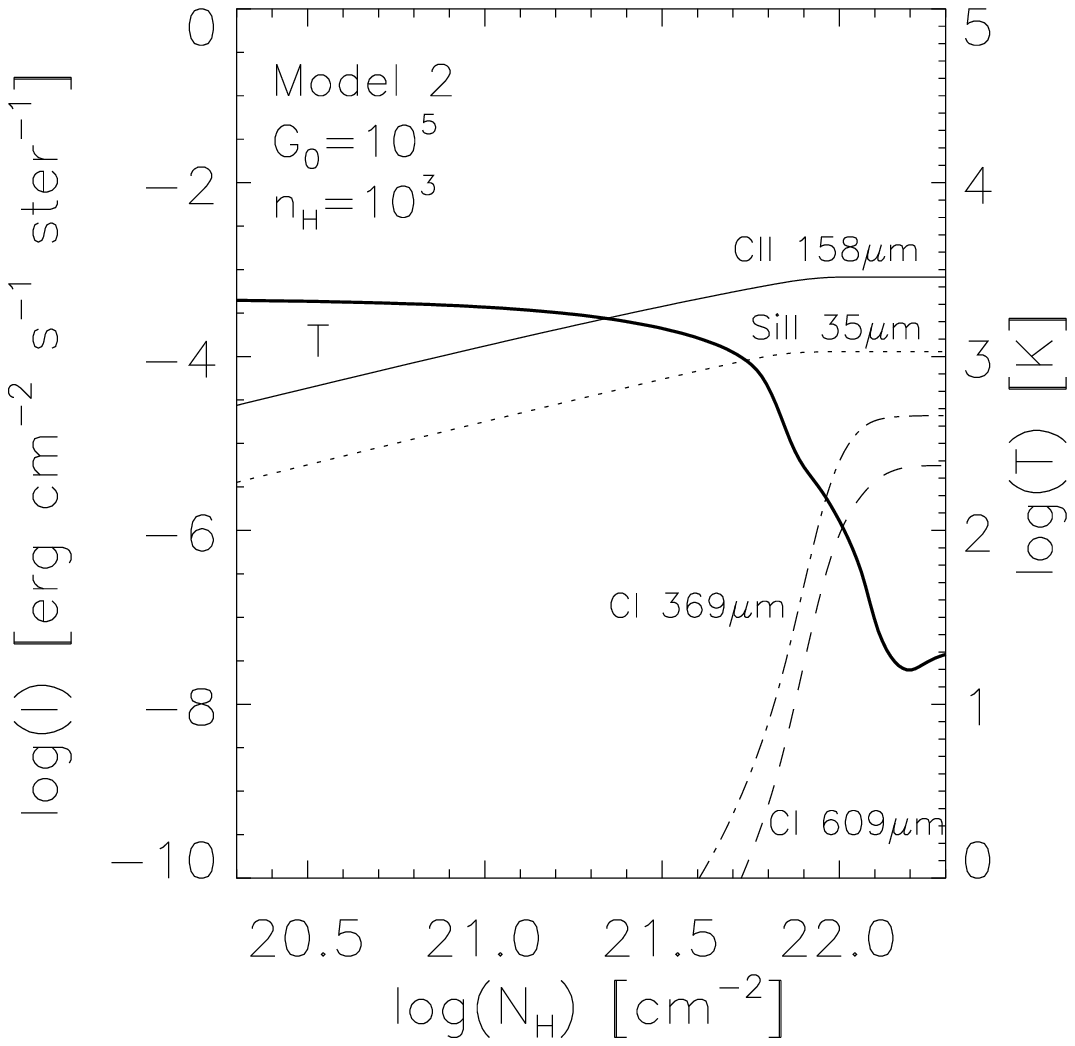}}
\end{minipage}
\hfill
\begin{minipage}[t]{11cm}
\resizebox{9.4cm}{!}{\includegraphics*{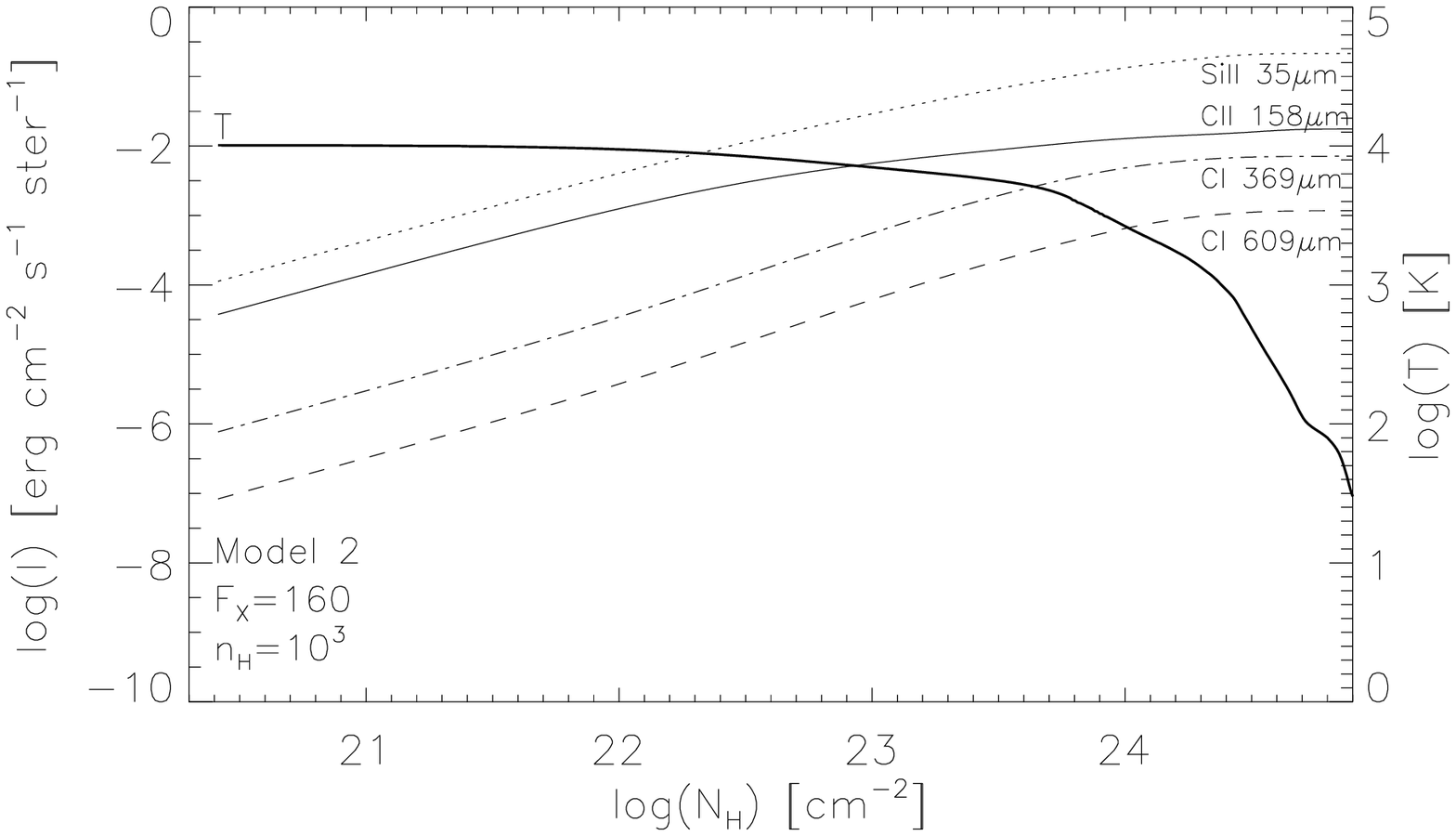}}
\end{minipage}
\hfill
\begin{minipage}[b]{5cm}
\resizebox{5.7cm}{!}{\includegraphics*{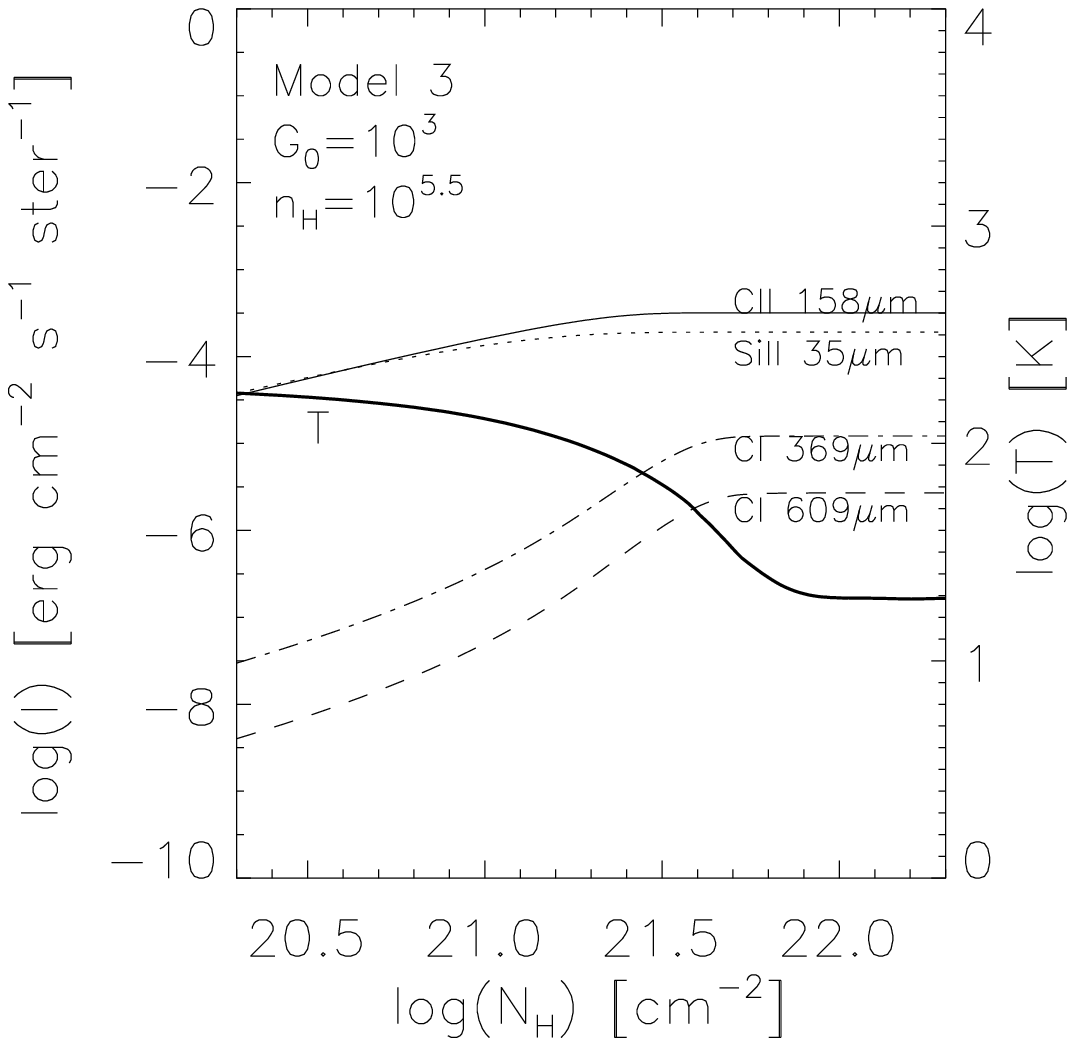}}
\end{minipage}
\hfill
\begin{minipage}[t]{11cm}
\resizebox{9.4cm}{!}{\includegraphics*{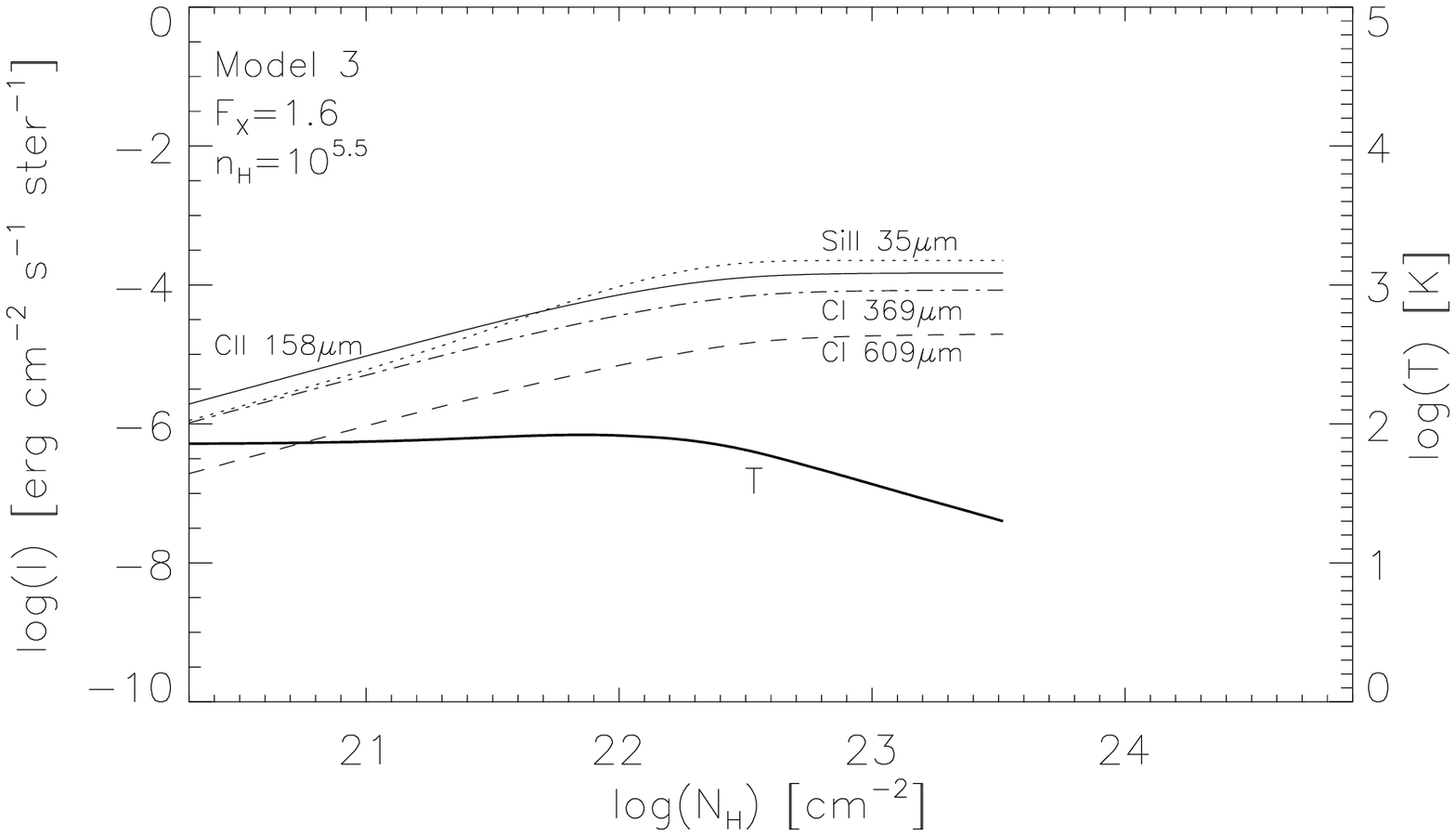}}
\end{minipage}
\hfill
\begin{minipage}[b]{5cm}
\resizebox{5.7cm}{!}{\includegraphics*{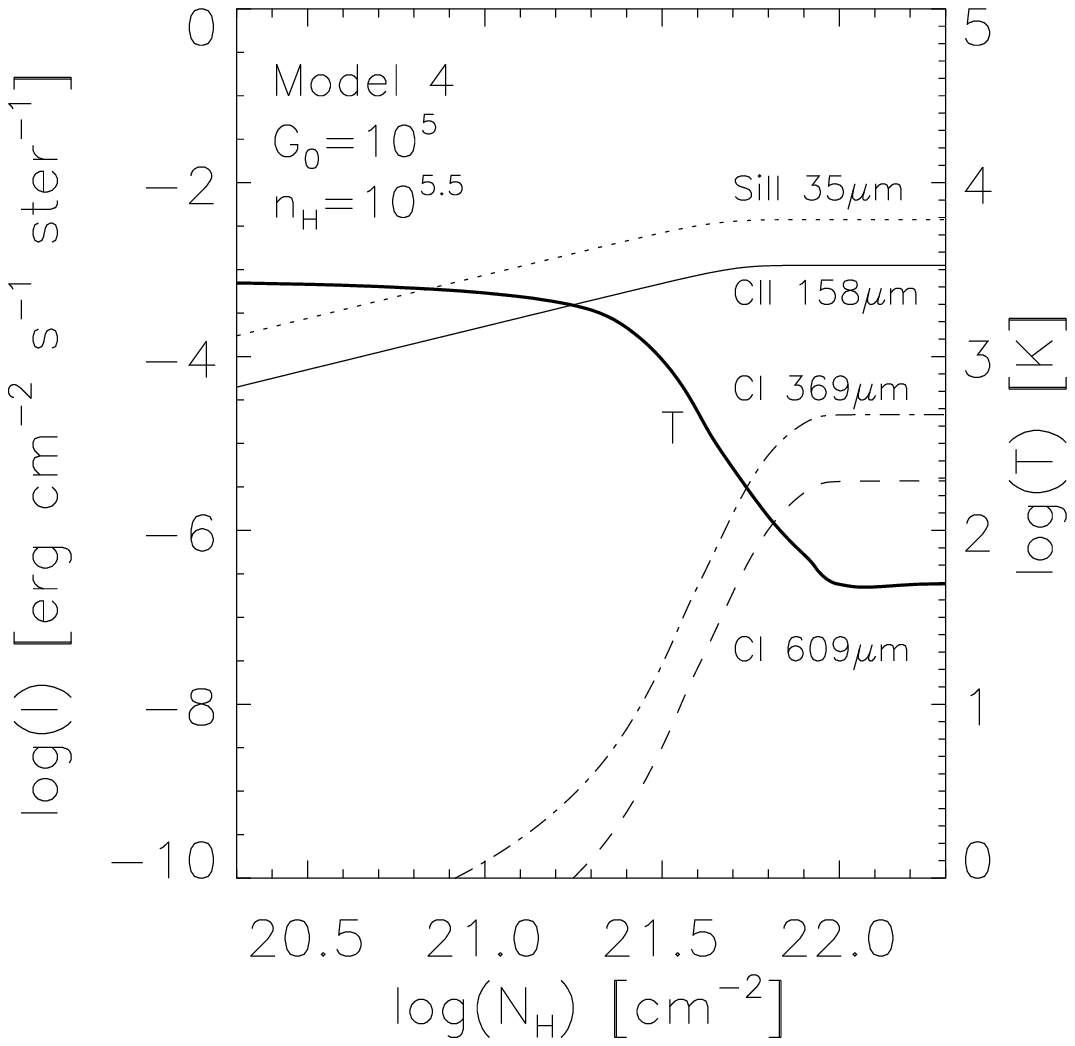}}
\end{minipage}
\hfill
\begin{minipage}[t]{11cm}
\resizebox{9.4cm}{!}{\includegraphics*{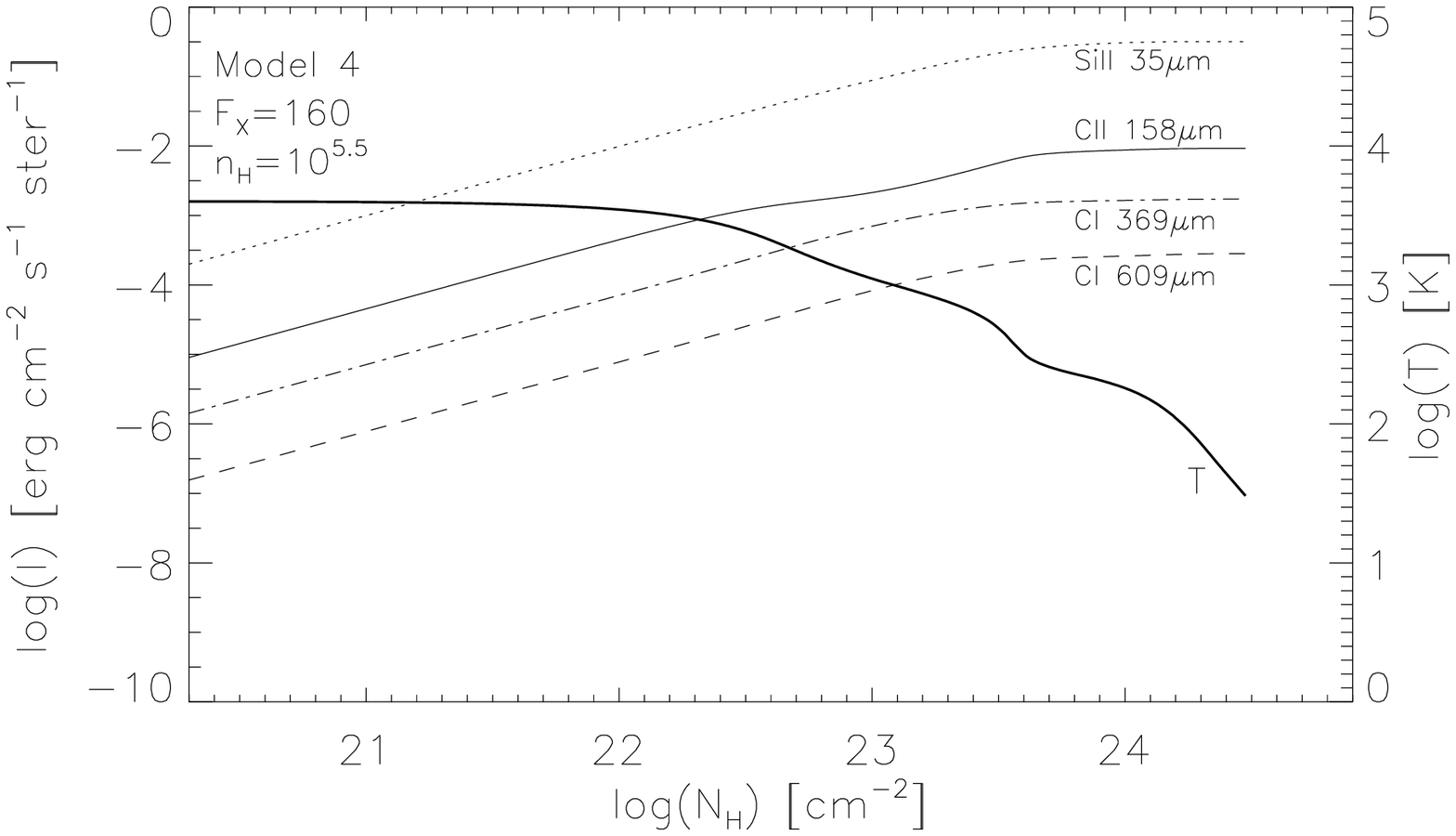}}
\end{minipage}
\caption[] {Cumulative line intensities of [CII] 158 (solid), [SiII]
  34.8 (dotted), [CI] 609 (dashed) and 369 $\mu$m (dashed), for PDR
  (left) and XDR (right) models.}
\label{Lines1}
\end{figure*}

\begin{figure*}[!ht]
\unitlength1cm
\begin{minipage}[b]{5cm}
\resizebox{5.05cm}{!}{\includegraphics*{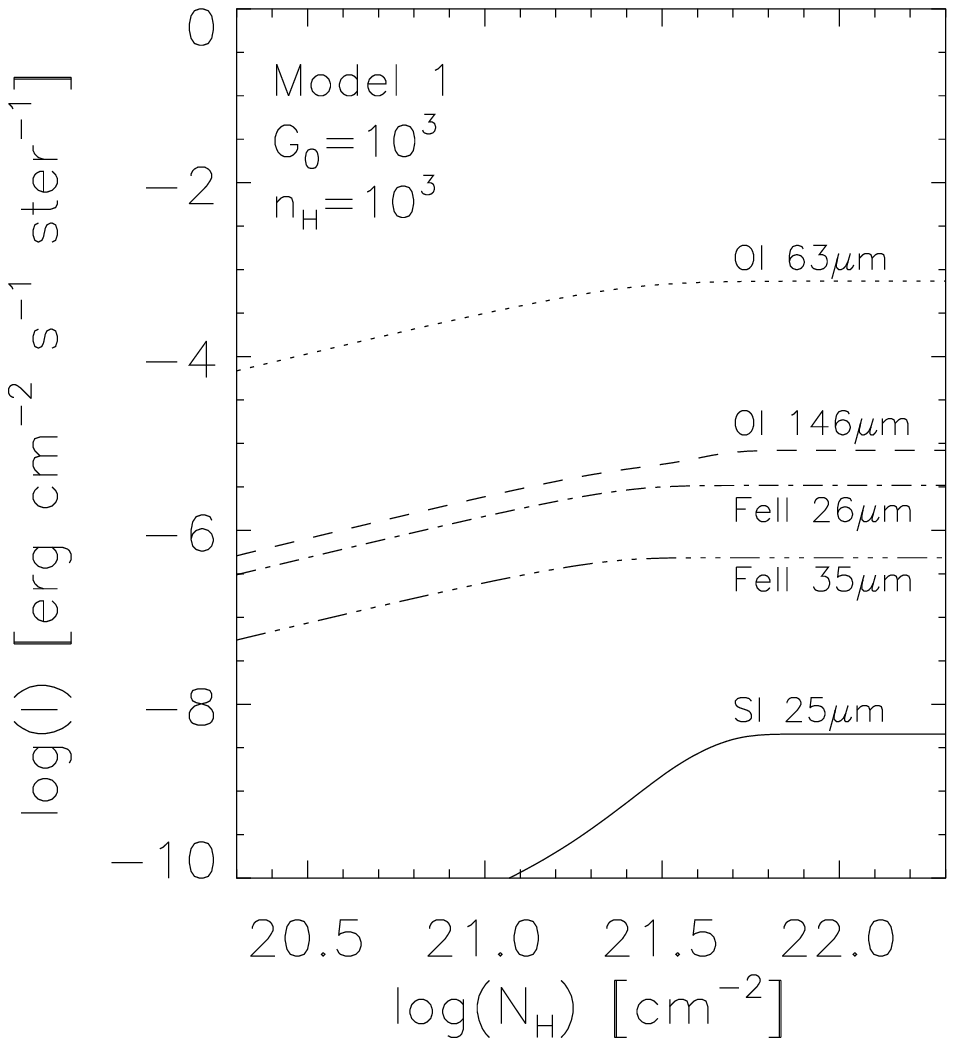}}
\end{minipage}
\hfill
\begin{minipage}[b]{11cm}
\resizebox{9.05cm}{!}{\includegraphics*{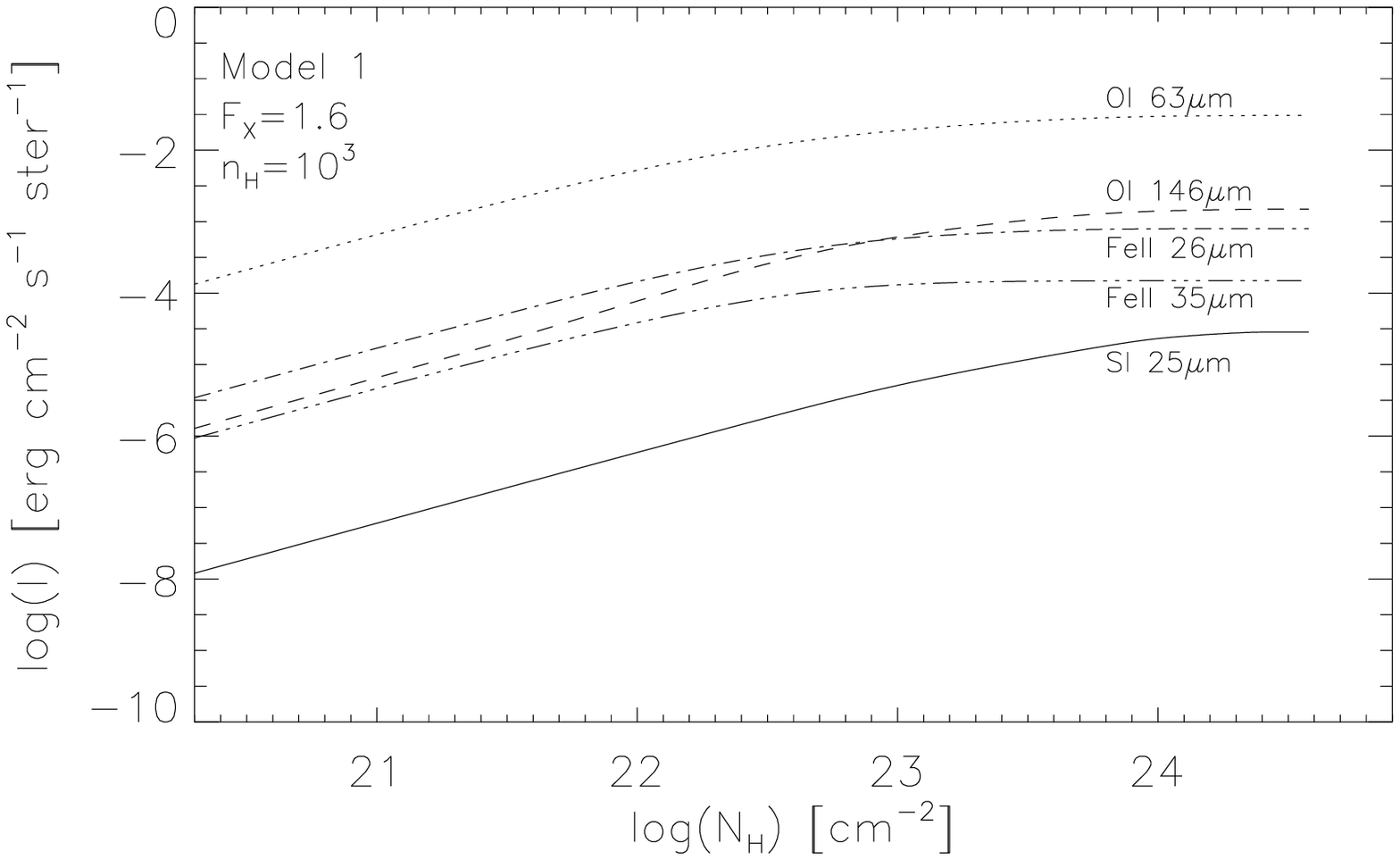}}
\end{minipage}
\hfill
\begin{minipage}[t]{5cm}
\resizebox{5.05cm}{!}{\includegraphics*{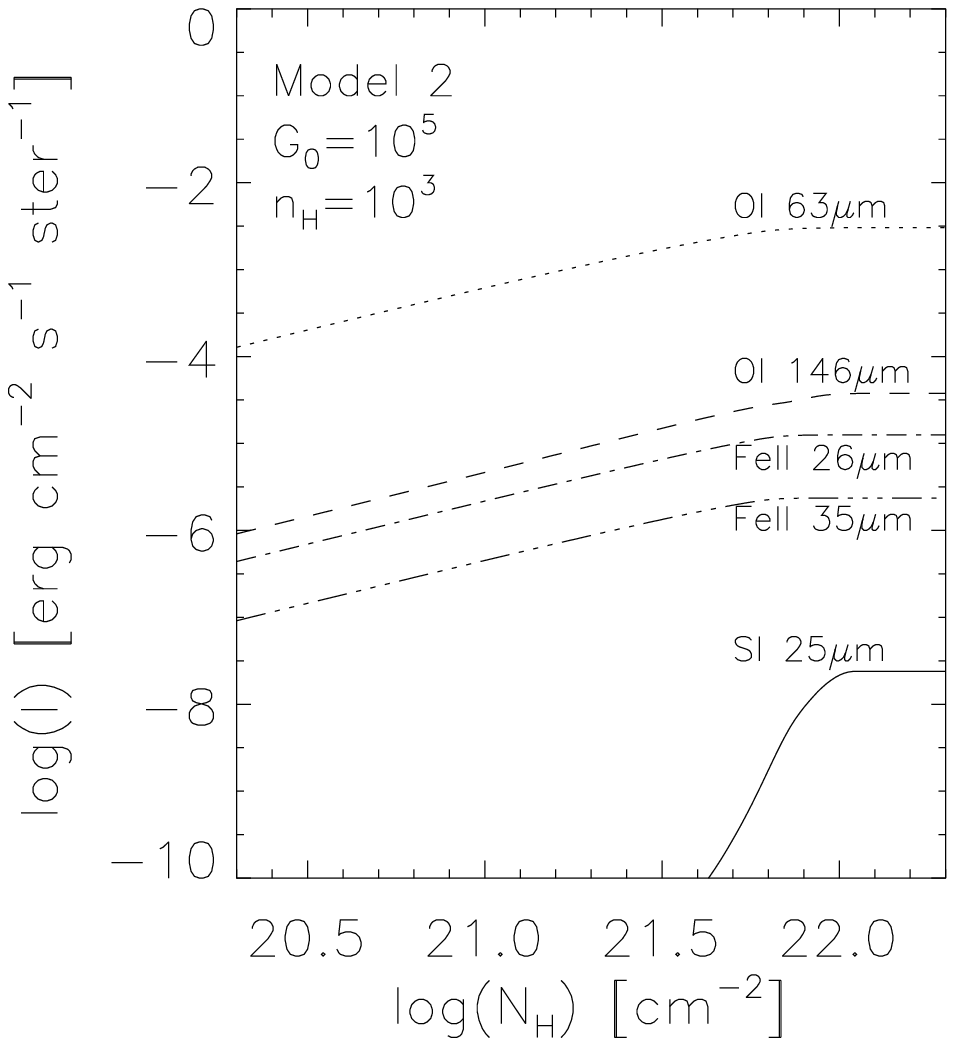}}
\end{minipage}
\hfill
\begin{minipage}[t]{11cm}
\resizebox{9.05cm}{!}{\includegraphics*{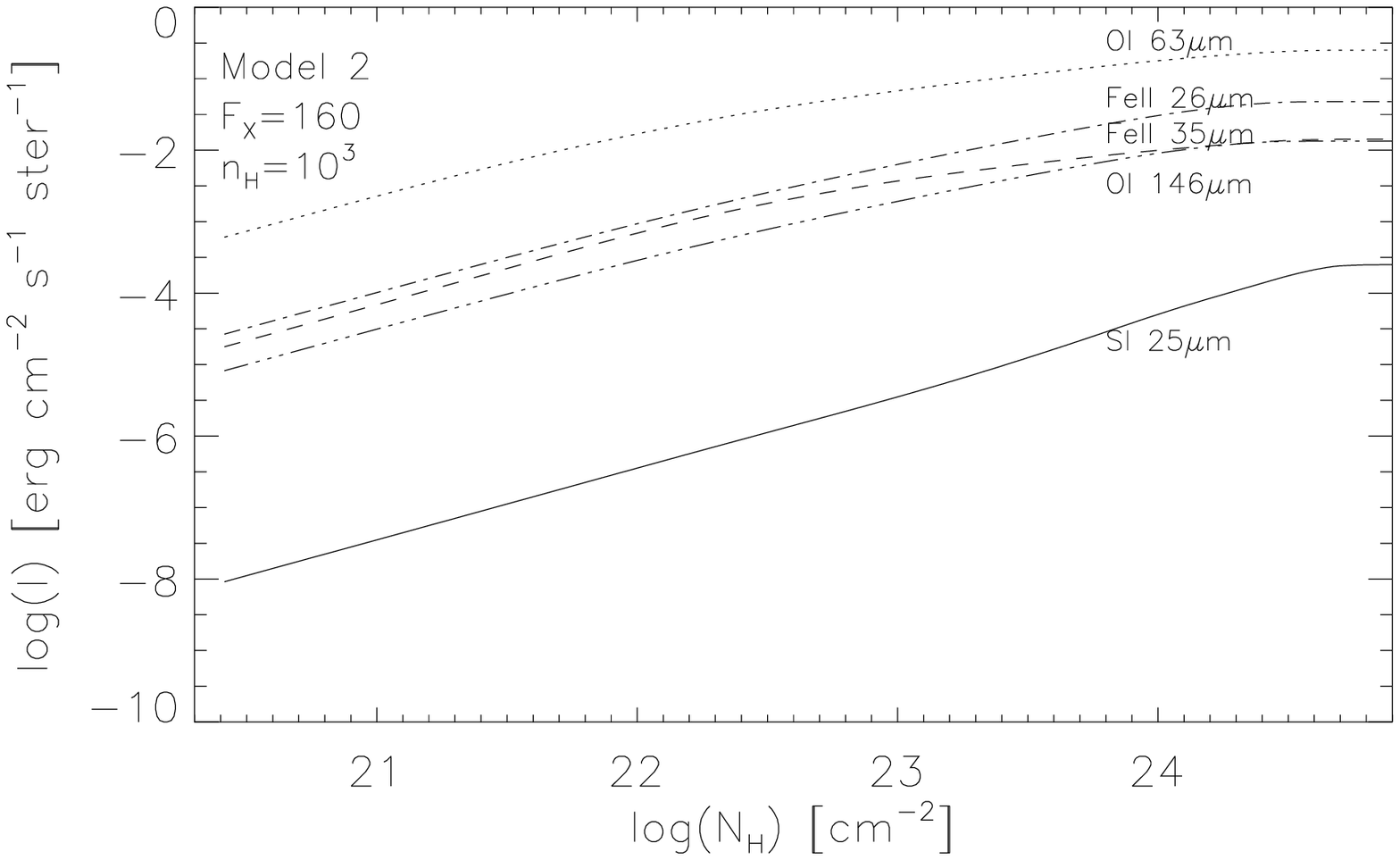}}
\end{minipage}
\hfill
\begin{minipage}[b]{5cm}
\resizebox{5.05cm}{!}{\includegraphics*{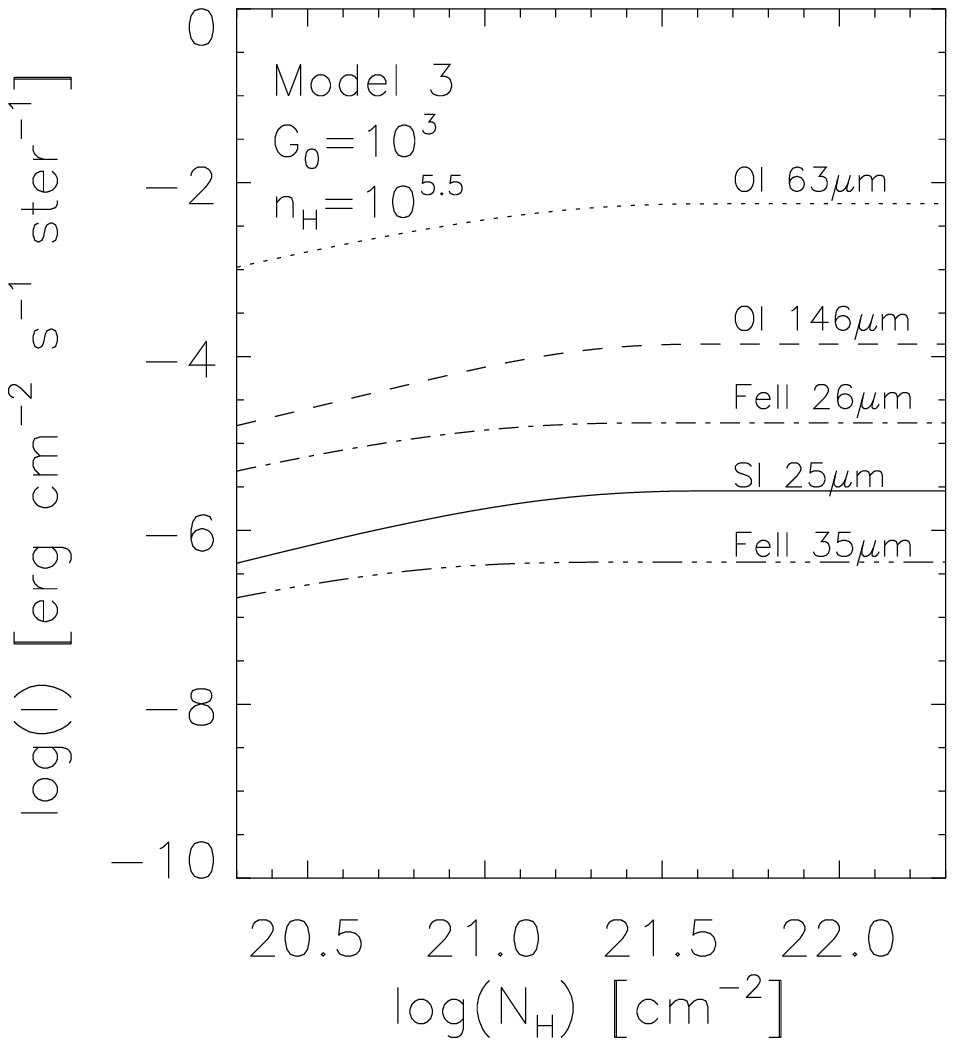}}
\end{minipage}
\hfill
\begin{minipage}[b]{11cm}
\resizebox{9.05cm}{!}{\includegraphics*{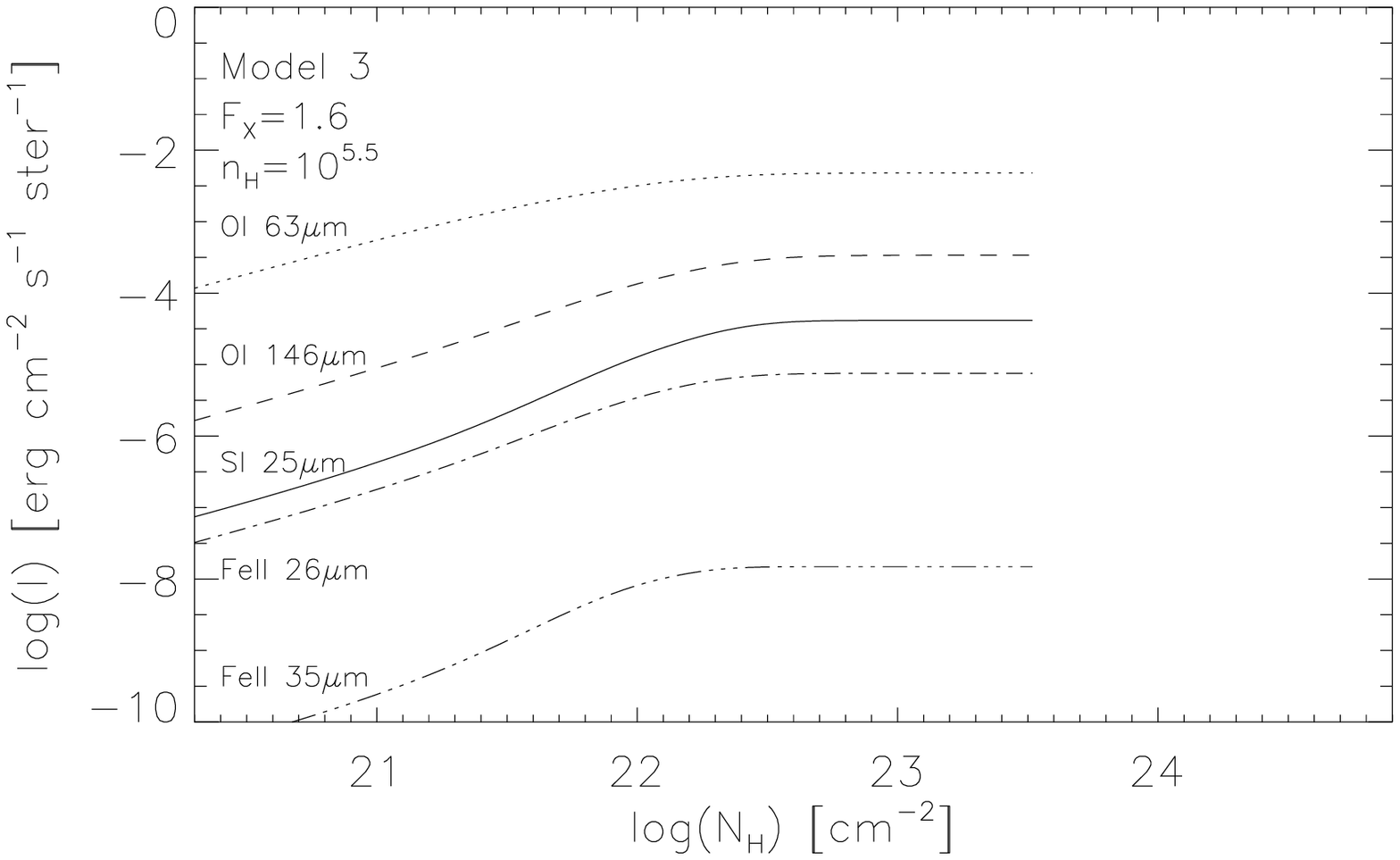}}
\end{minipage}
\hfill
\begin{minipage}[t]{5cm}
\resizebox{5.05cm}{!}{\includegraphics*{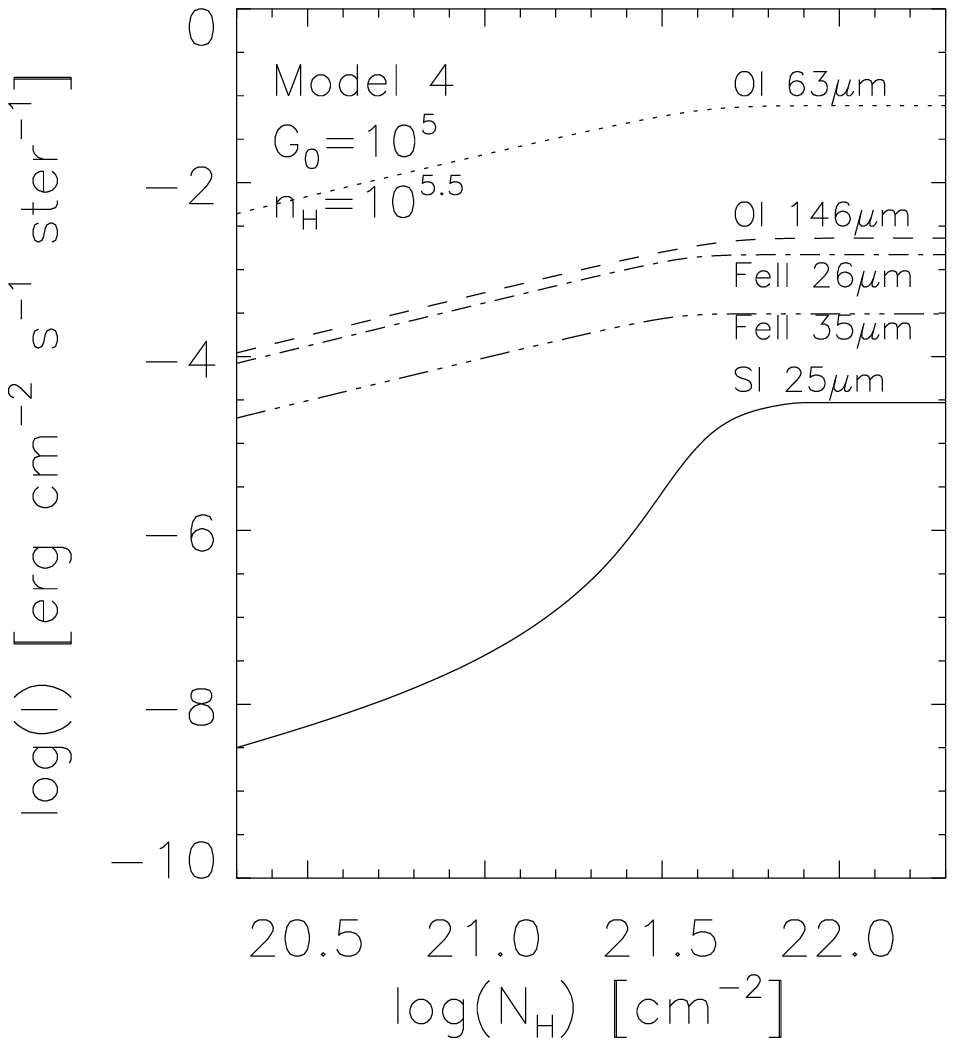}}
\end{minipage}
\hfill
\begin{minipage}[t]{11cm}
\resizebox{9.05cm}{!}{\includegraphics*{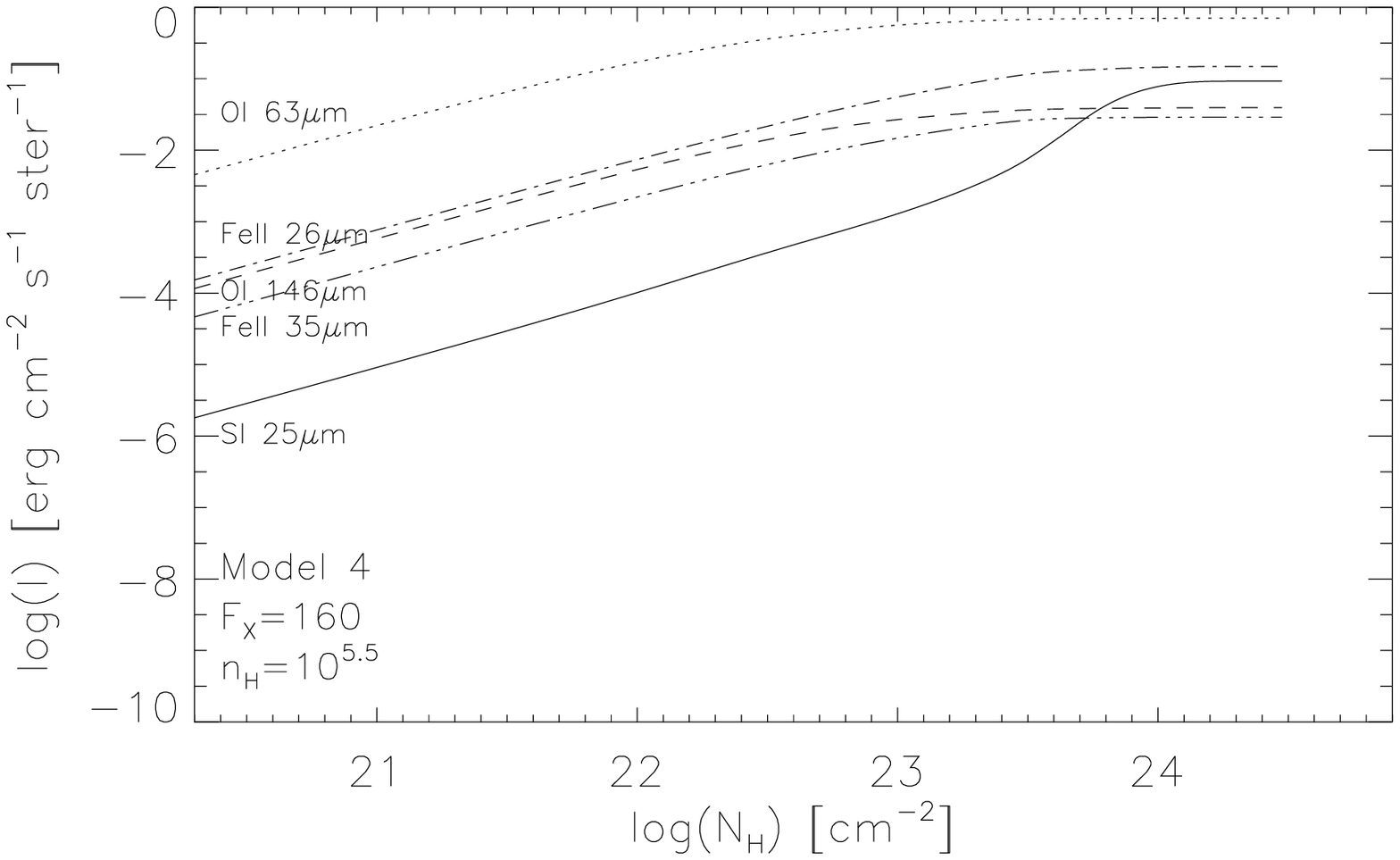}}
\end{minipage}
\caption[] { Cumulative line intensities of [SI] 25.2 (solid), [OI]
63.2 (dotted), 145.6 (dashed), [FeII] 26.0 (dot-dashed) and 35.4
$\mu$m (dotted-dashed), for PDR (left) and XDR (right) models.  }
\label{Lines2}
\end{figure*}

In Figs. \ref{Lines1} and \ref{Lines2}, we show cumulative line
intensities for fine-structure lines at every column density, i.e.,
the emergent intensity arising from the edge of the cloud to column
density $N_{\rm H} = n_{\rm H} z$:

\begin{eqnarray}
I(z)= \frac{1}{2\pi}\int^z_0 \Lambda(z') dz'.
\end{eqnarray}

\noindent Although the total [CII] 158 $\mu$m line intensity is higher
in the XDR, the flux originating from the edge to $N_{\rm H}\approx
10^{22}\ {\rm cm^{-2}}$ is higher in the PDR except when the XDR is
characterised by very high $H_X/n$ values which is the case in Model
2. In all PDR models, all carbon is in C$^+$ at the edge, while a
large part of the carbon is neutral in XDR Models 1,3 and 4. In all
models, oxygen is mostly in atomic form. The [OI] 63 $\mu$m line
intensity to $N_{\rm H}\approx 10^{22}\ {\rm cm^{-2}}$ is larger in
the low-density XDR models, which is possible due to higher electron
abundances. The intensity is lower in the low radiation, high density
XDR Model 3, since the temperature is higher in the PDR. For Model 4
they are about the same, since the density where the line gets
thermalised is almost reached. In the XDR, all line intensities
increase more or less steadily with increasing column density. PDRs,
however, primarily affect cloud surfaces causing more sudden
changes. The line intensities of [CI] 609 $\mu$m and 369 $\mu$m arise
from a more or less well defined part part of the cloud and start to
increase at column densities $N_{\rm H} \ge 10^{21.5}$ cm$^{-2}$.  The
line intensities of [CII] 158 $\mu$m are larger than those of [SiII]
35 $\mu$m in the PDRs except in Model 4. This is in contrast to the
XDR models, where the [SiII] 35 $\mu$m line intensity is always
stronger. The fact that [SiII] 35 $\mu$m lines are quite strong in
XDRs was already noted by \citet{Maloney1996}. The line intensities
for [FeII] 26 $\mu$m and 35 $\mu$m are larger for the XDR models
except again for Model 3.

\begin{figure*}[!ht]
\unitlength1cm
\begin{minipage}[b]{5cm}
\resizebox{4.9cm}{!}{\includegraphics*{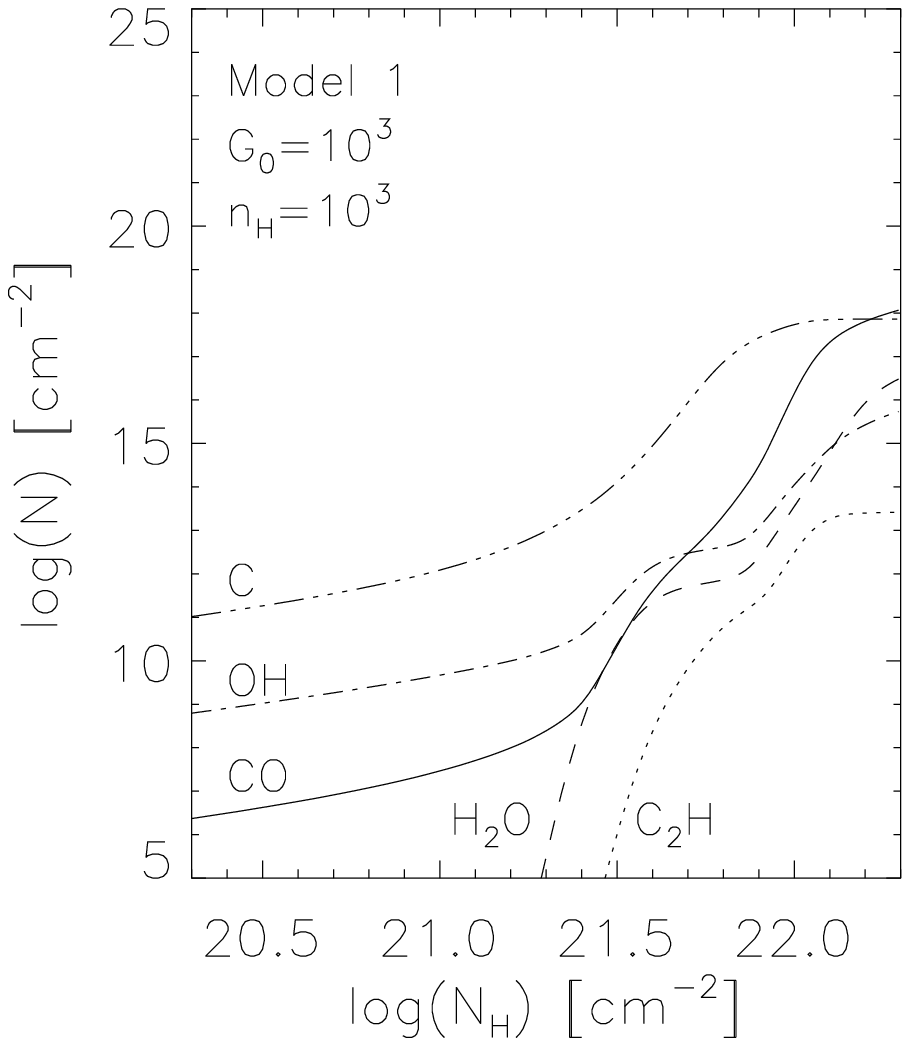}}
\end{minipage}
\hfill
\begin{minipage}[b]{11cm}
\resizebox{9.05cm}{!}{\includegraphics*{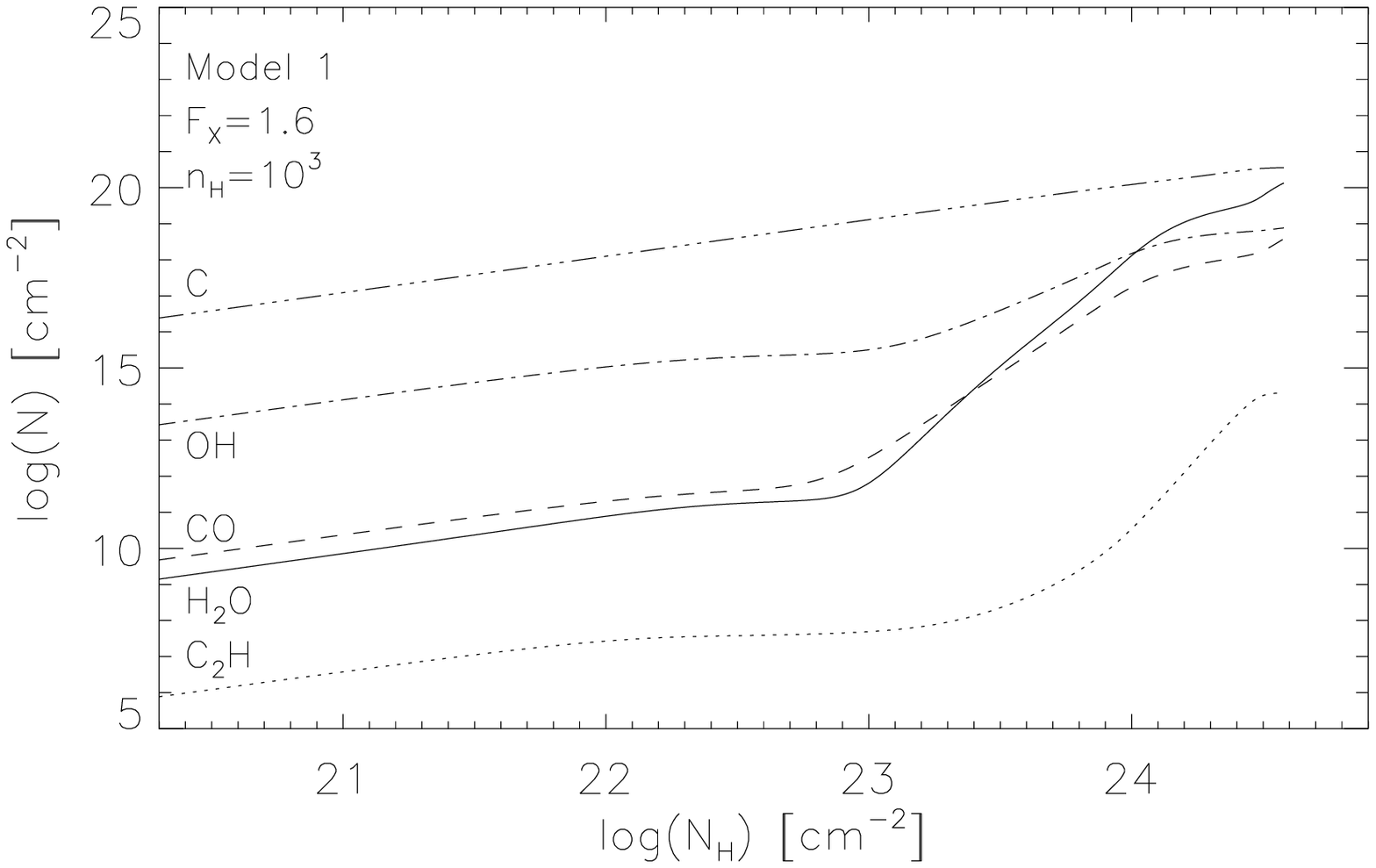}}
\end{minipage}
\hfill
\begin{minipage}[t]{5cm}
\resizebox{4.9cm}{!}{\includegraphics*{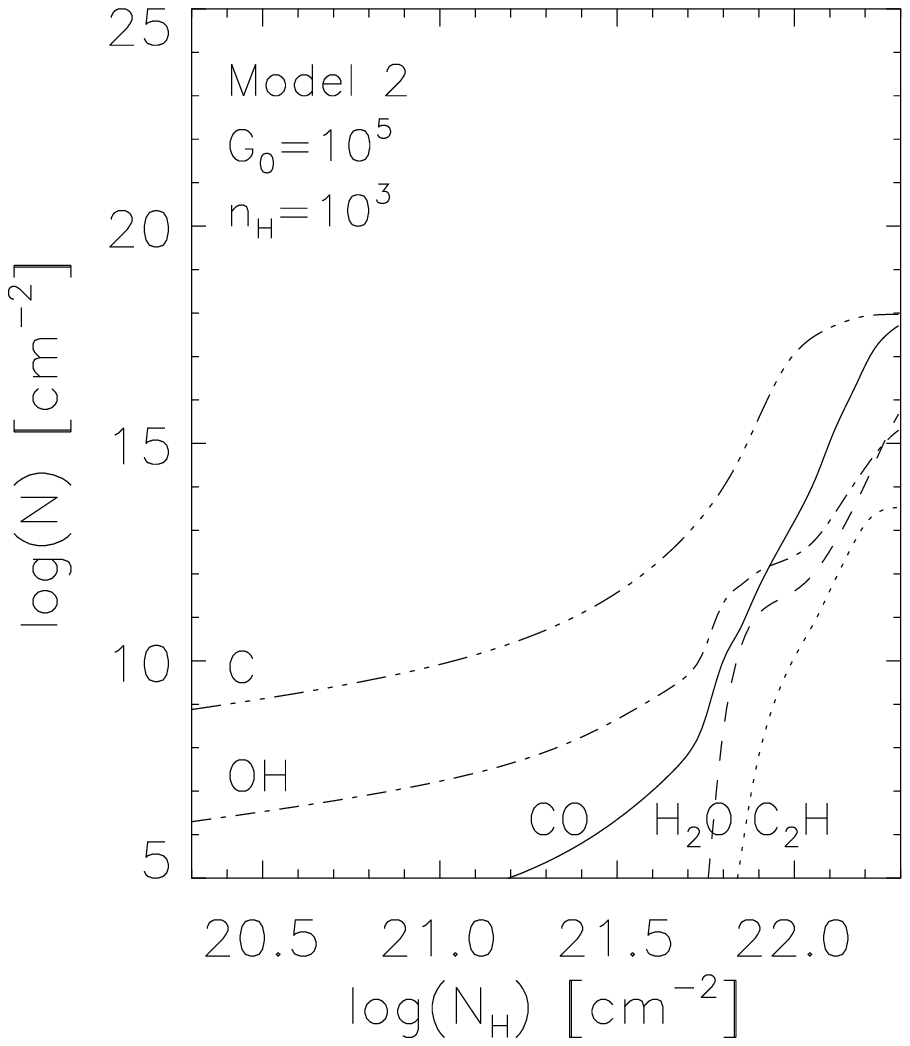}}
\end{minipage}
\hfil
\begin{minipage}[t]{11cm}
\resizebox{9.05cm}{!}{\includegraphics*{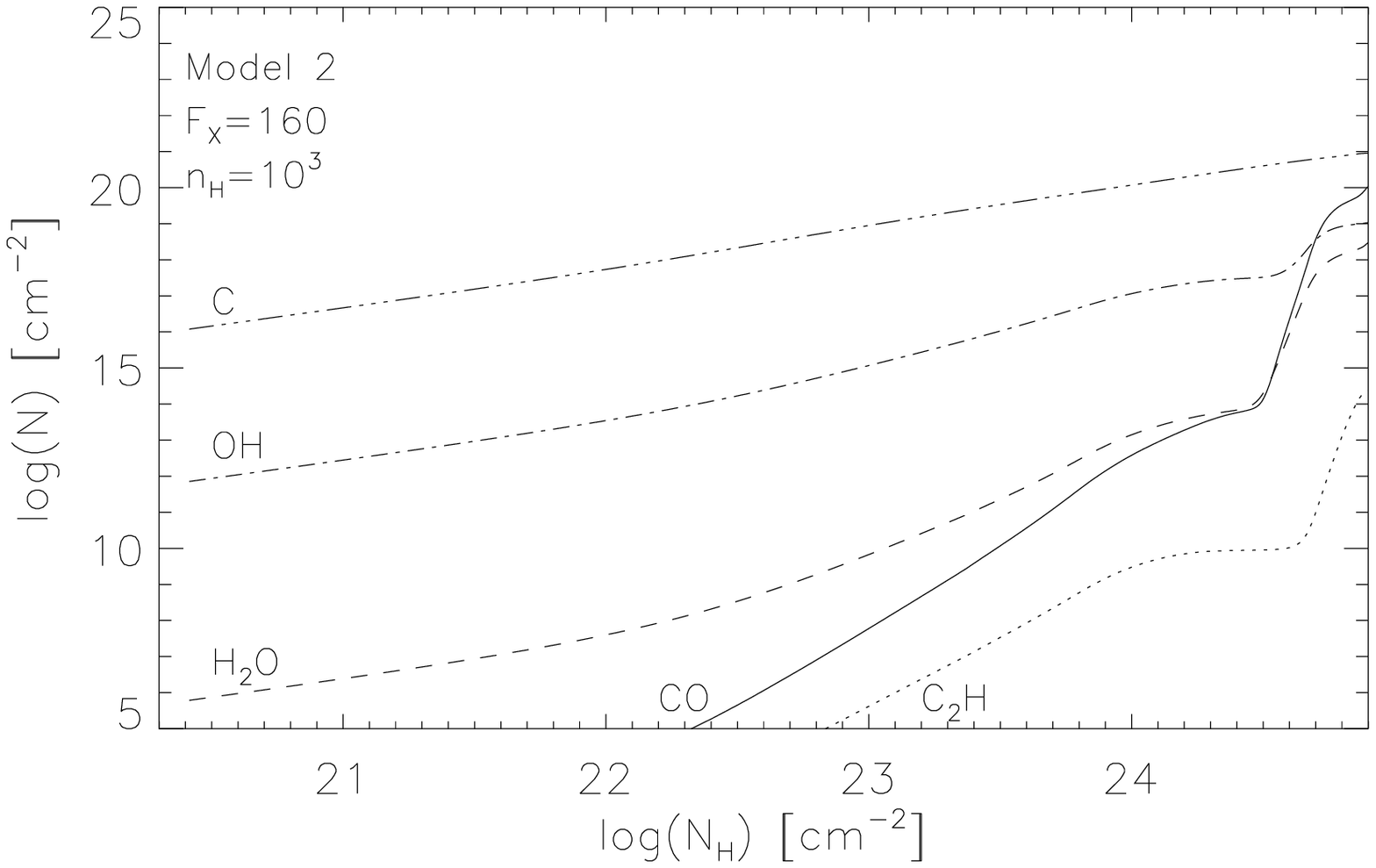}}
\end{minipage}
\hfill
\begin{minipage}[b]{5cm}
\resizebox{4.9cm}{!}{\includegraphics*{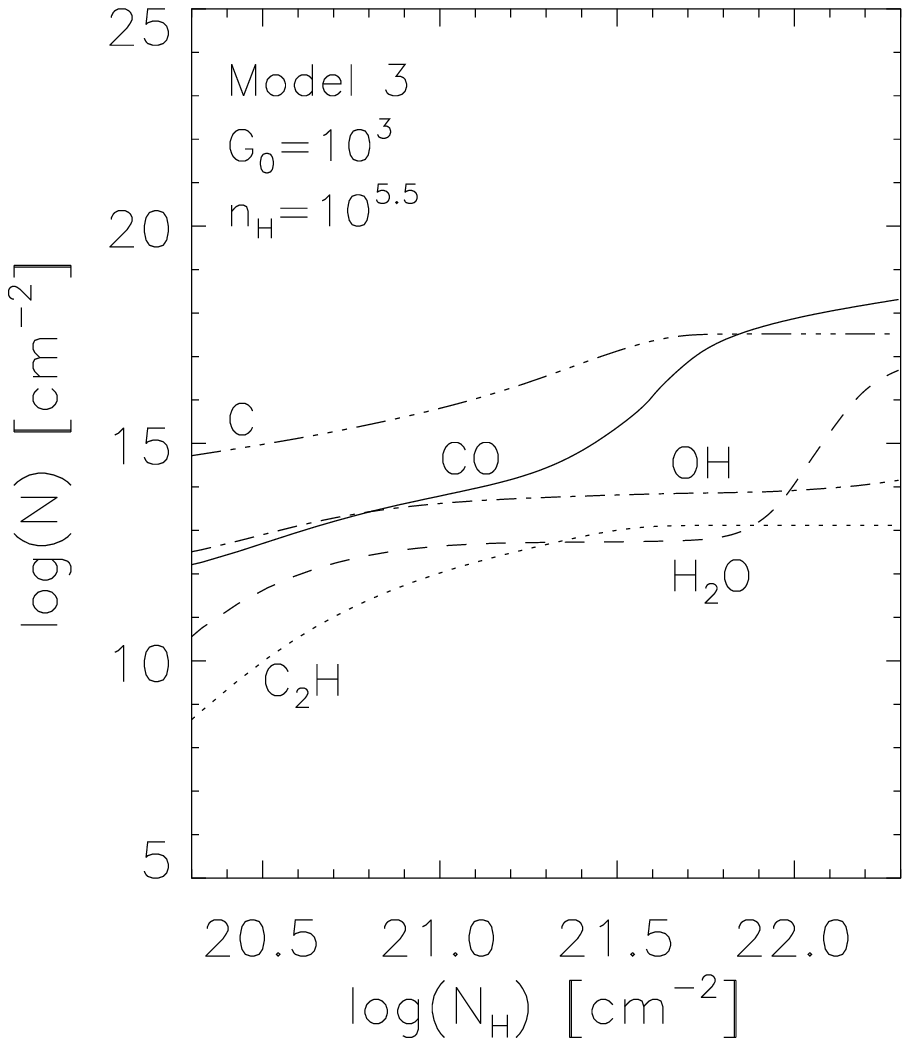}}
\end{minipage}
\hfill
\begin{minipage}[b]{11cm}
\resizebox{9.05cm}{!}{\includegraphics*{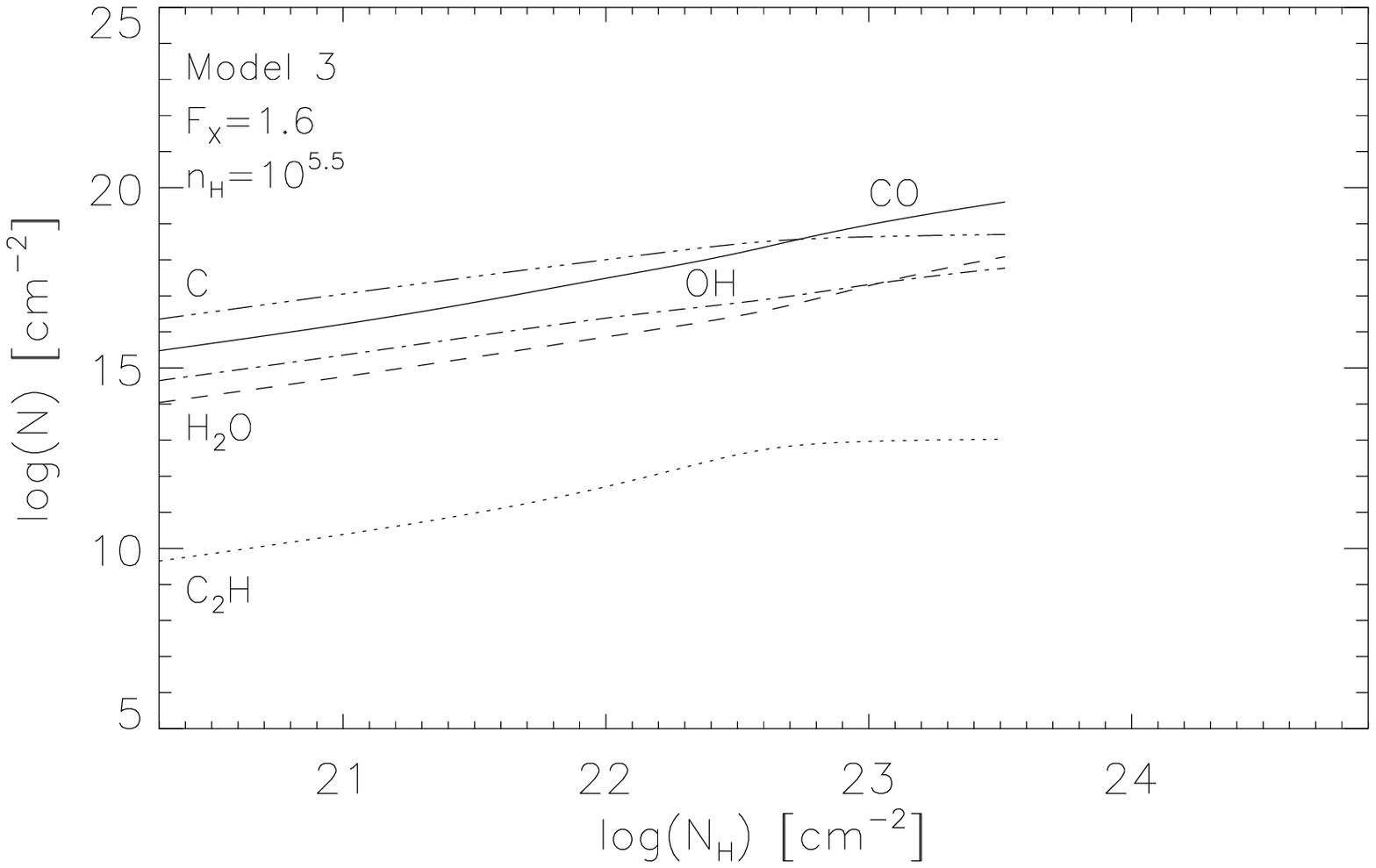}}
\end{minipage}
\hfill
\begin{minipage}[t]{5cm}
\resizebox{4.9cm}{!}{\includegraphics*{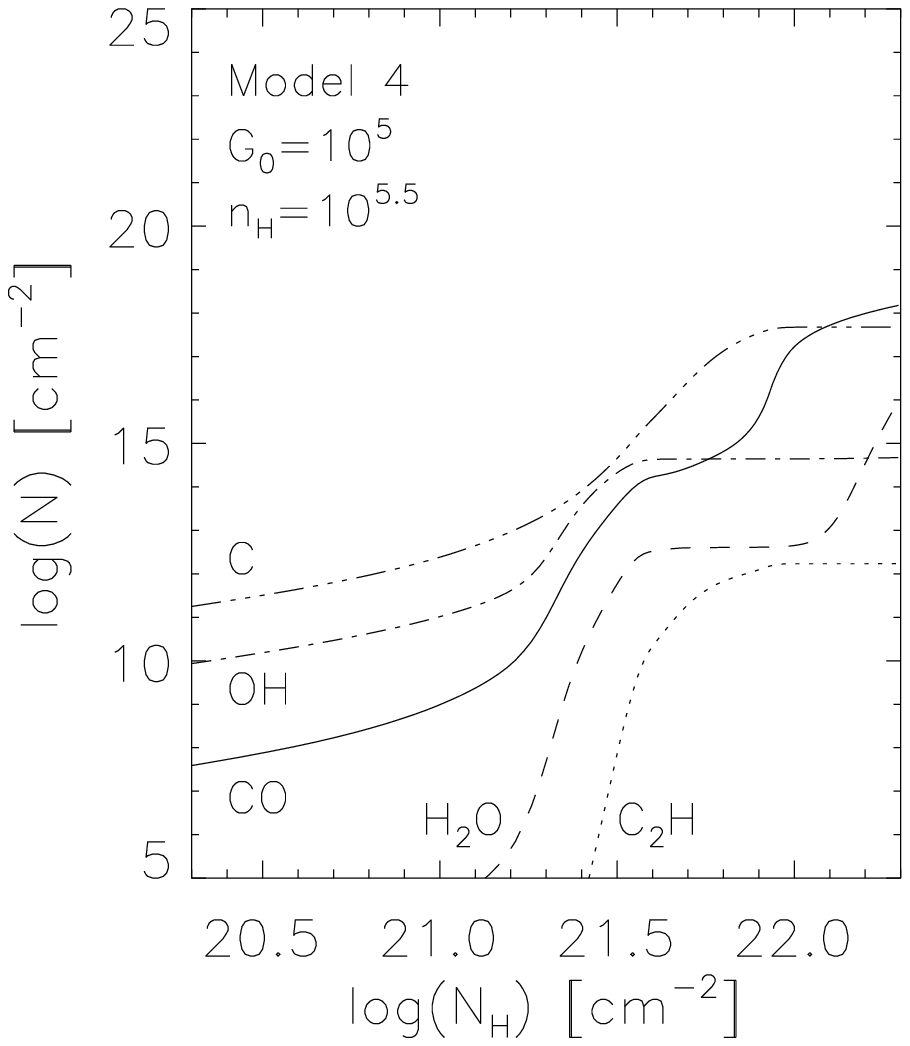}}
\end{minipage}
\hfill
\begin{minipage}[t]{11cm}
\resizebox{9.05cm}{!}{\includegraphics*{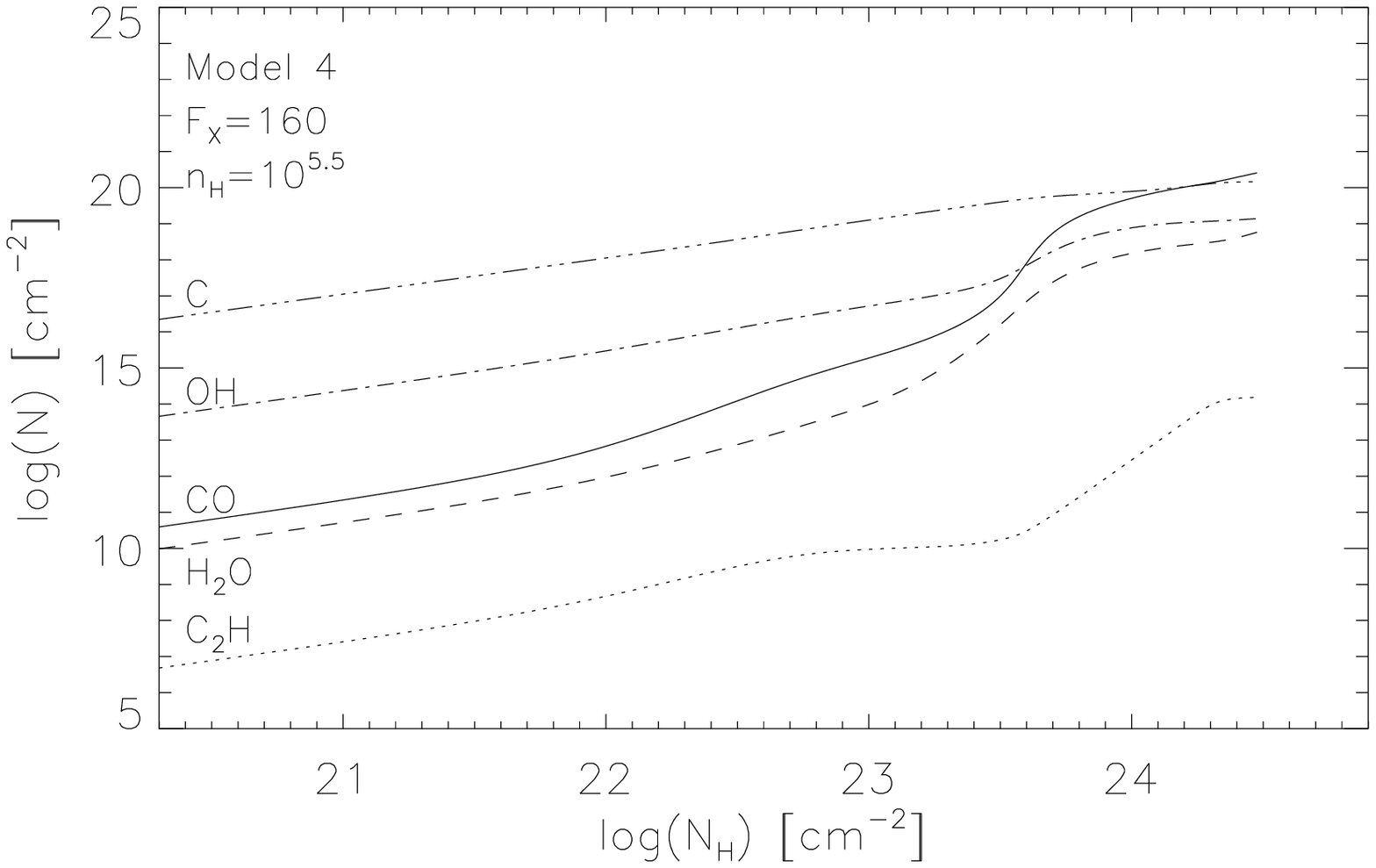}}
\end{minipage}
\caption[] {Cumulative column densities of C (dotted-dashed), CO
(solid), C$_2$H (dotted), H$_2$O (dashed) and OH (dot-dashed), for PDR
(left) and XDR (right) models.}
\label{coldens1}
\end{figure*}

\begin{figure*}[!ht]
\unitlength1cm
\begin{minipage}[b]{5cm}
\resizebox{4.9cm}{!}{\includegraphics*{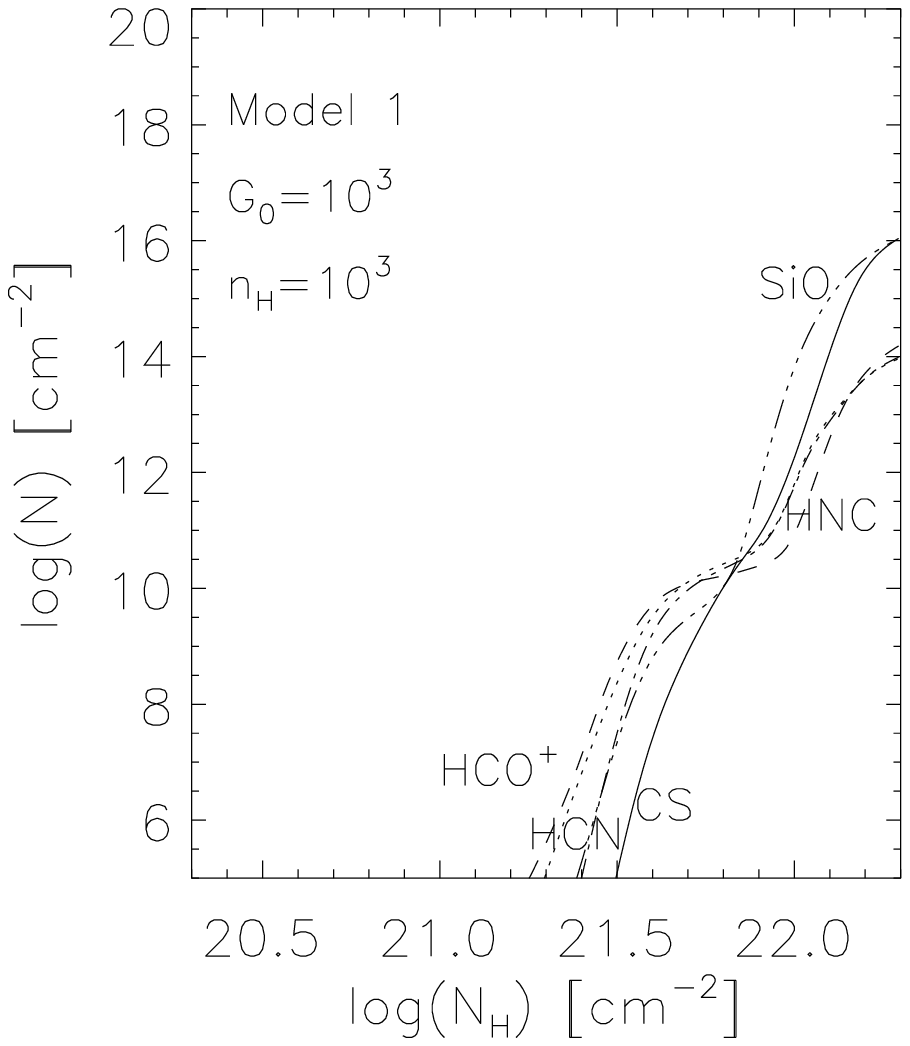}}
\end{minipage}
\hfill
\begin{minipage}[b]{11cm}
\resizebox{9.05cm}{!}{\includegraphics*{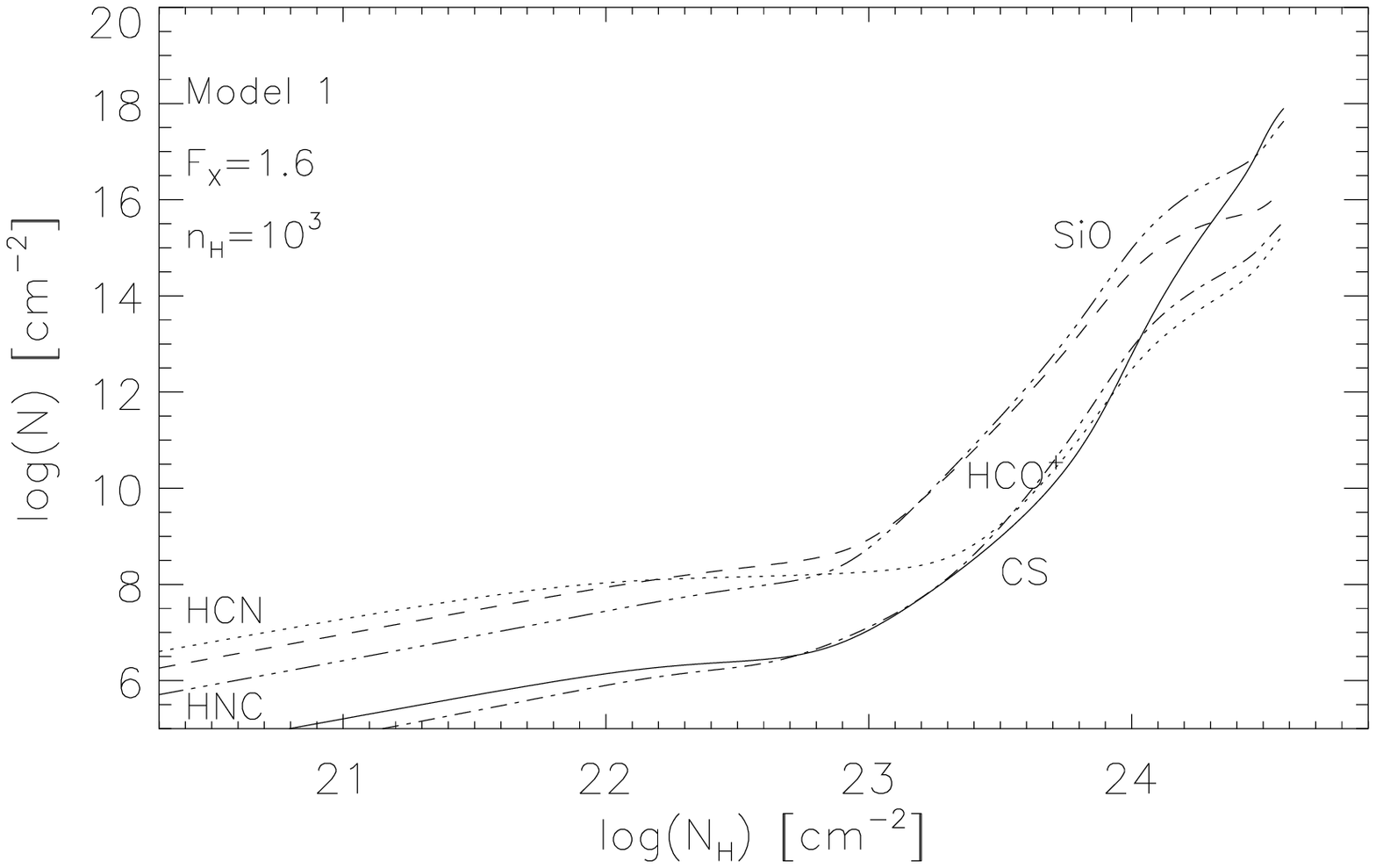}}
\end{minipage}
\hfill
\begin{minipage}[t]{5cm}
\resizebox{4.9cm}{!}{\includegraphics*{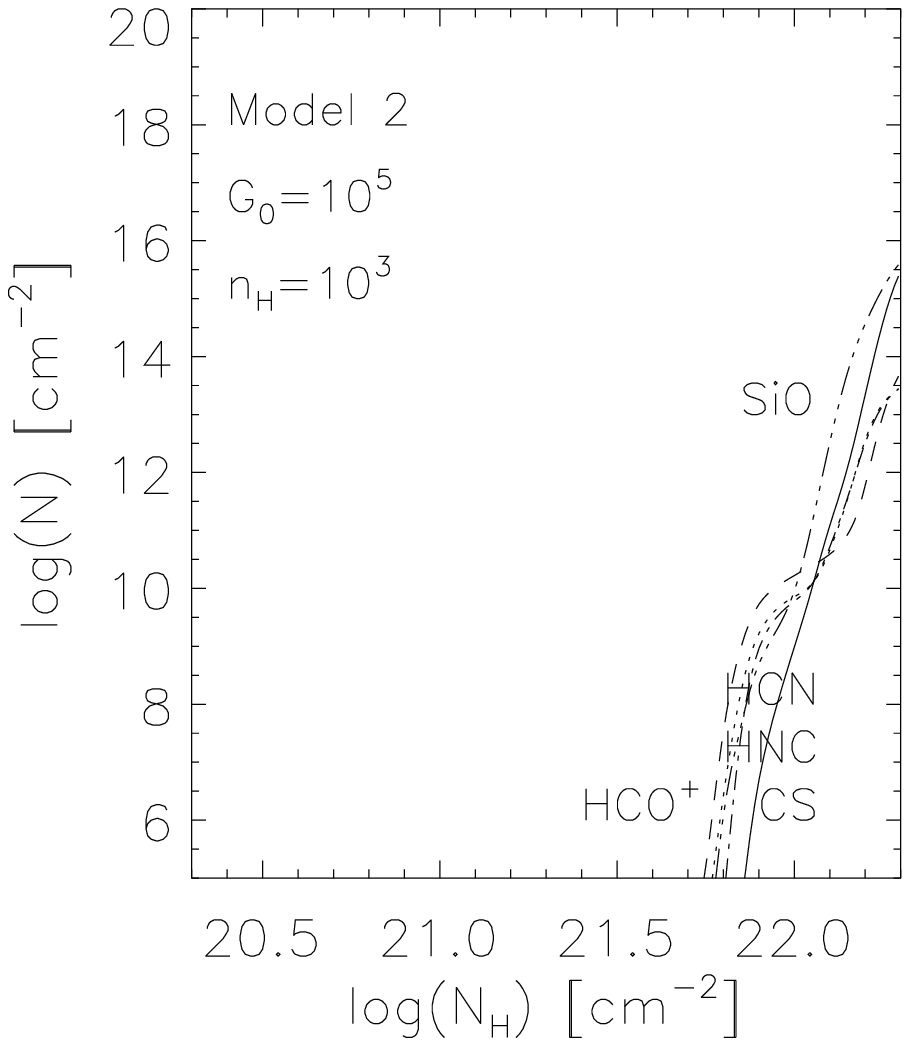}}
\end{minipage}
\hfill
\begin{minipage}[t]{11cm}
\resizebox{9.05cm}{!}{\includegraphics*{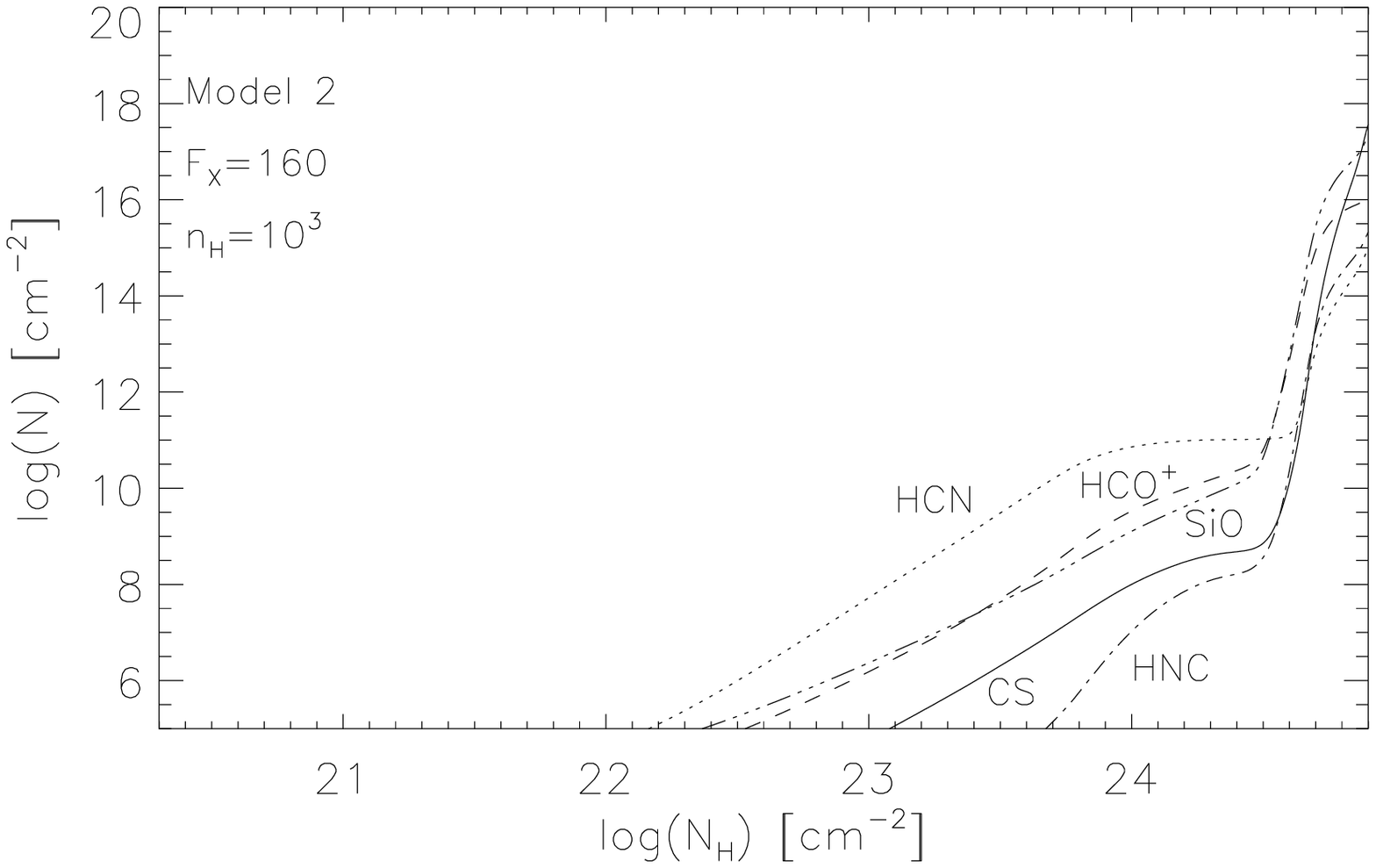}}
\end{minipage}
\hfill
\begin{minipage}[b]{5cm}
\resizebox{4.9cm}{!}{\includegraphics*{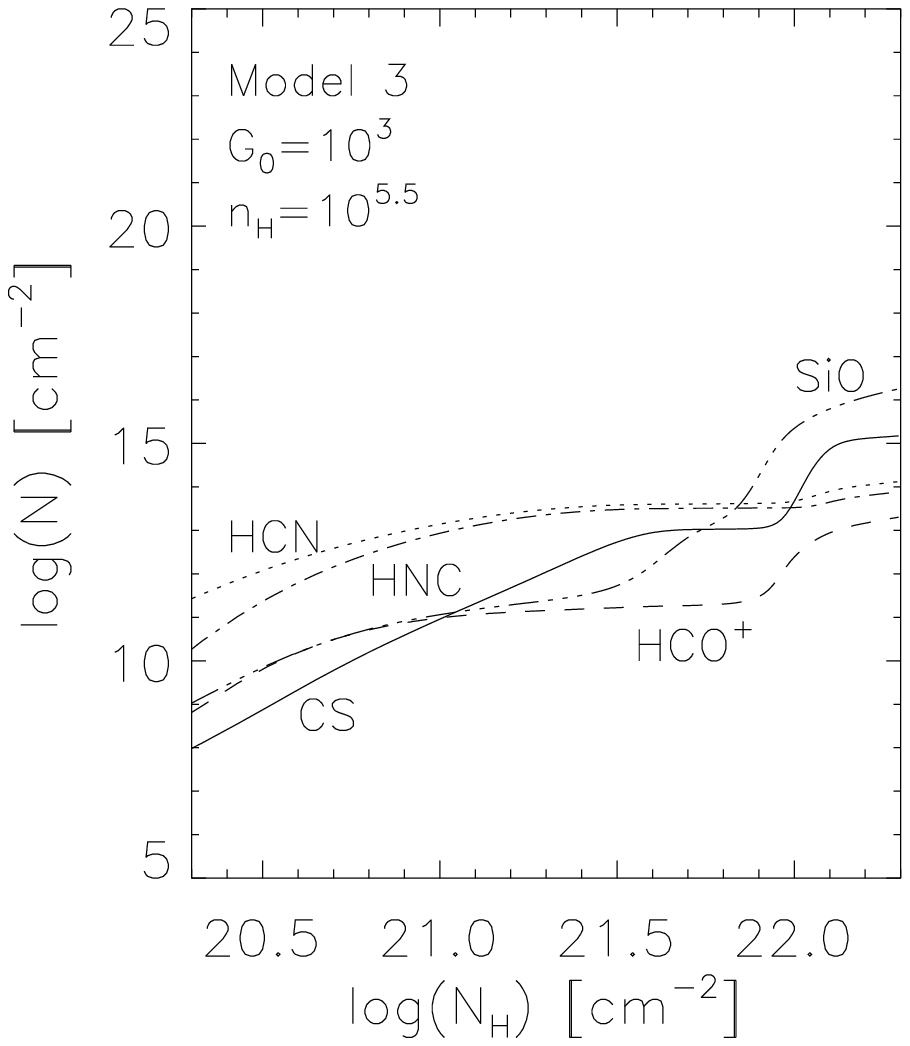}}
\end{minipage}
\hfill
\begin{minipage}[b]{11cm}
\resizebox{9.05cm}{!}{\includegraphics*{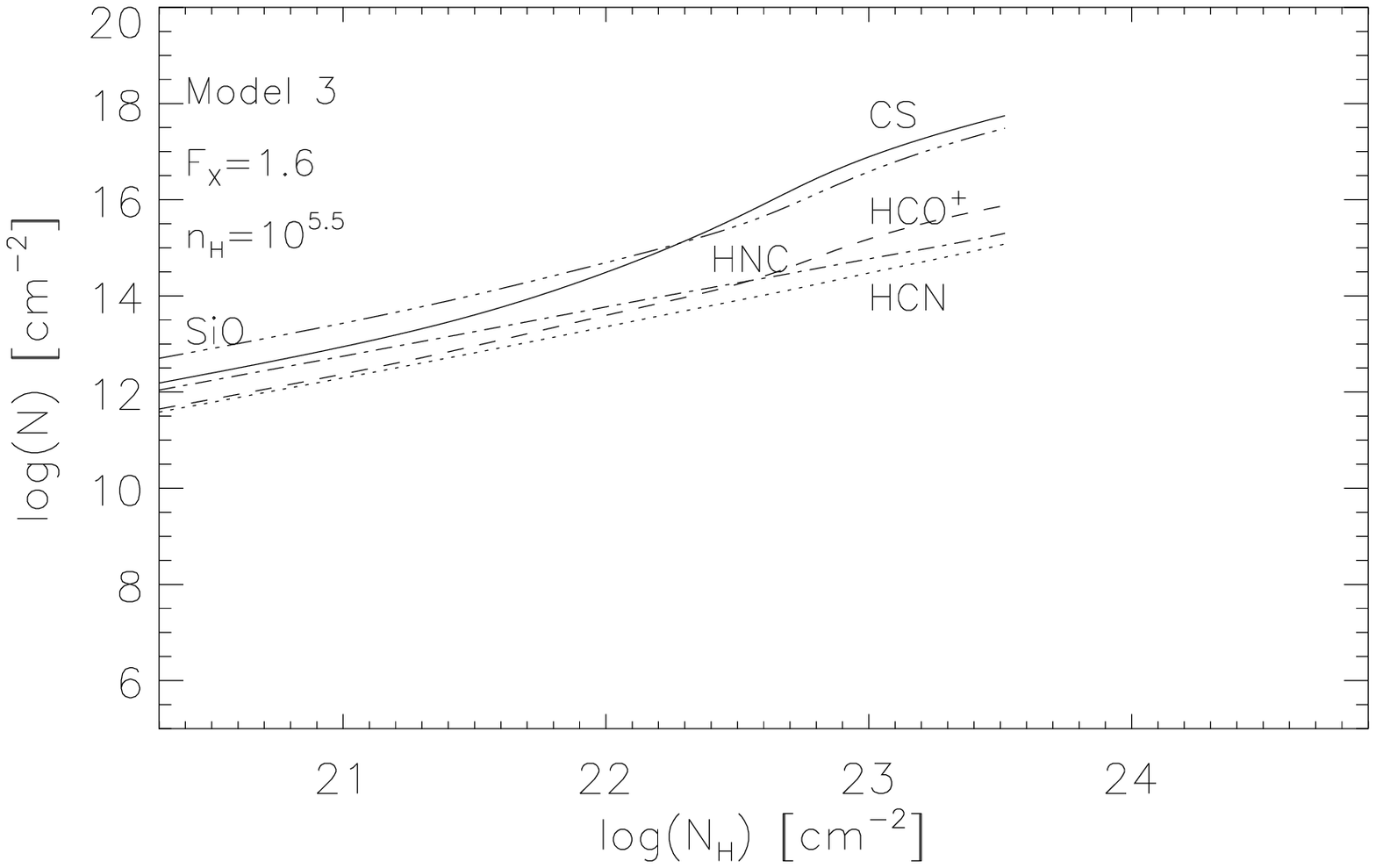}}
\end{minipage}
\hfill
\begin{minipage}[t]{5cm}
\resizebox{4.9cm}{!}{\includegraphics*{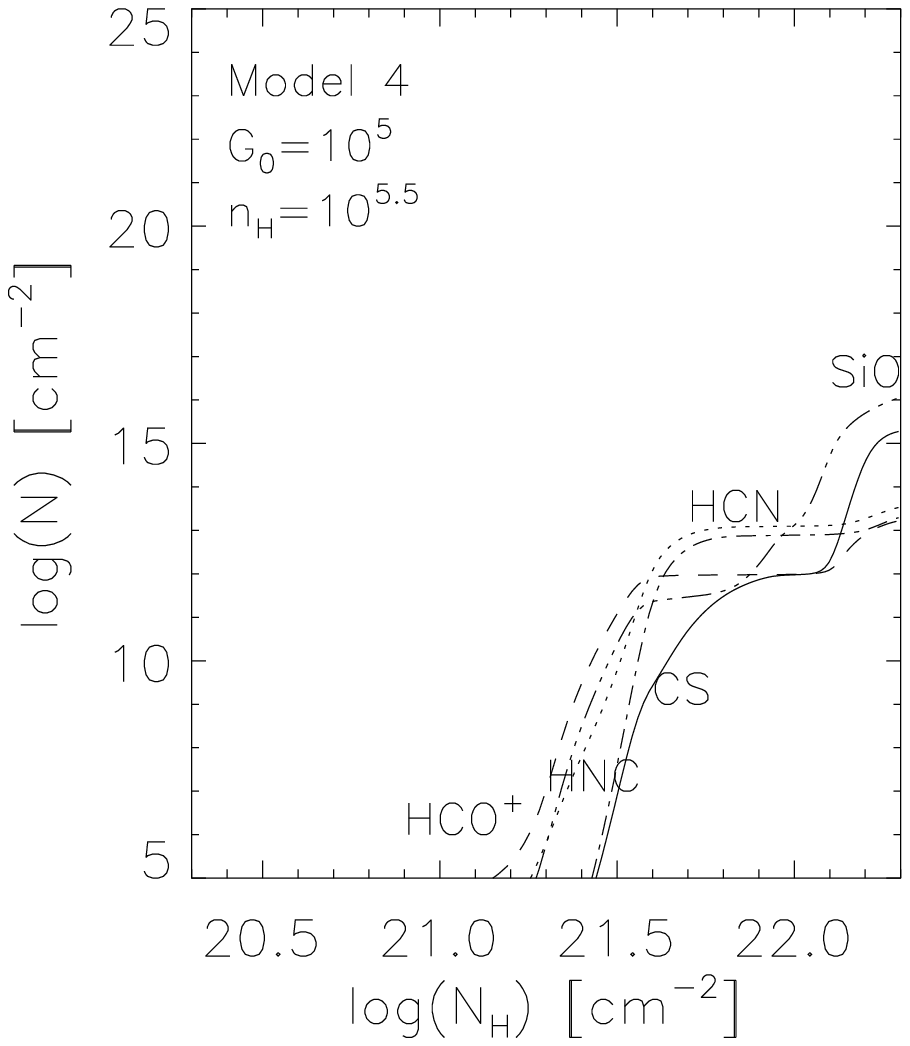}}
\end{minipage}
\hfill
\begin{minipage}[t]{11cm}
\resizebox{9.05cm}{!}{\includegraphics*{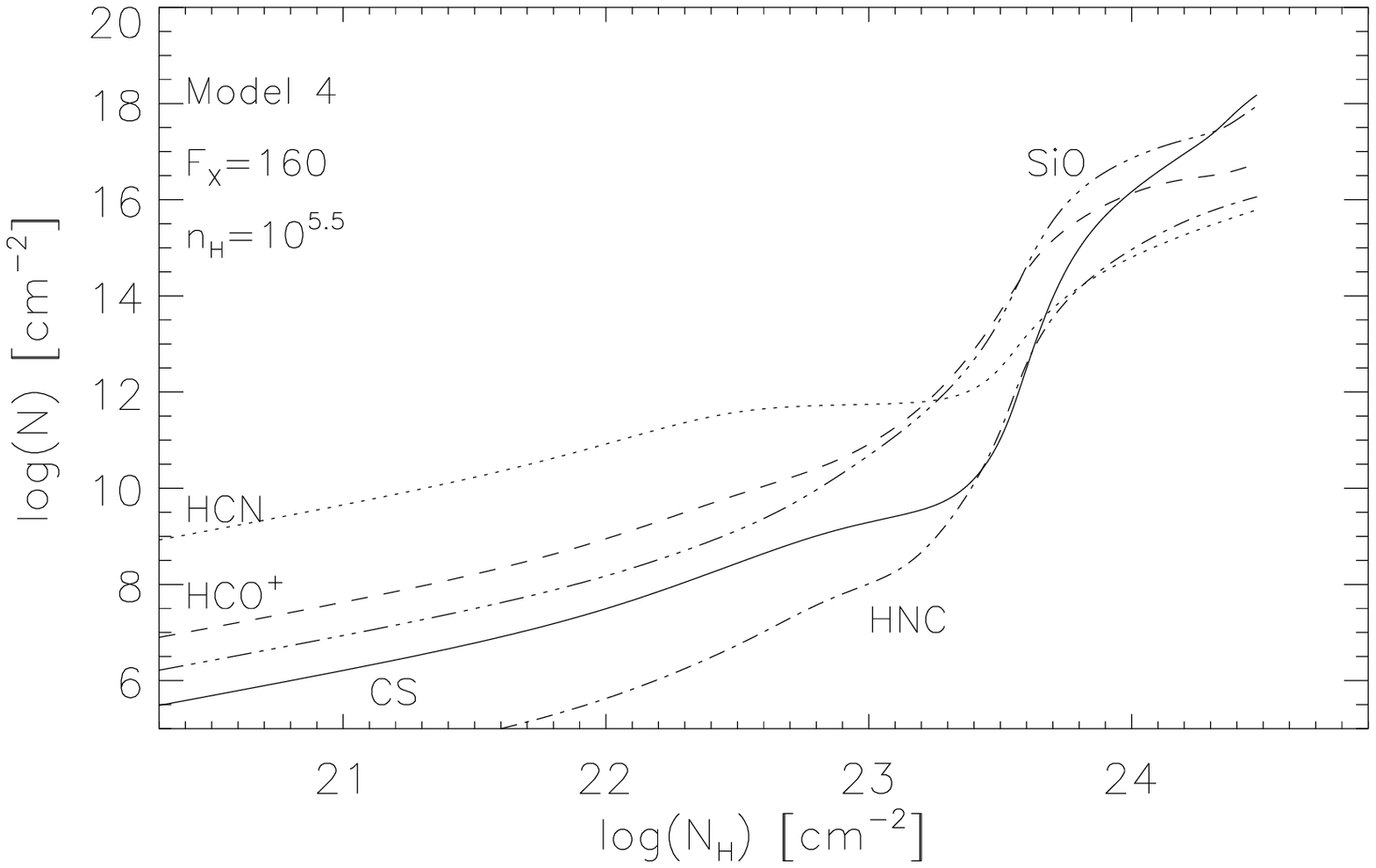}}
\end{minipage}
\caption[] {Cumulative column densities of CS (solid), HCN (dotted),
HCO$^+$ (dashed), HNC (dot-dashed) and SiO (dotted-dashed), for PDR
(left) and XDR (right) models.}
\label{coldens2}
\end{figure*}

In Figs. \ref{coldens1} and \ref{coldens2}, we show cumulative column
densities for selected species. They illustrate again that XDRs affect
whole cloud volumes and PDRs create layered structures. In PDRs, the
increase in column densities are very sudden for all species. For
example, C and CO show this due to the very distinct C$^+$/C/CO
transition. In the XDRs, however, the increases in column density
are much more gradual. The only sudden change in XDRs is where the
H/H$_2$ transition occurs.

\begin{figure*}[!ht]
\unitlength1cm
\begin{minipage}[b]{5cm}
\resizebox{4.9cm}{!}{\includegraphics*{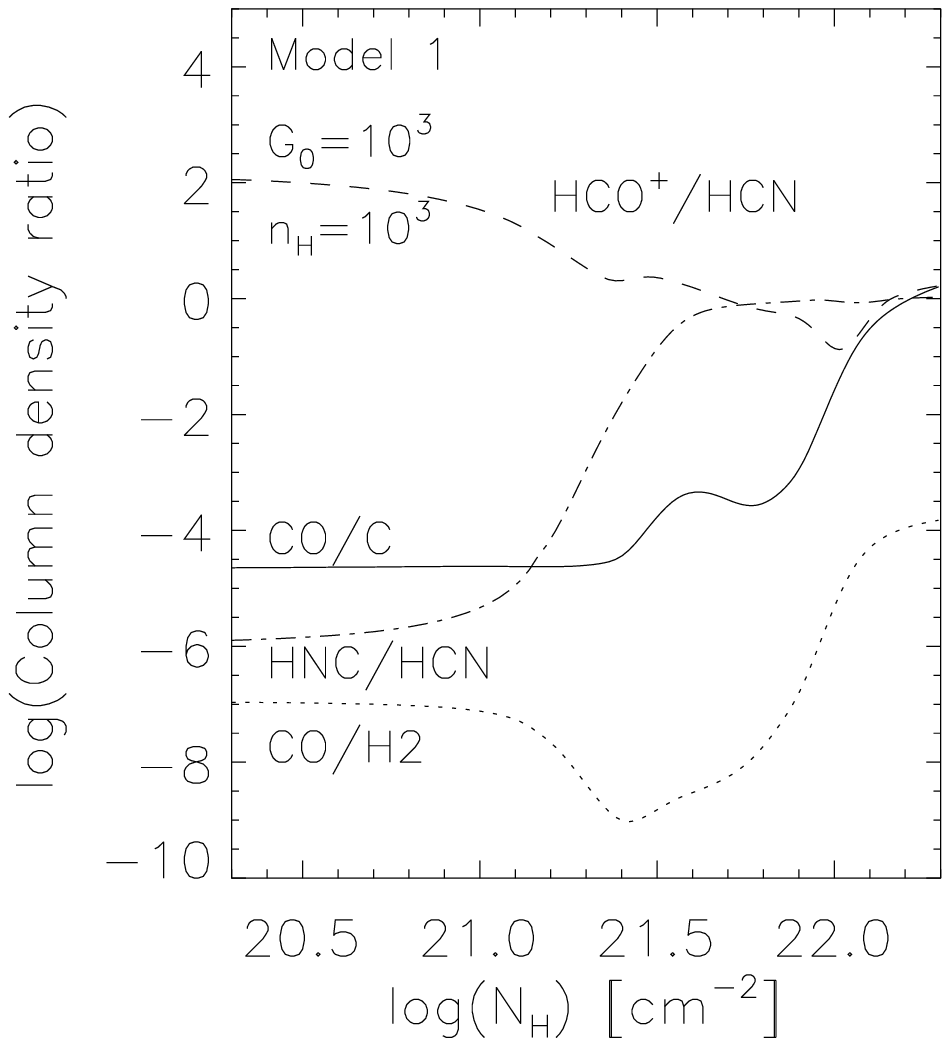}}
\end{minipage}
\hfill
\begin{minipage}[b]{11cm}
\resizebox{9.05cm}{!}{\includegraphics*{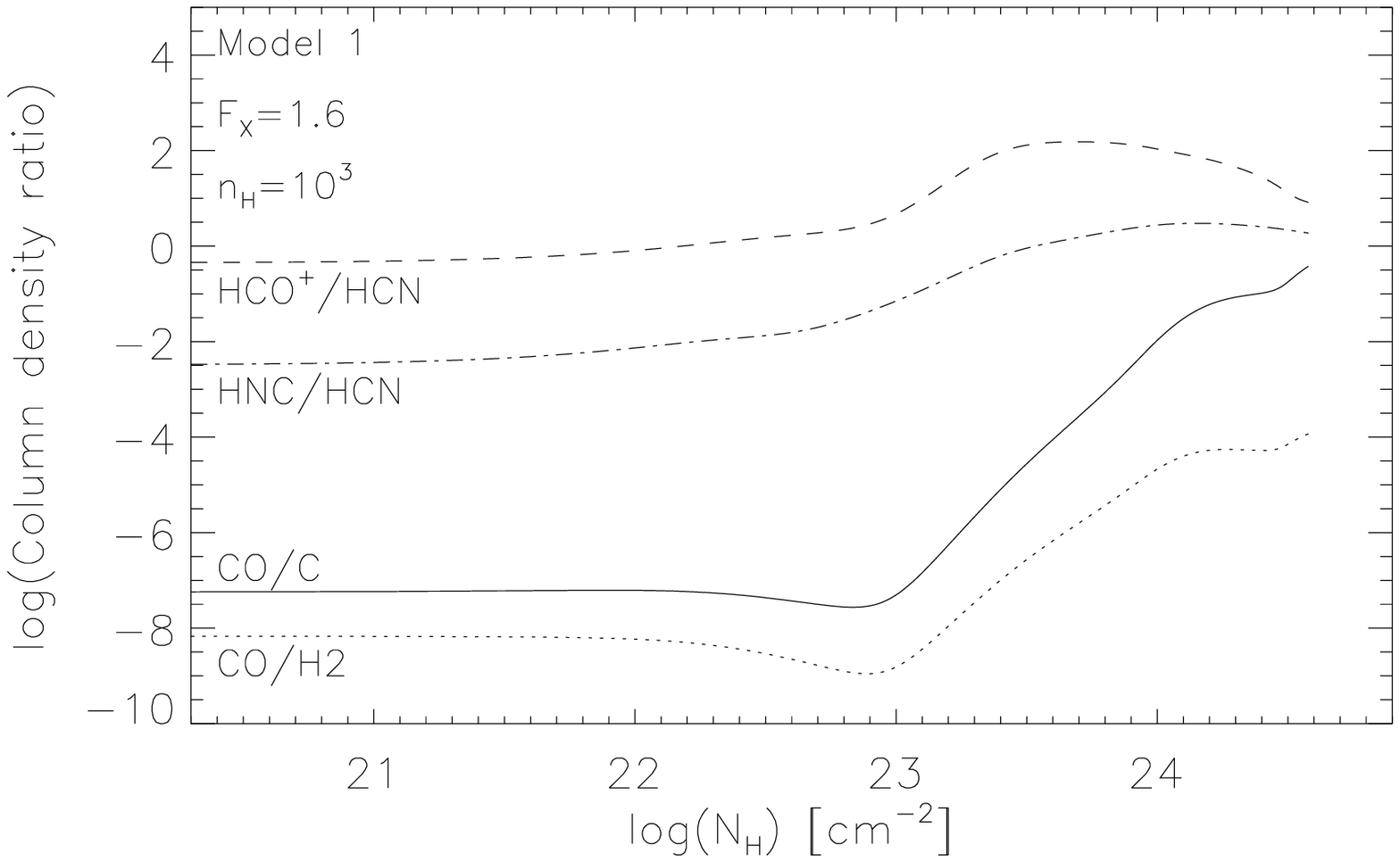}}
\end{minipage}
\hfill
\begin{minipage}[t]{5cm}
\resizebox{4.9cm}{!}{\includegraphics*{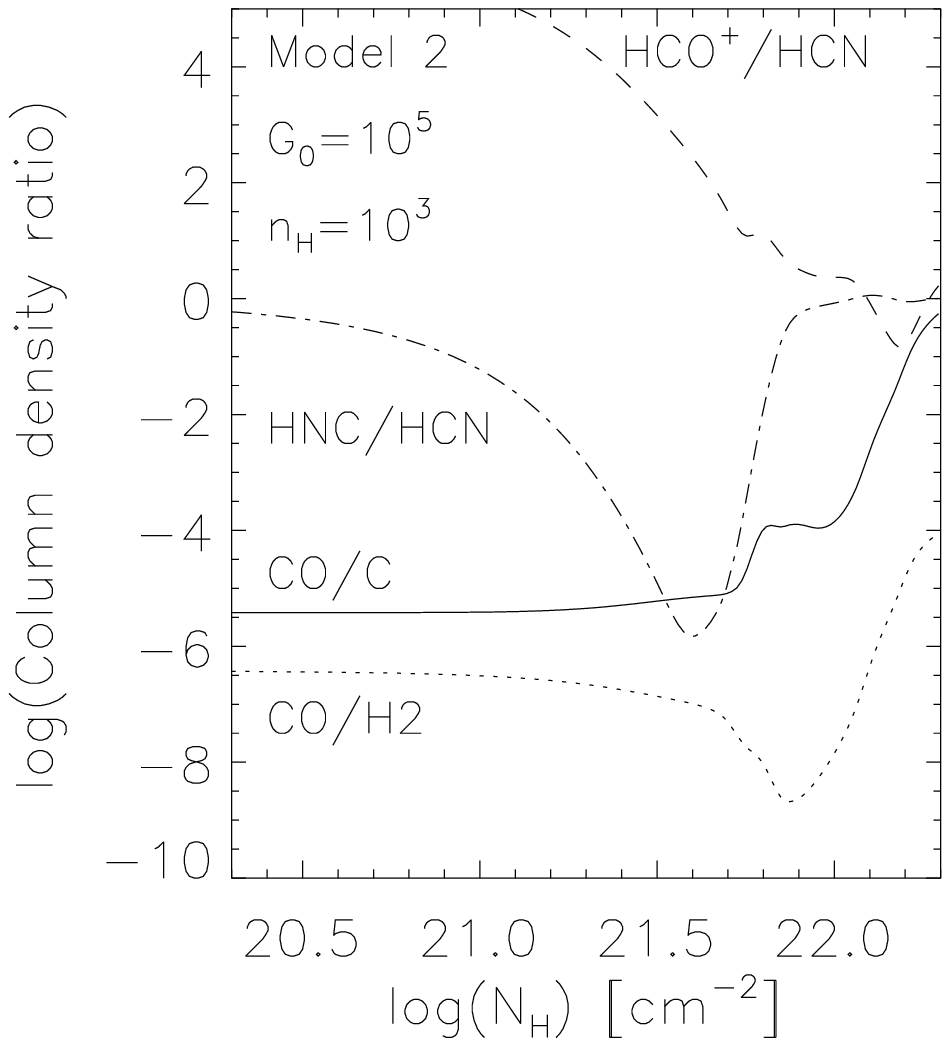}}
\end{minipage}
\hfill
\begin{minipage}[t]{11cm}
\resizebox{9.05cm}{!}{\includegraphics*{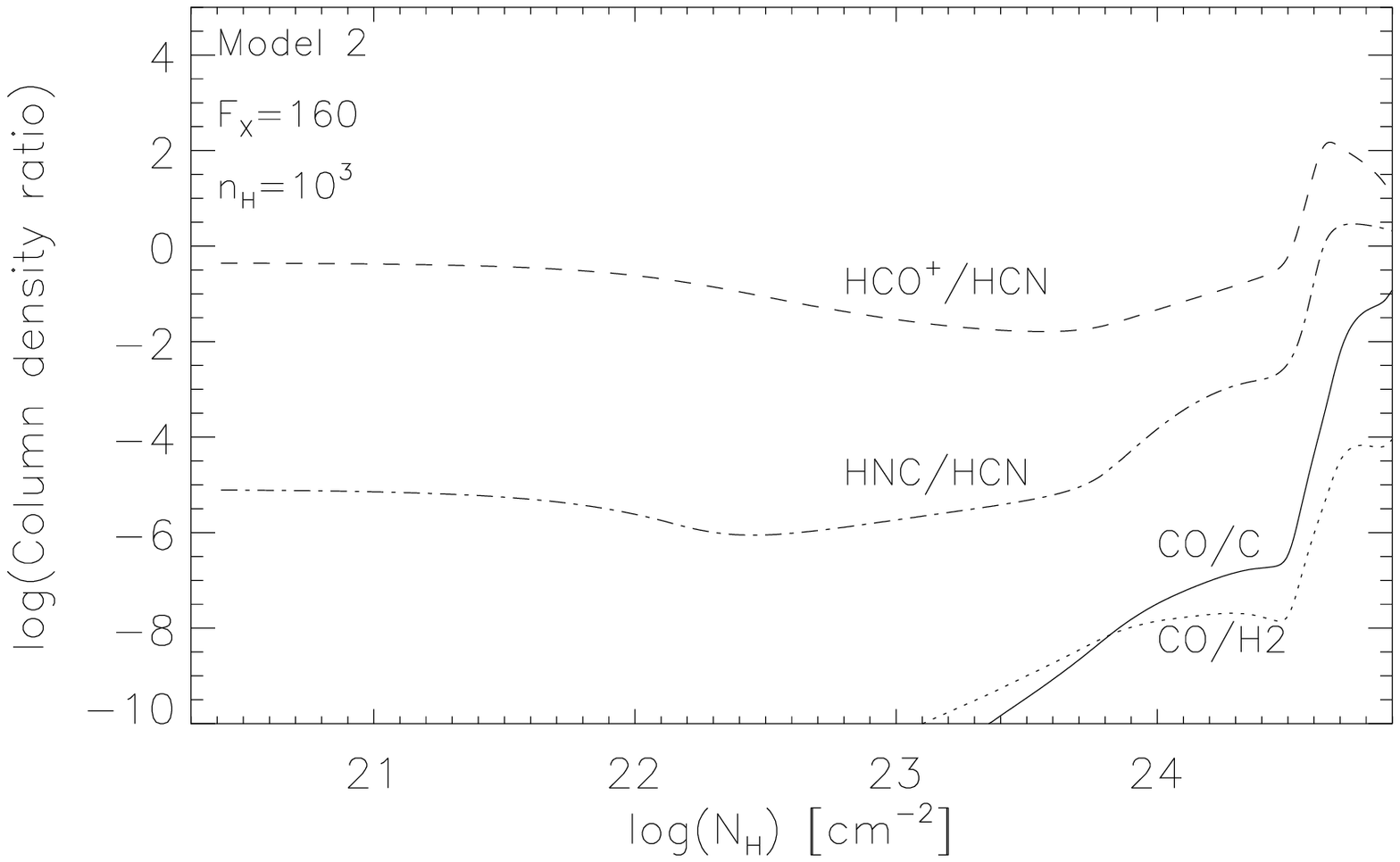}}
\end{minipage}
\hfill
\begin{minipage}[b]{5cm}
\resizebox{4.9cm}{!}{\includegraphics*{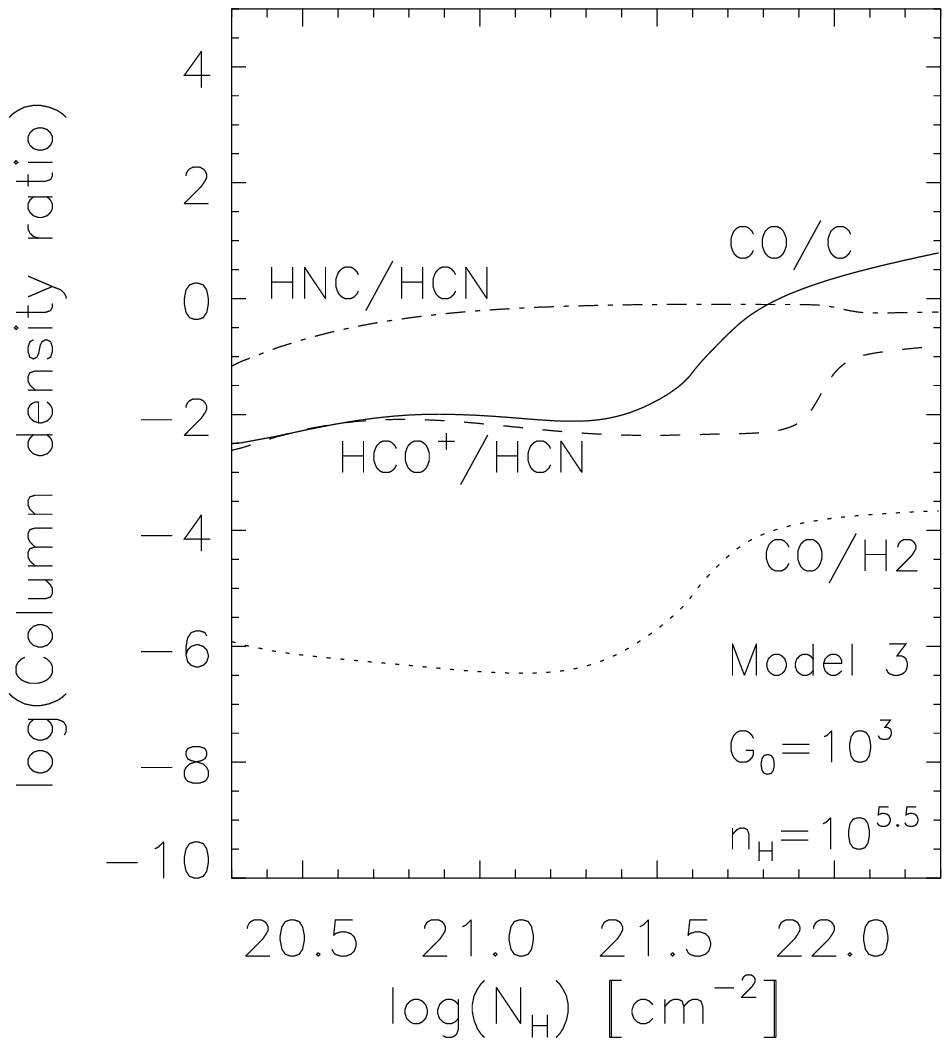}}
\end{minipage}
\hfill
\begin{minipage}[b]{11cm}
\resizebox{9.05cm}{!}{\includegraphics*{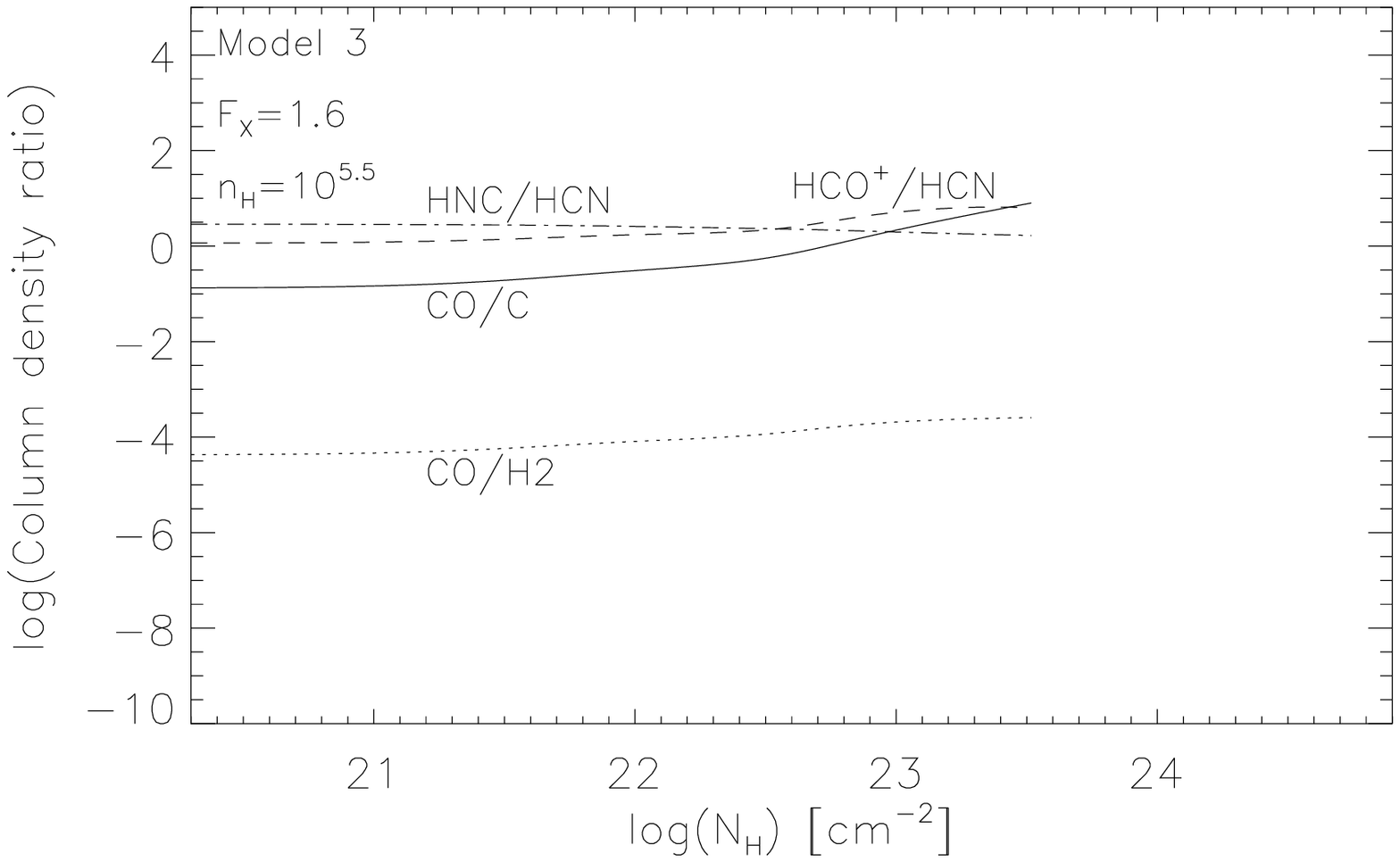}}
\end{minipage}
\hfill
\begin{minipage}[t]{5cm}
\resizebox{4.9cm}{!}{\includegraphics*{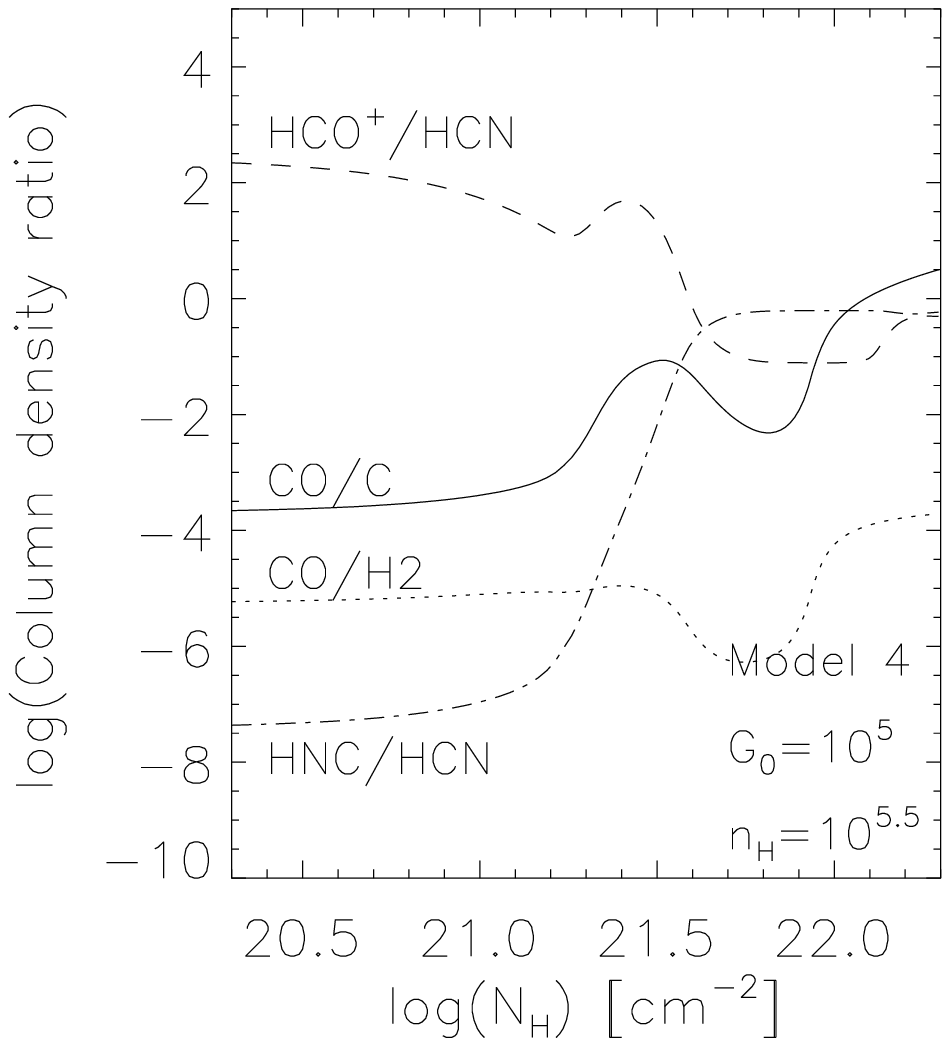}}
\end{minipage}
\hfill
\begin{minipage}[t]{11cm}
\resizebox{9.05cm}{!}{\includegraphics*{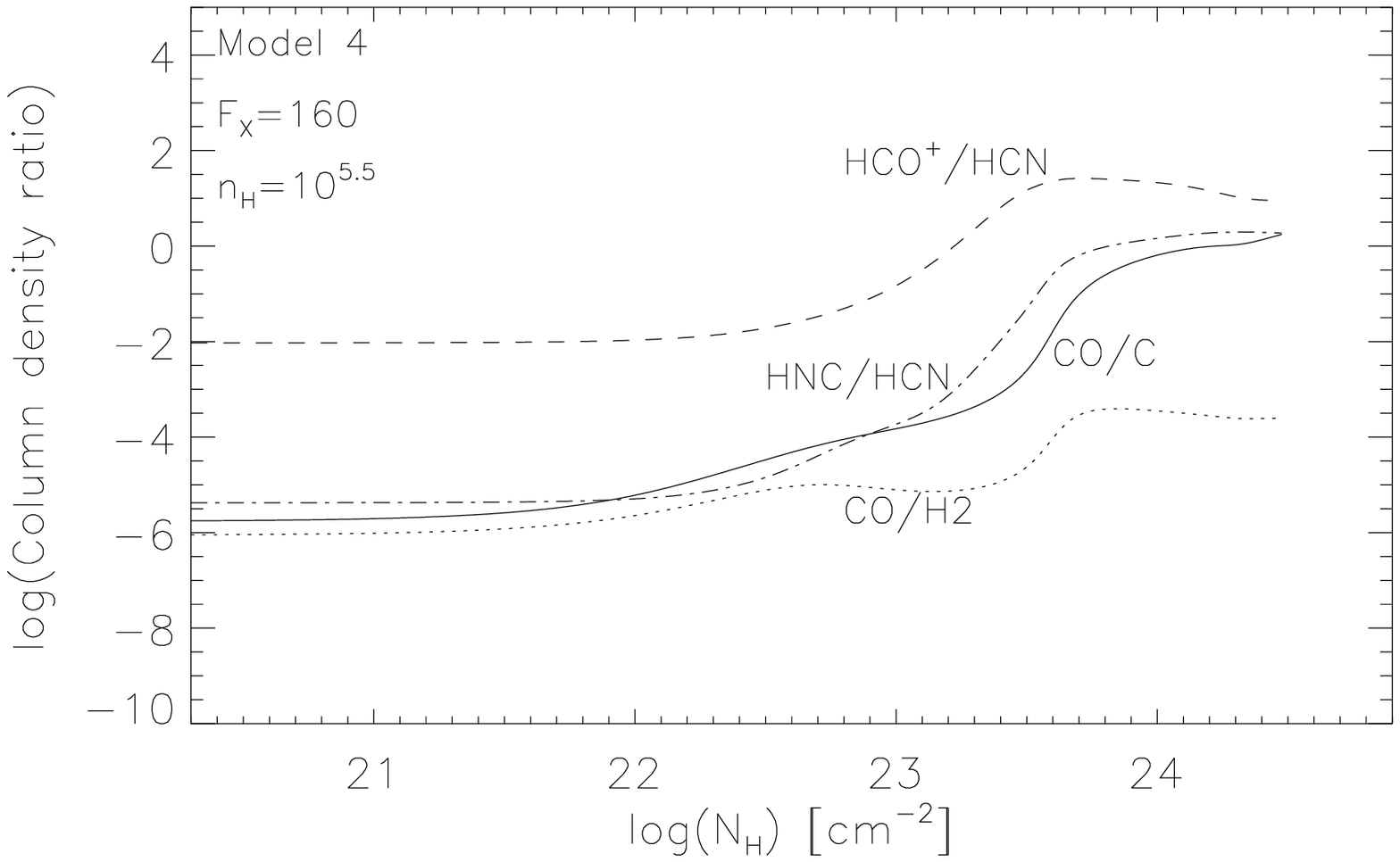}}
\end{minipage}
\caption[] {Column density ratios CO/C (solid), CO/H$_2$ (dotted),
HCO$^+$/HCN (dashed) and HNC/HCN (dot-dashed), for PDR (left) and XDR
(right) models.}
\label{ratios}
\end{figure*}

In Fig. \ref{ratios}, the cumulative column density ratios for
CO/H$_2$, CO/C, HNC/HCN, and HCO$^+$/HCN are shown as a function of
total hydrogen column density. The ratios for the XDRs
are almost constant upto $N_{\rm H} \approx 10^{22}\ {\rm cm^{-2}}$,
unlike those in PDR models. In PDRs, CO/C ratios increase by
approximately four orders of magnitude from the edge ($\le 10^{-4}$)
to $N_{\rm H}=10^{22.3}\ {\rm cm^{-2}}$ ($\ge 1$). In XDRs, this ratio
is constant to $N_{\rm H} \approx 10^{22}\ {\rm cm^{-2}}$ and then
increases slowly. For each cloud size, while keeping the energy input the
same, CO/C ratios increase at higher densities. The ratios go down
for higher radiation fields. For the same density and energy input,
CO/C is lower when the cloud is irradiated by X-ray photons, with the
exception for Model 3 where this is only valid at $N_{\rm H} >
10^{21.7}\ {\rm cm^{-2}}$. CO/H$_2$ is somewhat more complex. When
only the energy input is increased in PDRs, this ratio is higher when
$N_{\rm H} < 10^{21}\ {\rm cm^{-2}}$. For $N_{\rm H} = 10^{22.3}\ {\rm
cm^{-2}}$, the ratios are about the same. There is also a minimum
where the H/H$_2$ transition occurs. This minimum is more prominent
for higher radiation fields. In XDRs, the CO/H$_2$ ratio is lower when
the radiation field is higher. In PDRs and XDRs, the CO/H$_2$ ratios
are higher when the density is increased. When the cloud is irradiated
by X-ray photons, CO/H$_2$ ratios are lower, with the exception for
Model 3 again at $N_{\rm H} < 10^{21.5}\ {\rm cm^{-2}}$. In PDR Models
1, 2, and 3, significant column densities for HCN, HNC and HCO$^+ $
are reached between $N_{\rm H} = 10^{21.5}$ and $10^{22}\ {\rm
cm^{-2}}$. Therefore, the HNC/HCN and HCO$^+$/HCN ratios discussed are
for column densities $N_{\rm H} > 10^{22}\ {\rm cm^{-2}}$. In PDRs,
HNC/HCN is lower when the density is higher. No significant changes
are seen for different radiation fields at these columns. HNC/HCN is
generally lower for high $H_X/n$ in XDRs. At high column densities,
where $H_X/n$ is low, HCN/HNC ratios are equal or somewhat higher than
those for the PDR. HCO$^+$/HCN and HNC/HCN are of the same order in
PDRs, but in XDRs HCO$^+$/HCN is higher in most cases.

\begin{acknowledgements} 
We are indebted to Frank Israel, for initiating this project, and for
his helpful suggestions and remarks. We are greatful to Ewine van
Dishoeck for useful discussions on PDR and XDR physics. We thank the
anonymous referee for his careful reading of the manuscript and his
constructive comments.
\end{acknowledgements} 

\bibliographystyle{aa} 

\appendix

\section{Heating processes in PDRs}

\subsection{Photo-electric emission}
 
In PDRs the photo-electric emission from (small) dust grains and PAHs
is the dominant heating source. 
We use the analytical expression given by \citet{Bakes1994} 
which is given by

\begin{equation}
\Gamma_{\rm grain} = 10^{-24}\ \epsilon\ G'_{0,\rm dust}\ n_{\rm H} \
{\rm erg\ cm^{-3}\ s^{-1}},
\end{equation}

\noindent where $G'_{0,\rm dust}$ is the radiation field attenuated by
dust absorption \citep{Black1977} and the heating efficiency $\epsilon
$ is given by

\begin{eqnarray}
\epsilon & = & \frac{4.87 \cdot 10^{-2}}{[1 + 4 \cdot 10^{-3}(G_0 T_{\rm k}^{1/2}/n_{\rm e})^{0.73}]} + \nonumber\\ 
         &   & \frac{3.65 \cdot 10^{-2} (T_{\rm k}/10^4)^{0.7}}{[1 + 2 \cdot 10^{-4}(G_0 T_{\rm k}^{1/2}/n_{\rm e})]}.
\end{eqnarray}



\noindent Note that the efficiency depends on the ratio $G_0\ T_{\rm
k}^{1/2} / n_{\rm e}$, which is the ratio of the ionisation and
recombination rates. $A_V$ is the visual extinction at optical
wavelengths caused by interstellar dust. \citet{Bohlin1978} relate the
total column density of hydrogen, $N_{\rm H} = N({\rm H}) + 2 N({\rm
H_2})$ to colour excess, $E(B-V)$:

\begin{eqnarray}
\frac{N_{\rm H}}{E(B-V)} = 5.8 \times 10^{21}\ {\rm cm^{-2}\ mag^{-1}}.  
\end{eqnarray}

\noindent The visual extinction then follows consequently: $A_V = 3.1
E(B-V)$ and $A_V=5.34 \times 10^{-22}N_{\rm H}$. Note that the results
of \citet{Savage1977} are often used, but in this paper only H$_2$
(and not \hi) is taken into account. 


\subsection{Carbon ionisation heating}

At the edge of the cloud, most of the carbon is singly ionised. The
photo-electron energy released in an ionisation is $\Delta E_{\rm C} =
1.06$~eV. The ionisation rate, at a certain point in the cloud, is
given by $\kappa_{\rm ion} = 1.76 G'_{0, \rm carbon}\ \rm s^{-1}$. The
heating rate due to the ionisation of carbon is then given by

\begin{eqnarray}
\Gamma_{\rm C} = \kappa_{\rm ion}\ n({\rm C}) \ \Delta E_{\rm C}.  
\end{eqnarray}

\noindent After substitution of numerical values we get the following
heating rate for the local radiation field $G'_{0,\rm carbon}$:

\begin{eqnarray}
\Gamma_{\rm C} = 2.79 \times 10^{-22}\ n({\rm C})\ G'_{0,\rm carbon}\
{\rm erg\ cm^{-3}\ s^{-1}},
\end{eqnarray}

\noindent where this time $G'_{0,\rm carbon}$ is the radiation field
attenuated by dust absorption \citep{Black1977}, carbon
self-absorption \citep{Werner1970} and H$_2$ \citep{deJong1980}.





\subsection{H$_2$ photo-dissociation heating}

Absorption of Lyman-Werner band photons leads to the excitation of
H$_2$. About 10\% of the excitations leads to decay into the continuum
of the ground electronic state \citep{Field1966, Stecher1967}. The
heating related to this dissociation is given by

\begin{eqnarray}
\Gamma = 0.1\ \kappa_{\rm exc.} < E_{\rm diss} >,
\end{eqnarray}

\noindent where $<E_{\rm diss}>$ is the mean kinetic energy of the H
atoms and is set to 0.4~eV \citep{Spaans1996}. The excitation rate of
H$_2$ is given by $\kappa_{\rm exc.}=3.4 \times 10^{-10}\ G'_{0, \rm
H_2}$, where $G'_{0, \rm H_2}$ is the local radiation field given by

\begin{eqnarray}
G'_{0, \rm H_2} = \beta(\tau) G_0\ {\rm exp}(-2.5\ A_V). 
\end{eqnarray} 

\noindent Self-shielding is explicitly taken into account for the
excitation of H$_2$, by the introduction of the shielding factor
$\beta(\tau)$ (see \ref{shielding}). After substitution of numerical
values we get a heating rate of

\begin{eqnarray}
\Gamma_{\rm H_2}& =& 2.2 \times 10^{-23}\ \beta(\tau)\ G_0\\ \nonumber
                &&{\rm exp}(-2.5\ A_V)\ {\rm erg\ cm^{-3}\ s^{-1}}.
\end{eqnarray}

\subsection{H$_2$ collisional de-excitation heating}

FUV excitation is followed by decay to ro-vibrational levels in the
ground state. Collisional de-excitation leads to gas heating. This
cascade process is very complicated, but we simplify this process by
using a two-level approximation (see \ref{H2excit}). The resulting
heating rate is given by

\begin{eqnarray}
\Gamma_{\rm H_2} = [n({\rm H}) \gamma^{\rm H}_{10} + n({\rm H_2}) \gamma^{\rm H_2}_{10}] n({\rm H_2V})E_*\ {\rm erg\ cm^{-3}\ s^{-1}},
\end{eqnarray} 

\noindent where the coefficients are given by \citet{Hollenbach1979}

\begin{eqnarray}
\gamma^{\rm H}_{10} &=& 10^{-12}\ T_{\rm k}^{0.5} {\rm exp}(-1000 / T_{\rm k})  \\
\gamma^{\rm H_2}_{10} &=& 1.4 \times 10^{-12} T_{\rm k}^{0.5} {\rm exp}(-18100 / (T_{\rm k} + 1200)).
\end{eqnarray}

\noindent Both of the above expression are in units of ${\rm cm^3\
s^{-1}}$.

\subsection{Gas-grain collisional heating}\label{gas_grain_heating}

When gas and grains differ in temperature they can transfer heat
through collisions. The heating rate of the gas is given by
\citep{Hollenbach1979,Hollenbach1989}

\begin{eqnarray}
\Gamma_{\rm coll.} &=& 1.2\times 10^{-31} n^2\left(\frac{T_{\rm k}}{1000}\right)^{1/2}\left(\frac{100\ {\rm \AA}}{a_{\rm min}}\right)^{1/2}\\ \nonumber
                   & &\times[1-0.8\exp(-75/T_{\rm k})](T_{\rm d} - T_{\rm k}).
\end{eqnarray}

\noindent The minimum grain size is set at $a_{\rm min}=10$~\AA\ and the
dust temperature $T_{\rm d}$ is given by

\begin{eqnarray}
T_{\rm d} &=& (8.9 \times 10^{-11} \nu_0 G_0 \exp(-1.8 A_V) + \\ \nonumber
    & &2.7^5 + 3.4 \times 10^{-2} [0.42 - \ln(3.5 \times 10^{-2} \tau_{100} T_0)] \\ \nonumber 
    & & \times \tau_{100} T_0^6)^{0.2}, 
\end{eqnarray}

\noindent based on the results of \citet{Hollenbach1991}


\subsection{Gas-grain viscous heating}

Radiation pressure accelerates grains relative to the gas and the
resulting drag contributes viscous heating to the gas. Grain
acceleration time scales are short compared to other time scales, and
therefore the grains may be considered moving at their local drift
velocity, $v_{\rm d}$. All the momentum is transferred to the gas,
predominantly by Coulomb forces. For drift velocities $v_d
<10^3$~cm~s$^{-1}$ \citep{Spitzer1978}, no significant gas-grain
separation takes place. In the following we take $v_d =
10^2$~cm~s$^{-1}$. The heating rate is given by

\begin{eqnarray}
\Gamma_{\rm visc.} &=& 8 \pi e^4 n_{\rm d} Z_{\rm d}^2 (kT_{\rm k})^{-1} (\ln \Lambda) v_{\rm d} \nonumber \\ 
       & & [n({\rm C^+}) G(y_{\rm C^+}) + n_{\rm e} G(y_{\rm e})],
\end{eqnarray}

\noindent where $n_{\rm d}$ is the grain volume density, $Z_{\rm d}$
is the grain charge, $n({\rm C}^+)$ and $n_{\rm e}$ are the respective
$\rm C^+$ and electron volume densities and the functions $\Lambda$
and $G(y)$ are given by

\begin{eqnarray}
\Lambda &=& 1.5\ Z_{\rm d}^{-1} e^{-3} (kT_{\rm k})^{1.5} (\pi n_{\rm e})^{-0.5} \\
G(y)    &=& \frac{1}{2y^2}\lbrace {\rm erf}(y) - \frac{2}{\pi^{0.5}} y e^{-y^2} \rbrace,
\end{eqnarray}

\noindent where $y=v_{\rm d}/v_{\rm th}$ and $v_{\rm th}$ the thermal
velocity of C$^+$ ions and electrons. The error function ${\rm
erf}(y)$ is given by

\begin{eqnarray}
{\rm erf}(y) = \int_0^y e^{-t^2}dt
\end{eqnarray}

\subsection{Cosmic-ray heating}


At large column densities, cosmic ray heating can become
important. \citet{Glassgold1973} and \citet{Cravens1978} calculated
that the amount of heat deposited in a molecular gas is about 8~keV
per primary ionisation. Then, \citet{Tielens1985} find for the total
heating rate, including helium ionisation

\begin{equation}
\Gamma_{\rm CR} = 1.5 \times 10^{-11}\ \zeta\ n({\rm H_2})\ {\rm erg\ cm^{-3}\ s^{-1}}, 
\end{equation}

\noindent where $\zeta$ is the cosmic ray ionisation rate per H$_2$ molecule.

\section{Heating processes in XDRs}

\subsection{Heating due to Coulomb interactions}

When X-rays are absorbed, fast electrons are produced. These fast
electrons lose part of their energy through Coulomb interactions with
thermal electrons, so the X-ray heating is given by

\begin{eqnarray}
\Gamma_{X} = \eta\ {\rm n}\ H_X,
\end{eqnarray}

\noindent where $\eta$ is the heating efficiency, depending on the
H$_2$/H ratio and the electron abundance $x$. We use the results of
\citet{Dalgarno1999}. Their calculated heating efficiency $\eta$ in an
ionised gas mixture is given by

\begin{eqnarray}
\eta = \frac{10r\eta_{\rm H_2 He} + \eta_{\rm He H}}{10r + 1},
\end{eqnarray}

\noindent where $r=n({\rm H}_2)/n({\rm H})$. $\eta_{\rm H_2 H}$ and
$\eta_{\rm He H}$ are the heating efficiencies for the ionised pure He
and H$_2$ mixture and the He and H mixture, respectively. Both are
parametrised through

\begin{eqnarray}
\eta' = 1 + (\eta_0 - 1) / (1 + cx^{\alpha}).
\end{eqnarray}

\noindent The values of $\eta_0$, $c$ and $\alpha$ are given in Table
7 of \citet{Dalgarno1999}, and $x$ is the electron fractional
abundance. It has to be modified when the H$_2$-He mixture is
considered:

\begin{eqnarray}
x' = \frac{1.83 x}{1 + 0.83 x}.
\end{eqnarray}

\subsection{Heating due to H$_2$ ionisation}\label{ion_heat}

H$_2$ ionisation can lead to gas heating \citep{Glassgold1973}. When
H$_2$ is ionised by a fast electron and subsequently recombines
dissociatively, about 10.9 eV ($1.75 \times 10^{-11}$ erg) of the
ionisation energy can go into kinetic energy. H$_2^+$ can also charge
transfer with H. This is an exothermic reaction, with an energy yield
of 1.88 eV, of which we assume half, 0.94 eV ($1.51 \times 10^{-12}$
erg), to go into heating. H$_2^+$ can also react to H$_3^+$, and
subsequently recombine dissociatively or react with other
species. \citet{Glassgold1973} argued that for every H$_3^+$ ion
formed 8.6 eV ($1.37\times 10^{-11}$ erg) goes into gas heating. The
H$_2$ ionisation rate cooling is then given by

\begin{eqnarray}
\Gamma_{\rm H_2\ ion} & = & \frac{17.5 k_{\rm e} x_{\rm e} + 1.51 k_{\rm H} x_{\rm H} + 13.7 k_{\rm H_2} 
x_{\rm H_2}}{k_{\rm e} x_{\rm e} + k_{\rm H} x_{\rm H} + k_{\rm H_2} x_{\rm H_2}} \times \\  \nonumber
&   & 10^{-12} \zeta_{\rm H_2}\ x_{\rm H_2}\ n \ {\rm erg\ cm^{-3} s^{-1}},
\end{eqnarray}

\noindent where $k_{\rm e}$, $k_{\rm H}$ and $k_{\rm H_2}$ are the rates of
dissociative recombination, charge transfer with hydrogen and the
reaction to H$_3^+$, respectively.

\subsection{Gas-grain collisional heating}

We use the results of Sect. \ref{gas_grain_heating}. The dust
temperature was found by \citet{Yan1997}:

\begin{eqnarray}
T_{\rm d} = 1.5 \times 10^4 (H_X/x_{\rm d})^{0.2}\ \rm K,
\end{eqnarray}

\noindent where $x_{\rm d} = 1.6\times10^{-8}$ is the grain abundance
and $H_X$ in erg~s$^{-1}$.

\subsection{H$_2$ vibrational heating/cooling}\label{H2vibcool}

When the vibrational levels of H$_2$ are populated by non-thermal
processes, thermal collisional quenching and excitation can result in
a net heating despite downward radiations. When non-thermal reactions
are not important, H$_2$ can be an important coolant. The resulting
collisional vibrational heating or cooling is given by

\begin{eqnarray}
\Gamma_{\rm H_2 vib, col} & = & \Sigma_{vj} n_{vj} \times \\ \nonumber
& &\Sigma_{v'j'} C(vj \rightarrow v'j') \times \\ \nonumber
& &(E_{vj} - E_{v'j'})\ {\rm erg\ cm^{-3}\ s^{-1}}
\end{eqnarray}

\noindent Where $C(vj \rightarrow v'j')$ is the total collision rate
from level $vj$ to $v'j'$ in units of s$^{-1}$. Radiative cooling due
to downward decay of the vibrational levels is given by

\begin{eqnarray}
\Lambda_{\rm H_2 vib, rad} = \Sigma_{vj} A(vj\rightarrow v'j') n_{vj}\ {\rm erg\ cm^{-3}\ s^{-1}}
\end{eqnarray}

\noindent The population of the vibrational levels is discussed in
Sect. \ref{vibpop}.

\section{Cooling processes}

\subsection{Fine-structure line cooling}\label{fine_cool}

Since most of the gas is atomic in the radical region, the dominant
coolants are the atomic fine-structure lines. The most prominent
cooling lines are the [CII] 158 $\mu$m and [OI] 63 $\mu$m and 146
$\mu$m lines. For the calculation of the thermal balance we also take
into account Si$^+$, C, Si, S, Fe and Fe$^+$. We use a compilation for
the collisional data from \citet{Sternberg1995},
\citet{Hollenbach1989}, \citet{Sampson1994}, \citet{Dufton1994},
\citet{Johnson1987}, \citet{Roueff1990}, \citet{Schroder1991},
\citet{Mendoza1983}, \citet{Chambaud1980} and \citet{Jaquet1992}. We
take into account collisions with electrons, H$^+$, H and H$_2$ (ortho
and para) for the excitation of the species to different levels. In
the PDRs, collisions with H$^+$ are not the dominant excitation source
but in XDRs the ionised fraction of hydrogen can be as large as ten
percent and become important for the excitation of some levels.

\subsection{Metastable-line cooling}\label{metacool}

We included the metastable cooling lines of C, C$^+$, Si, Si$^+$, O,
O$^+$, S, S$^+$, Fe and Fe$^+$. All the data is taken from
\citet{Hollenbach1989} except for Si$^+$ \citep{Dufton1994}, C$^+$
\citep{Sampson1994} and O$^+$ \citep{McLaughlin1993}.

\subsection{Recombination cooling}

At temperatures higher than $\sim 5000$~K, cooling due to
recombination of electrons with grains (PAHs) is important. The
cooling depends on the recombination rate which is proportional to the
product $n_{\rm e}\ n_{\rm H}$. The cooling rate increases when $G_0
T_{\rm k}^{0.5}/n_{\rm e}$ goes up, due to an increase in charge and
hence Coulomb interaction. \citet{Bakes1994} calculated numerically
the recombination cooling for a variety of physical conditions. An
analytical fit to the data is given by

\begin{eqnarray}
\Lambda &=& 3.49 \cdot 10^{-30} T_{\rm k}^{\alpha} (G_0 T_{\rm k}^{1/2}/n_{\rm e})^{\beta}\\ \nonumber
        & & n_{\rm e} n_{\rm H}\ {\rm erg\ s^{-1}\ cm^{-3}}
\end{eqnarray}

\noindent where $\alpha = 0.944$ and $\beta = 0.735/T_{\rm k}^{0.068}$. 

\subsection{Molecular cooling by H$_2$, CO and H$_2$O}

For the rotational and vibrational cooling of H$_2$, CO and H$_2$O, we
use the fitted rate coefficients of \citet{Neufeld1993} and
\citet{Neufeld1995}. They present a cooling rate for species $i$ through:

\begin{eqnarray}
\Lambda = L\ n(x_i)\ n({\rm H_2})\ {\rm erg\ cm^{-3}\ s^{-1}},
\end{eqnarray}

\noindent where $n({\rm H_2})$ and $n(x_i)$ are the densities of H$_2$
and species $x_i$, respectively. $L$ is given by

\begin{eqnarray}
\frac{1}{L} = \frac{1}{L_0} + \frac{\rm n(H_2)}{L_{LTE}} + 
\frac{1}{L_0} \left[\frac{{\rm n(H_2)}}{\rm n_{1/2}}\right]^\alpha 
\left(1 - \frac{{\rm n_{1/2}}}{L_{LTE}}\right).
\end{eqnarray}

\noindent We interpolate in the tables given by \citet{Neufeld1993}
and \citet{Neufeld1995}, to find the values $L_0$, $\rm n_{1/2}$ and
$L_{LTE}$ and $\alpha$. $L_0$ is the cooling rate coefficient in the
low density limit and $n_{1/2}$ is the H$_2$ density where $L$ has
fallen by a factor of two below $L_0$. $\alpha$ is chosen to minimize
the maximal fractional error in the fit at other densities. $L_0$ is a
function of temperature, and $L_{LTE}$, $n_{1/2}$, and $\alpha$ are
functions of temperature and the optical depth parameter
$\tilde{N}(x_i)$, which is given by the gradient $N(x_i)/\delta v_{\rm
d}$. $N(x_i)$ is the column density of the species $x_i$. To take into
account collisional excitation by electrons and atomic hydrogen, we
follow \citet{Yan1997} and replace $n({\rm H_2})$ by $n_{\rm rot}$ and
$n_{\rm vib}$. For H$_2$ rotational and vibrational cooling, $n_{\rm
rot}$ and $n_{\rm vib}$ are given by

\begin{eqnarray}
n_{\rm rot}({\rm H_2}) = n_{\rm vib}({\rm H_2}) = n({\rm H_2}) + 7 n({\rm H}) + 16 n({\rm e}).
\end{eqnarray}

\noindent For rotational cooling by CO, $n_{\rm rot}$ is given by

\begin{eqnarray}
n_{\rm rot}(\rm CO) &=& n({\rm H_2}) + 1.414n({\rm H})\sigma_{\rm H} / \sigma_{\rm H_2} \\ \nonumber
&&              + 1.3 \times 10^{-8}n({\rm e}) / \sigma_{\rm H_2} v,
\end{eqnarray}

\noindent where $\sigma_{\rm H}=2.3 \times 10^{-15}\ {\rm cm^{-2}}$,
$\sigma_{\rm H_2}=3.3\times 10^{-16}(T_{\rm k}/10^3))^{-1/4}\ {\rm cm^{-2}}$
and $v=1.03\times10^4 T_{\rm k}^{0.5}\ {\rm cm\ s^{-1}}$. For H$_2$O
rotational cooling, $n_{\rm rot}$ is given by

\begin{eqnarray}
n_{\rm rot}({\rm H_2O})& =& n({\rm H_2}) + 10 n({\rm H}) + \\ \nonumber
                       &  & n({\rm e})k_e(1,20,1.9,T_{\rm k})/k_{\rm H_2},
\end{eqnarray}

\noindent where $k_{\rm H_2}=7.4 \times 10^{-12} T_{\rm k}^{0.5}{\rm cm^3
s^{-1}}$ and $k_{\rm e}(i,b,d,T_{k})$ are the H$_2$ and electron impact
excitation rate coefficients, respectively. $k_e(i,b,d,T_{\rm k})$ for the excitation from level $i\rightarrow i+1$ in units of
$\rm cm^3\ s^{-1}$ is given by

\begin{eqnarray}
k_{\rm e}(i,b,d,T_{\rm k}) &=& \frac{3.56\times 10^{-6} d^2}{T_{\rm k}^{0.5} [2 - 1 / (i+1)]} 
\exp(\beta\Delta E) \times \\ \nonumber
&&\ln\left[C\Delta E + \frac{C}{\beta}\exp\left(\frac{-0.577}{1 + 2\beta\Delta E}\right)\right]
\end{eqnarray}

\noindent where $b$ is the rotational constant in $\rm cm^{-1}$, $d$
the dipole moment in Debye, $\beta=11600/T_{\rm k}$, $\Delta E = 2.48 \times
10^{-4} b(i+1)$ and $C$ is given by

\begin{eqnarray}
C&=&\frac{9.08 \times 10^3}{b(i+1)}\qquad \qquad \qquad \qquad d \le 1.53 \\ \nonumber
C&=&\frac{1.93 \times 10^4}{db(i+1)} \exp(-1.18 / d^3) \ \ \quad  d > 1.53
\end{eqnarray}

\noindent For CO vibrational cooling, $n_{\rm vib}$ is given by

\begin{eqnarray}
n_{\rm vib}({\rm CO}) = n({\rm H_2}) + 50 n({\rm H}) + n({\rm e}) L_{\rm CO,e} / L_{\rm CO,0}
\end{eqnarray}

\noindent where 

\begin{eqnarray}
L_{\rm CO,e} &=& 1.03 \times 10^{-10} (T_{\rm k}/300)^{0.938} \exp(-3080 / T_{\rm k})\\ \nonumber
L_{\rm CO,0} &=& 1.14 \times 10^{-14} \exp(-68.0 / T_{\rm k}^{1/3}) \exp(-3080 / T_{\rm k})
\end{eqnarray}

\noindent For H$_2$O vibrational cooling, $n_{\rm vib}$ is given by

\begin{eqnarray}
n_{\rm vib}(\rm H_2O) = n({\rm H_2}) + 10 n({\rm H}) + n({\rm e}) L_{\rm H_2O,e} / L_{\rm H_2O,0}
\end{eqnarray}

\noindent where 

\begin{eqnarray}
L_{\rm H_2O,e} &=& 2.6 \times 10^{-6} T_{\rm k}^{-0.5} \exp(-2325 / T_{\rm k})\\ \nonumber
L_{\rm H_2O,0} &=& 0.64 \times 10^{-14} \exp(-47.5 / T_{\rm k}^{1/3}) \exp(-2325 / T_{\rm k})
\end{eqnarray}

In the XDR models, H$_2$ vibrational cooling is treated differently,
since non-thermal processes play an important role, which is discussed
in Sect. \ref{H2vibcool}.


\subsection{Cooling by electron impact with H}

The cooling due to the excitation of hydrogen is important at
temperatures $T > 5000$ K. The cooling rate is given by
\citet{Spitzer1978}:

\begin{eqnarray}
\Lambda_{\rm {\bf e-H}} &=& 7.3\times 10^{-19} n_{\rm e}\ n({\rm H}) \\ \nonumber
&&\times \exp(-118400 / T_{\rm k})\ {\rm erg\ cm^{-3} s^{-1}}.  
\end{eqnarray}







\section{Chemistry}

For most of the chemical reaction rates, we make use of the UMIST
database for astrochemistry by \citet{leTeuff2000}. In de PDR model we
use a network containing all the species with a size up to 6
atoms. For the XDR model we use all species with sizes up to 3 atoms
and some of 4 atoms. These species are taken from
\citet{Yan1997}. Below we discuss the additional reactions.


\subsection{H$_2$ formation on dust grains}

The formation of H$_2$ is very efficient over a wide range of
temperatures. It was already shown by \citet{Gould1963} that H$_2$ is
not formed efficiently in the gas phase. Most of the formation, which
is still not very well understood, takes place on grain surfaces
\citep{Hollenbach1971}. Recently, \citet{Cazaux2002, Cazaux2004}
developed a model for the formation of hydrogen under astrophysically
relevant conditions. They compared their results with the laboratory
experiments by \citet{Pirronello1999} and \citet{Katz1999}. They find
a recombination rate of

\begin{eqnarray} 
R_{\rm H_2} & = & 0.5\ n_{\rm H}\ v_{\rm H}\ n_{\rm d}\ \sigma_{\rm d}\ \epsilon_{\rm H_2} S_{\rm H}(T_{\rm k}) \\ \nonumber  
        & \approx & 6 \times 10^{-17}\ (T_{\rm k}/300)^{0.5} n_{\rm H}\ n\ \epsilon_{\rm H_2} S(T_{\rm k})\ {\rm cm}^{-3}\ {\rm s}^{-1},
\label{eqTH}
\end{eqnarray}

\noindent where $n_{\rm d}$ and $\sigma_{\rm d}$ are the volume density and cross
section of dust grains and $n_{\rm H}$, $v_{\rm H}$ and $S(T_{\rm k})$ are the volume
density, thermal velocity and thermally averaged sticking coefficient
of hydrogen atoms. We use the sticking coefficient given by
\citet{Hollenbach1979}

\begin{eqnarray}
S(T_{\rm k}) & = [1 + 0.4 (T_{\rm k} + T_{\rm d})^{0.5} + & 2 \times 10^{-3}\ T_{\rm k} + \\ \nonumber
     &                              & 8 \times 10^{-6}\ T_{\rm k}^2]^{-1},
\end{eqnarray}

\noindent where $T_{\rm d}$ is the dust temperature. Eq. (\ref{eqTH}) is the
same as eq. (4) in \citet{Tielens1985}, except for the term
$\epsilon_{\rm H_2}$, the recombination efficiency, which is given by

\begin{eqnarray}
\epsilon_{\rm H_2} = \left(\frac{\mu F}{2 \beta_{\rm H_2}} + 1 + \frac{\beta_{{\rm H}_p}}{\alpha_{pc}}\right)^{-1},
\end{eqnarray}

\noindent where $\mu$ is the H$_2$ fraction that stays on the surface
after formation, $\beta_{\rm H_2}$ and $\beta_{{\rm H}_p}$ are the
desorption rates of molecular hydrogen and physisorbed hydrogen atoms,
respectively, $F$ is the flux of hydrogen atoms and $\alpha_{pc}$ is
the evaporation rate from physisorbed to chemisorbed sites. These
three terms dominate in different temperature regimes. See
\citet{Cazaux2002,Cazaux2004} for a more detailed discussion.

\subsection{Additional reactions in PDR models}

\subsubsection{Recombination on PAHs}

Collisions of electrons and ions with grains can become an important
recombination process in dense clouds of low ionisation. We include
reactions with PAHs following Sect. 5 of \citet{Wolfire2003} in the
PDR models. The C$^+$/C transition occurs at larger column densities
when PAHs are included. Deep into the cloud, the electron abundance is
reduced by several orders of magnitude.

\subsubsection{Vibrational excitation of H$_2$}\label{H2excit}

In PDRs, molecular hydrogen can be excited by absorption of FUV
photons in the Lyman-Werner bands. Fluorescence leads to dissociation
in about $10\%$ in of the cases (see Field et
al. \citeyear{Field1966}; Stecher \& Williams \citeyear{Stecher1967}),
and in the remaining $90\%$ of the cases to a vibrationally excited
state of the ground electronic state \citep{Black1976}. To simplify
matters, we treat the electronic ground state as having a vibrational
ground state and a single excited vibrational
state. \citet{London1978} found that the effective quantum number for
this pseudo-level is $v = 6$, and the effective energy is $E_*/\rm k =
2.6\ {\rm eV}/\rm k = 30163$~K. We treat excited molecular hydrogen,
H$_2$V, as a separate species in our chemistry. H$_2$V can be
destroyed by direct FUV dissociation, radiative decay or collisional
de-excitation, and chemical reactions with other species. Since
vibrational decay is a forbidden process, a large abundance of H$_2$V
can be maintained. H$_2$V can react with other species with no
activation barrier or a reduced one. In the UMIST database, the rates
for a reaction between two species are parameterised as

\begin{eqnarray}
R = \alpha \ (T_{\rm k}/300)^{\beta} \ \exp(-\gamma\ /T_{\rm k})\ {\rm cm^3\ s^{-1}}.
\end{eqnarray}

\noindent For reactions with H$_2$V, $\gamma$ is replaced by
$\gamma^*$ = max(0.0, $\gamma$ - 30163). When reactions have an
activation barrier lower than 2.6~eV, the barrier is set to zero. When
the barrier is larger than 2.6~eV, the barrier is reduced by
2.6~eV. \citet{Tielens1985} state that for important reactions such as

\begin{eqnarray}
{\rm H_2V + C^+ \rightarrow CH^+ + H} \nonumber
\end{eqnarray}

\noindent and

\begin{eqnarray}
{\rm H_2V + O \rightarrow OH + H}, \nonumber
\end{eqnarray}

\noindent this is a good approximation since the activation barrier of
$\sim 0.5$~eV is a lot smaller than the vibrational excitation energy
of $2.6$~eV. For reactions with barriers of the same order or larger
one can overestimate the reaction rates.

\subsubsection{Shielding of H$_2$ and CO}\label{shielding}

In PDRs, the photo-dissociation rate of both H$_2$ and CO is
influenced by line as well as continuum absorption. The dissociation
rate of H$_2$ is decreased by self-shielding. For an H$_2$ line
optical depth $\tau \leq 10$, we adopt the self-shielding factor given
by \citep{Shull1978}.
When the line absorption is dominated by the Doppler cores
or the Lorentz wings (i.e., $\tau > 10$), we use the self-shielding
factor as given by \citet{deJong1980}.
The CO photo-dissociation rate is decreased by both CO self-shielding
and H$_2$ mutual shielding. We use Table 5 of \citet{vanDishoeck1988},
to determine the shielding factor as a function of column densities
$N({\rm H_2})$ and $N({\rm CO})$. 


\subsection{Additional reactions in XDRs}

\subsubsection{Primary ionisations}

In the XDRs we do not use the photo-ionisation rates from
UMIST. X-rays are absorbed in K-shell levels releasing an electron. An
electron from a higher level may fill the empty spot and with the
energy surplus another so called Auger electron is ejected. This
process leads to multiply ionised species. Due to charge transfer with
H, H$_2$ and He, they are quickly reduced to the doubly ionised
state. We therefore assume that the ionisation by an X-ray photon
leads to a doubly ionised species, as does absorption of an X-ray
photon by a singly ionised species. When rates for charge transfer
with H and He are very fast, elements are quickly reduced to singly
ionised atoms, which is the case for O$^{2+}$, Si$^{2+}$ and
Cl$^{2+}$. Therefore, we add only O$^{2+}$ to the chemical network to
represent them. We assume that Si and Cl get singly ionised after
absorbing an X-ray photon. We also include C$^{2+}$, N$^{2+}$,
S$^{2+}$ and Fe$^{2+}$. The direct (or primary) ionisation rate
of species $i$ at a certain depth $z$ into the cloud is given by

\begin{eqnarray}
\zeta_{i,\rm prim} = \int_{E_{\rm min}}^{E_{\rm max}} \sigma_i(E) \frac{F(E,z)}{E} dE,
\end{eqnarray}

\noindent where the ionisation cross sections $\sigma_i$ are taken
from \citet{Verner1995}.

\subsubsection{Secondary ionisations}

Part of the kinetic energy of fast photoelectrons is lost by
ionisations. These secondary ionisations are far more important for H,
H$_2$ and He than direct ionisation. \citet{Dalgarno1999} calculate
the number of ions $N_{\rm ion}$ produced for a given species $i$. For
a given electron energy $E$, $N_{\rm ion}$ is given by

\begin{eqnarray}
N_{\rm ion} = E/W,
\end{eqnarray}

\noindent where $W$ is the mean energy per ion
pair. \citet{Dalgarno1999} calculated $W$ for pure ionised H-He and
H$_2$-He mixtures and parameterised $W$ as:

\begin{eqnarray}
W=W_0(1+cx^\alpha),
\label{WorkFunction}
\end{eqnarray}
 
\noindent where $W_0$, $c$ and $\alpha$ are given in Table 4 of their
paper. The corrected mean energies for ionisation in the H-H$_2$-He
mixture are given by

\begin{eqnarray}
W(\rm H^+) = W_{H,He}(H^+)\left[[1+1.89\frac{n(\rm H_2)}{n(\rm H)}\right],\\
W(\rm H_2^+) = W_{H_2,He}(H_2^+)\left[1+0.53\frac{n(\rm H)}{n(\rm H_2)}\right].
\end{eqnarray} 

\noindent The ionisation rate at depth $z$ into the cloud for species
$i$ is then given by

\begin{eqnarray}
\zeta_{i,\rm sec} &=& \int_{E_{\rm min}}^{E_{\rm max}} \sigma_{\rm pa}(E) F(E,z) \frac{E}{W} dE\ \ {\rm s^{-1}\ per\ H\ nucleus} \\ \nonumber
\end{eqnarray}

\noindent We rewrite this to a rate dependent on the fractional abundance of the species $x_i$:

\begin{eqnarray}
\zeta_{i,\rm sec} &=& \int_{E_{\rm min}}^{E_{\rm max}} \sigma_{\rm pa}(E) F(E,z) \frac{E}{W x_i} dE\ \ {\rm s^{-1}\ per\ species}\ i,
\end{eqnarray}

\noindent where $x_i$ is the fraction of species $i$. Since we
integrate over the range 1-10 keV and $W$ goes to a limiting value, we
use the parameters applicable to the 1 keV electron. The
ionisation rate then simplifies to:

\begin{eqnarray}
\zeta_{i,\rm sec} &=& \frac{1\ {\rm keV}}{W(1\ {\rm keV}) x_i} \int_{E_{\rm min}}^{E_{\rm max}} \sigma_{\rm pa}(E) F(E,z) dE \\ \nonumber
        &=& \frac{1\ {\rm keV}}{W(1\ {\rm keV}) x_i} H_X \ {\rm s^{-1}\ per\ species}\ i.
\end{eqnarray}

\noindent We also include secondary ionisations for C, N, O, Si, S,
Cl, Fe, C$^+$, N$^+$, O$^+$, S$^+$ and Fe$^+$. We scale the ionisation
rate of these species to that of atomic hydrogen by

\begin{eqnarray}
\zeta_i = \zeta_{\rm H} \frac{\sigma_{{\rm ei},i}}{\sigma_{{\rm ei},{\rm H}}}\ {\rm s^{-1}}.
\end{eqnarray}

\noindent We integrate over the range 0.1-1.0 keV to get an average
value of the electron impact ionisation cross section
$\sigma_{\rm ei}$. Using the experimental data fits of
\citet{Lennon1988}. The scaling factors $\sigma_{{\rm ei},i}/\sigma_{{\rm ei},{\rm
H}}$ for C, N, O, Si, S, Cl, Fe, C$^+$, N$^+$, O$^+$, S$^+$, and
Fe$^+$ are 3.92, 3.22, 2.97, 6.67, 6.11, 6.51, 4.18, 1.06, 1.24, 1.32,
1.97, and 2.38, respectively.

\subsubsection{FUV photons from secondary electrons}

When energetic electrons created in X-ray ionisations collide with
atomic and molecular hydrogen, H$_2$ Lyman-Werner and H Lyman $\alpha$
photons are produced, which can significantly affect the
chemistry. The photoreaction rate $R_i$ per atom or molecule of
species $i$ is given by

\begin{eqnarray}
R_i = \frac{x_{\rm H_2} \zeta_{\rm H_2} p_m + x_{\rm H} \zeta_{\rm H} p_a}{1 - \omega}\ {\rm s^{-1}}.
\end{eqnarray}

\noindent The values of $p_a$ are taken from table 4.7 of
\citet{Yan1997} and values of $p_m$ are the rates for cosmic-ray
induced reactions from \citet{leTeuff2000}. There is an exception for
CO, however, where we take the rate, corrected for self-shielding,
given by \citet{Maloney1996}:

\begin{eqnarray}
R_{\rm CO} = 2.7 x_{\rm CO}^{-1/2} (T_{\rm k}/1000)^{0.5} \zeta_{\rm H_2} x_{\rm H_2}\ {\rm s^{-1}}.
\end{eqnarray}

\subsubsection{Vibrationally excited H$_2$}\label{vibpop}

Vibrationally excited H$_2$ can enhance reactions with an activation
barrier and also be an important heating or cooling source. To
calculate the populations of the vibrational levels of H$_2$, we take
into account:

\begin{itemize}
\item{Collisions with fast electron produced by X-ray photo-ionisation.}
\item{Collisions with thermal electrons, H, H$_2$ and He.}
\item{Chemical destruction and production in chemical reactions.}
\item{Radiative decay.}
\end{itemize}

\noindent We use the results of \citet{Dalgarno1999} to calculate the
X-ray induced excitation to the vibrational levels $v=1$ and
$v=2$. The ratio of the yields Y($v=2$)/Y($v=1$) is about
0.070. Excitation to higher levels is not taken into account, since
the yield to higher levels decreases very rapidly. First we calculate
the mean energy for excitation, $W$, in the H$_2$-He mixture. The
parameters are listed in Table 5 of \citet{Dalgarno1999}. The function
$W$ has the same form as equation (\ref{WorkFunction}). The mean
energy for excitation also depends on the abundances of H and
H$_2$. The yield has to be corrected with a factor $C$(H,H$_2$), which
is given by

\begin{eqnarray} 
C(\rm H, H_2) &=& \frac{2n({\rm H_2})}{n({\rm H}) + 2n({\rm H_2})},\qquad x \ge 10^{-4} \\ \nonumber
C(\rm H, H_2) &=& \frac{n({\rm H_2})/n({\rm H})}{n({\rm H_2})/n({\rm H}) + a(x)},\ 10^{-7} < x < 10^{-4}
\end{eqnarray}

\noindent where $a(x) = 0.5 (x/10^{-4})^{0.15}$. The rates
for excitation by thermal electrons are taken from \citet{Yan1997},
who finds that the transitions rates for H$_2(v=0)$ to H$_2(v=1,2)$
are given by

\begin{eqnarray}
R(0\rightarrow 1) &=& 9.7 \times 10^{-11} (T_{\rm k}/300)^{0.87} \exp(-6140 / T_{\rm k}) \\ 
R(0\rightarrow 2) &=& 7.5 \times 10^{-12} (T_{\rm k}/300)^{0.91} \exp(-11900 / T_{\rm k}).
\end{eqnarray}

\noindent The excitation rate for the transition $v \rightarrow v+1$
is taken to be $v$ times the $0\rightarrow 1$ rate. Excitations with
$\Delta v > 1$ are not taken into account. The quenching rates are
calculated through detailed balance. The quenching rates from $v
\rightarrow v'$ by atomic hydrogen are given in table 4.2 of
\citet{Yan1997} and are of the form:

\begin{eqnarray}
R(v\rightarrow v') = \alpha (T_{\rm k}/300)^\beta \exp(-\gamma / T_{\rm k})\ {\rm cm^3 s^{-1}}.
\label{eqnform}
\end{eqnarray}

\noindent The excitation rates are obtained by detailed balance. For
the molecular excitation and quenching rates we use the results of
\citet{Tine1997}. Collisions where either before or after one of the
H$_2$ molecules is in the $v=0$ state are considered. The rate
coefficients are of the form:

\begin{eqnarray}
\log_{10}R(v_1,v'_1; v_2, v'_2) = A + B / T_{\rm k} + C \log_{10} T_{\rm k},
\end{eqnarray}

\noindent and are given in table 1 of \citet{Tine1997}, who also
considered collisions with He. They give a rate coefficient for the
$v=1\rightarrow0$ transition:

\begin{eqnarray}
\log_{10}R(1\rightarrow0) &=& -8.8T_{\rm k}^{-1/3} - 16.5  \qquad  T_{\rm k} \le 90 \rm K\\ \nonumber
                            &=& -18.9T_{\rm k}^{-1/3} - 14.2 \quad  90 < T_{\rm k} \le 230 \rm K\\ \nonumber
                            &=& -47.4T_{\rm k}^{-1/3} - 9.4  \qquad  T_{\rm k} > 230 \rm K. 
\end{eqnarray}

\noindent For the other transitions with $\Delta v=1$, the same rates
are used. The upward transitions can be obtained by detailed balance.
\citet{Yan1997} also calculated the dissociation and ionisation rates
by thermal electrons and since the ionisation threshold is much higher
than the vibrational energies one rate is used for all vibrational energies:

\begin{eqnarray}
R_{\rm e, diss} = 7.03 \times 10^{-9} (T_{\rm k} / 300)^{0.41} \exp(-118600 / T_{\rm k}) \\ 
R_{\rm e, ion} =  8.9 \times 10^{-10} (T_{\rm k} / 300)^{0.57} \exp(-179400 / T_{\rm k})
\end{eqnarray}

\noindent The dissociation rates by atomic hydrogen are given in table
4.3 of \citet{Yan1997}, which are of the same form as equation
(\ref{eqnform}).  For the dissociation rates by H$_2$ we use the results
of \citet{Lepp1983}, which are given by

\begin{eqnarray}
R_{\rm H_2, diss} &=& 6.29 \times 10^{-15} \times \\ \nonumber
         & & A \exp(1.44 v - 0.037 v^2)f(T_{\rm k}) / f(4500\rm K),
\end{eqnarray}

\noindent where $A=1.38$, $f(T_{\rm k})=T_{\rm k}^{0.5} \alpha \exp(-\alpha)$,
$\alpha=[1 + (E_{\rm th} + 1) / kT_{\rm k}]$ and $E_{\rm th} = 4.48 eV - E(v)$. For
the dissociative attachment reaction:

\begin{eqnarray}
\rm H_2 + e \rightarrow H + H^-,\nonumber
\end{eqnarray}

\noindent we use the results of \citet{Wadehra1978} and the reaction
rates have the same form as equation (\ref{eqnform}). Vibrationally
excited H$_2$ can be destroyed in chemical reactions. Endothermic
reactions with vibrationally excited H$_2$ can lower the activation
barrier, by using the energy of the vibrational level. The barrier is
reduced, but cannot become negative: $E' = {\rm min}(0, E - E(v))$.
When H$_2$ is formed in chemical reactions which are exothermic, part
of the formation energy goes into the excitation of the vibrational
levels. Formation of H$_2$ on grains can play a very significant
role. H$_2$ has a binding energy of 4.48 eV. Following
\citet{Sternberg1989}, we assume one third of this energy to be
distributed statistically over all the vibrational levels:

\begin{eqnarray}
x({\rm H_2}(v)) = \frac{\exp(-E(v) / 1.493)}{\Sigma_v \exp(-E(v) / 1.493)},
\label{statweight}
\end{eqnarray}

\noindent where $x({\rm H_2}(v))$ is the fraction of H$_2$ formed in
vibrational state $v$. When H$_2$ formation reactions are endothermic,
all the H$_2$ is in the ground vibrational state. When they are
exothermic part of the energy is distributed statistically following
equation (\ref{statweight}). The Einstein A coefficients for radiative
decay are taken from \citet{Turner1977}. We take a weighted average
over the rotational levels of each vibrational level, which we assume
to be thermalised, to get an Einstein A coefficient for the decay from
$v \rightarrow v'$.

\section{Energy deposition rate per hydrogen nucleus}\label{energy_dep}

The photon energy absorbed per hydrogen nucleus, $H_{X}$, is given by

\begin{eqnarray}
H_X = \int_{E_{\rm min}}^{E_{\rm max}} \sigma_{\rm pa}(E) F(E,z) dE.
\end{eqnarray}

\noindent The interval [$E_{\rm min}$,$E_{\rm max}$] is the spectral
range where the energy is emitted. The photoelectric absorption cross
section per hydrogen nucleus, $\sigma_{\rm pa}$, is given by

\begin{eqnarray} 
\sigma_{\rm pa}(E) = \sum_i \mathcal{A}_i({\rm total}) \sigma_i(E).
\end{eqnarray}

\noindent \citet{Morrison1983} state that the X-ray opacity is
independent of the degree of depletion onto grains. Therefore, we take
the total (gas and dust) elemental abundances, $\mathcal{A}_i({\rm
total})$, as given in Table \ref{XDRabundances} to calculate
$\sigma_{\rm pa}$. The X-ray absorption cross sections, $\sigma_i$, are
taken from \citet{Verner1995}.
The flux $F(E,z)$ at depth $z$ into the cloud is given by

\begin{eqnarray} 
F(E,z) = F(E,z=0)\ \exp(-\sigma_{\rm pa}(E)\ N_{\rm H}),
\end{eqnarray}

\noindent where $N_{\rm H}$ is the total column of hydrogen nuclei and
$F(E,z=0)$ the flux at the surface of the cloud.


\end{document}